\documentclass{article}

\usepackage{arxiv}

\usepackage[utf8]{inputenc} 
\usepackage[T1]{fontenc}    
\usepackage{amsmath,amssymb,amsthm}
\usepackage{mathtools}
\usepackage{amsfonts}       
\usepackage[numbers]{natbib} 
\usepackage{url}            
\usepackage{booktabs}       
\usepackage{longtable}
\usepackage{nicefrac}       
\usepackage{microtype}      
\usepackage{graphicx}
\usepackage{placeins}
\usepackage{float}
\usepackage{hyperref}       
\usepackage[capitalise,noabbrev]{cleveref} 
\graphicspath{ {./images/} {./figures/} }

\newcommand{\R}{\mathbb{R}}
\newcommand{\E}{\mathbb{E}}
\newcommand{\Lgen}{\mathcal{L}}
\newcommand{\drift}{b}
\newcommand{\diff}{a}
\newcommand{\Kj}{K_j}

\newtheorem{assumption}{Assumption}

\title{Symbolic Weak-form Recovery of 2-D Stochastic Generators}

\author{
 Sai Sathvik Gullipalli \\
  Independent Researcher\\
  \texttt{saisathvik127@gmail.com} \\
   \And
 Eshwar R A \\
  Independent Researcher\\
  \texttt{eshwarra5@gmail.com} \\}

\begin{document}
\maketitle
\begin{abstract}
Recovering two-dimensional It\^o generators from trajectory data is difficult
because drift increments have low signal-to-noise, bivariate weak designs can be
ill-conditioned, and unconstrained tensor estimates need not be positive
semidefinite. We study WG-SINDy estimator combining
covariance-shaped spatial kernels, a ridge-stabilized local-polynomial
projection, adaptive-LASSO/STLSQ selection, one in-sample per-component
feasible diagonal GLS pass, and a PSD projection--Cholesky read-out with mild
isotropic shrinkage. The released estimator uses a data-dependent full-cloud smoother and one
in-sample per-component feasible diagonal GLS pass; accordingly, we do not
claim exact finite-sample martingale cancellation or a feasible-GLS efficiency
theorem for the reported implementation. We evaluate the
estimator on 29 synthetic two-dimensional systems: 19 meet their declared
per-system recovery contracts, eight are retained as named limits, and two
remain scoped reviews. Across the 19 PASS rows, the median central-grid drift
metric is 0.204 and the median tensor error is
0.0397. Among the six systems with a finite,
non-degenerate off-diagonal target, the median $a_{12}$ cosine is 0.997.
Positive-semidefinite validity is imposed by construction.  These results are
synthetic, in-sample sampled-region diagnostics and do not establish universal
or real-data recovery.
\end{abstract}


\section{Introduction}\label{sec:intro}
A central problem across quantitative finance, molecular dynamics, neuroscience,
and climate modelling is the \emph{inverse problem} for stochastic dynamics:
given a discretely sampled trajectory of a system, reconstruct the stochastic
differential equation that generated it. The object that fully characterises a
diffusion's local dynamics is its infinitesimal generator $\Lgen f =
\drift\!\cdot\!\nabla f + \tfrac{1}{2}\,\diff\!:\!\nabla^2 f$, encoding drift
$\drift$ and diffusion tensor $\diff = \sigma\sigma^\top$. Recovering $\Lgen$
symbolically (rather than as a black-box surrogate) exposes spectral gaps, escape
rates, leverage correlations, and drift-curl diagnostics directly.

The weak-form spatial-kernel estimator of Eshwar and
Honnavar~\citep{eshwar2026weak} addresses this problem in one dimension. For
fixed or independently constructed spatial weight functions, projecting
increments onto functions of the current state preserves the usual martingale
cancellation: the weight evaluated at $X_{t_n}$ is
$\mathcal{F}_{t_n}$-measurable, so its product with the next Brownian increment
has mean zero. The released estimator additionally learns its normalization,
centres, bandwidths, and local-polynomial projection from the sampled state
cloud; exact finite-sample cancellation is therefore not asserted for that
data-dependent smoother.

Extending this to genuinely two-dimensional, coupled, non-diagonal diffusions is
harder than it looks. A component-wise port of the 1D estimator to 2D (the
``naive port'' any reader would try first) fails on three fronts: the design
matrix built from raw bivariate monomials over a scale-disparate, localised state
cloud is severely ill-conditioned; the unweighted normal equations are
statistically inefficient at low drift signal-to-noise; and the recovered
diffusion tensor need not be positive semidefinite. We show these failures
explicitly (\Cref{sec:naive}) and address them with targeted modifications.

\paragraph{Contributions}
(i) Our algorithm is a weak-form 2D generator estimator that recovers both drift components and all three independent
entries of the diffusion tensor symbolically inside the declared scope (\Cref{sec:method}). (ii) We separate the
fixed-weight spatial martingale identity from the released estimator, whose standardization, kernel geometry,
local-polynomial projection, and feasible diagonal GLS weights are learned from the sampled state cloud. An
independently constructed smoother-and-weight variant would recover exact cancellation, but that variant is not used
in the reported experiments. (iii) The ablation identifies strong in-scope contributions from pooling, feasible GLS, and
local-polynomial smoothing; the remaining components provide structural guarantees or protection in stress regimes
(\Cref{sec:ablation}). (iv) We evaluate the frozen estimator on 29 2D systems with per-system datasheets
(\Cref{app:datasheets}) and an explicit falsification boundary.

\subsection{Scope of claims}\label{sec:scope}
We claim recovery over a broad class of \emph{identifiable} 2D It\^o diffusions: those whose drift and diffusion
lie in the chosen feature library, whose trajectories cover the state space, and whose design matrix has full
column rank. We do \emph{not} claim universal 2D recovery. We name, and do not paper over, the limits where
recovery fails: the risk-neutral log-price drift $\mu-\tfrac12 v$ of stochastic-volatility models
(a low per-step signal whose squared signal-to-noise can be near $10^{-4}$ in the simulated regime), near-singular and rank-deficient diffusion
tensors, Feller-violating boundaries, non-spanning libraries, and under-covered state spaces. The diffusion
tensor (not $\sigma$ itself) is the identifiable object, since $\sigma$ and $\sigma O$ for any orthogonal $O$
produce the same $\diff=\sigma\sigma^\top$.

\subsection{Code and data availability}\label{sec:code}

The repository contains the estimator, experiment code, checked-in aggregate CSVs, and manuscript assets. Full raw
campaign outputs and the external baseline snapshot are not included in this release; consequently, the current package
supports audit of the checked-in aggregates but not bit-for-bit regeneration of every historical result. The repository
is available at:

\href{https://github.com/Sathvik-Gullipalli/Symbolic-Weak-form-Recovery-of-2-D-Stochastic-Generators}{https://github.com/Sathvik-Gullipalli/Symbolic-Weak-form-Recovery-of-2-D-Stochastic-Generators}.

\section{Background}\label{sec:background}

\subsection{It\^{o} stochastic differential equations}\label{sec:ito_sde}

We consider a two-dimensional state $X_t = (X_t^{(1)}, X_t^{(2)})^\top \in \R^2$
evolving as an It\^{o} diffusion,
\begin{equation}
    \mathrm{d}X_t = \drift(X_t)\,\mathrm{d}t + \sigma(X_t)\,\mathrm{d}W_t,
    \label{eq:ito_sde}
\end{equation}
where $\drift:\R^2\to\R^2$ is the drift vector, $\sigma:\R^2\to\R^{2\times m}$
is the noise-loading matrix, and $W_t$ is an $m$-dimensional Brownian motion.
The drift governs the average direction of motion; the diffusion part controls
the random fluctuations around it.

The local noise covariance is the diffusion tensor
$\diff(x) = \sigma(x)\sigma(x)^\top$, which in two dimensions is the symmetric
matrix
\[
    \diff(x)
    =
    \begin{pmatrix}
        \diff_{11}(x) & \diff_{12}(x) \\
        \diff_{12}(x) & \diff_{22}(x)
    \end{pmatrix}.
\]
The diagonal entries $\diff_{11}$, $\diff_{22}$ are the local variances of the
two coordinates. The off-diagonal entry $\diff_{12}$ measures instantaneous
coupling between the two noise shocks; in a stochastic volatility model it
carries the leverage dependence between return and variance shocks.

Recovering the system from trajectory data means identifying five scalar
functions: $\drift_1$, $\drift_2$, $\diff_{11}$, $\diff_{12}$, and $\diff_{22}$.

\subsection{From one-dimensional to two-dimensional
  SDEs}\label{sec:one_d_vs_two_d}

In one dimension the recovery problem has only two unknowns: a scalar drift
$b(x)$ and a local variance $a(x) = \sigma^2(x)$. The two-dimensional case is
structurally different. The drift becomes a vector of two functions, each
potentially depending on both state variables, so the motion of one coordinate
can influence the other. The noise becomes a full $2\times 2$ tensor rather than
a scalar.

The off-diagonal entry $\diff_{12}$ has no analogue in scalar SDEs and cannot be
recovered from marginal variances alone. A method that works for a 1D SDE does
not automatically handle the 2D problem: the estimator must recover multiple
drift components, cross-variation terms, and a diffusion tensor that qualifies
as a covariance object.

\subsection{The infinitesimal generator}\label{sec:generator_background}

The infinitesimal generator of~\eqref{eq:ito_sde} acts on smooth test functions
$f:\R^2\to\R$ by
\begin{equation}
    \Lgen f
    =
    \drift_1\partial_1 f
    +
    \drift_2\partial_2 f
    +
    \tfrac{1}{2}
    \bigl(
        \diff_{11}\partial_{11}f
        +
        2\diff_{12}\partial_{12}f
        +
        \diff_{22}\partial_{22}f
    \bigr).
    \label{eq:generator}
\end{equation}
The first-order terms are controlled by the drift; the second-order terms by the
diffusion tensor. Recovering $\Lgen$ symbolically means identifying which
candidate library functions $\{\theta_k\}$ appear in each coefficient, under a
sparsity assumption on those coefficients.

\subsection{Why direct increment regression is
  difficult}\label{sec:direct_regression_difficulty}

The natural approach is to regress increments $\Delta X_n / \Delta t$ directly
on library features. The problem is that the Brownian noise term scales as
$1/\sqrt{\Delta t}$, so a single increment is a very noisy observation of the
drift, and the noise grows as the sampling interval shrinks. Products of
increments, used for the diffusion tensor, carry similar bias from finite-step
effects and observation noise.

Weak-form methods avoid this by averaging many increments against spatial test
functions. A properly chosen spatial weight reduces the martingale noise while
preserving the identity linking averaged increments to the drift and diffusion
coefficients.

\subsection{The 1D weak-form spatial-kernel
  estimator}\label{sec:weak_form_1d}

The estimator of \citet{eshwar2026weak} places Gaussian kernels $\Kj$ over the
observed state range. Each kernel acts as a soft local bin, weighting increments
by proximity to its centre. Drift and diffusion coefficients are then recovered
by solving a shared linear system built from these weighted sums.

The reason this works is adaptedness: $\Kj(X_{t_n})$ depends only on
information up to time $t_n$ and is $\mathcal{F}_{t_n}$-measurable. The future
Brownian increment is independent of $\mathcal{F}_{t_n}$, so
\begin{equation}
    \E\!\left[
        \Kj(X_{t_n})\sigma(X_{t_n})(W_{t_{n+1}}-W_{t_n})
        \,\middle|\,
        \mathcal{F}_{t_n}
    \right]
    =
    0.
    \label{eq:tower_property}
\end{equation}
This martingale cancellation removes the Brownian noise without the bias that
arises when test functions depend on future path values. In practice, finite-step
bias correction and lag-one observation-noise correction are also applied.

\subsection{What changes in two dimensions}\label{sec:two_dimensional_extension}

Extending to two dimensions is not simply running the same regression twice. The
off-diagonal coefficient $\diff_{12}$ controls instantaneous cross-noise
coupling and is invisible to methods that estimate only marginal variances. A
valid extension must preserve the martingale
identity~\eqref{eq:tower_property}, recover all five generator coefficient
fields, and return a diffusion tensor that is positive semidefinite.

Our algorithm keeps the spatial weak-form identity and extends the regression target
to the full two-dimensional tensor, remaining derivative-free, sparse, symbolic,
and generator-based.

\subsection{Related methods}\label{sec:related_methods}

Sparse equation discovery is closely associated with SINDy, which uses sparse
regression over a candidate function library to identify parsimonious governing
equations~\citep{brunton2016sindy}. Our regression stage also uses sequential
thresholding and adaptive $\ell_1$ regularization; the latter follows the
adaptive-LASSO principle of assigning coefficient-specific penalty weights from
an initial estimate~\citep{tibshirani1996lasso,zou2006adaptive}. The
local-polynomial projection used to reduce boundary bias is related to classical
local-polynomial regression and smoothing~\citep{fan1996local}.

For stochastic systems, Kramers--Moyal and stochastic-SINDy estimators use
conditional increment moments to estimate drift and
diffusion~\citep{boninsegna2018sparse}. These estimators
provide natural local-moment baselines, although single-step drift targets remain
noisy because the Brownian contribution dominates the $O(\Delta t)$ drift signal
at small sampling intervals. Weak SINDy reduces derivative sensitivity through
weak formulations~\citep{messenger2021weak}, but its standard formulation was
developed for deterministic dynamics and does not directly impose the geometry
of a full two-dimensional diffusion tensor.

Extended dynamic mode decomposition and generator-EDMD methods approximate
Koopman or generator action on a chosen
dictionary~\citep{klus2020gedmd}. Their primary target is an
operator representation rather than a jointly sparse symbolic drift vector and
positive-semidefinite diffusion tensor. The present method instead targets the
five generator coefficient fields while imposing a PSD tensor read-out.

The comparisons reported below use lightweight in-repository implementations of
these methodological families. They are diagnostic baseline proxies, not
reproductions certified by the original authors, and their trajectory budgets are
not identical. Consequently, the comparisons indicate where the proposed
estimator is competitive under the shipped configurations; they do not establish
universal method-level superiority.
\section{Methodology}\label{sec:method}

We develop a weak-form estimator for two-dimensional It\^{o} generators that
extends the spatial-kernel construction of \citet{eshwar2026weak} from scalar
diffusions to coupled, non-diagonal diffusion tensors. The method is motivated by
the martingale identity obtained for fixed or independently constructed spatial
weights. In the released implementation, however, the smoother is estimated from
the full sampled state cloud, so exact finite-sample adaptedness and unbiasedness
are not claimed. The principal two-dimensional challenge is the simultaneous
recovery of five coefficient fields,
\[
    \drift_1,\qquad \drift_2,\qquad
    \diff_{11},\qquad \diff_{12},\qquad \diff_{22},
\]
under scale disparity, heteroscedastic drift noise, finite-step bias, and the
positive-semidefiniteness constraint on $\diff=\sigma\sigma^\top$.

Throughout this section, we observe $R$ independent trajectories
$\{X^{(r)}_{t_n}\}_{n=0}^{N}$ of the diffusion
\begin{equation}
    \mathrm{d}X_t
    =
    \drift(X_t)\,\mathrm{d}t
    +
    \sigma(X_t)\,\mathrm{d}W_t,
    \qquad
    X_t\in\R^2,
    \label{eq:method_sde}
\end{equation}
at uniform step size $\Delta t$, where
\[
    \diff(x)=\sigma(x)\sigma(x)^\top
    =
    \begin{pmatrix}
        \diff_{11}(x) & \diff_{12}(x)\\
        \diff_{12}(x) & \diff_{22}(x)
    \end{pmatrix}.
\]
The infinitesimal generator is
\begin{equation}
    \Lgen f(x)
    =
    \drift_1(x)\partial_1 f(x)
    +
    \drift_2(x)\partial_2 f(x)
    +
    \frac12
    \left[
        \diff_{11}(x)\partial_{11}f(x)
        +
        2\diff_{12}(x)\partial_{12}f(x)
        +
        \diff_{22}(x)\partial_{22}f(x)
    \right].
    \label{eq:method_generator}
\end{equation}
The target is not $\sigma$ itself, which is identifiable only up to orthogonal
rotation, but the tensor $\diff=\sigma\sigma^\top$ and the drift vector
$\drift$.

We assume a sparse library representation. Let
\[
    \Theta(x)=
    \bigl[\theta_1(x),\ldots,\theta_K(x)\bigr]\in\R^{1\times K}.
\]
For each drift component and tensor entry,
\begin{equation}
    \drift_i(x)=\Theta(x)c^{(i)},\qquad i\in\{1,2\},
    \label{eq:drift_library}
\end{equation}
and
\begin{equation}
    \diff_{ij}(x)=\Theta(x)d^{(ij)},\qquad
    1\leq i\leq j\leq2.
    \label{eq:tensor_library}
\end{equation}
The coefficient vectors are sparse. Recovery therefore reduces to estimating the
five sparse coefficient vectors
\[
    c^{(1)},\quad c^{(2)},\quad
    d^{(11)},\quad d^{(12)},\quad d^{(22)}.
\]

\subsection{Standing assumptions}\label{sec:method_assumptions}

The analysis is stated under the following conditions.

\begin{assumption}[Identifiable two-dimensional diffusion]\label{ass:method_standing}
\leavevmode
\begin{enumerate}
    \item \emph{Ergodicity.}
    The diffusion~\eqref{eq:method_sde} admits a unique stationary measure $\mu$
    and is ergodic with finite mixing time relative to the observation horizon
    $N\Delta t$.

    \item \emph{Regularity.}
    The drift $\drift$ and noise loading $\sigma$ are locally Lipschitz with
    at most linear growth. The library functions $\theta_k$ and spatial kernels
    used below are bounded on the sampled region and sufficiently smooth for the
    Taylor expansions used in the local-polynomial projection.

    \item \emph{Library realizability.}
    The true drift and diffusion tensor entries lie in the span of the chosen
    library as in \eqref{eq:drift_library}--\eqref{eq:tensor_library}. If this
    fails, the estimator returns the best weak-projection approximation in the
    chosen library, not the exact generator. The PSD Cholesky read-out in
    \Cref{sec:method_cholesky} additionally requires library-spanned Cholesky
    factor fields for exact symbolic read-out through that stage; otherwise the
    final tensor should be interpreted as a PSD approximation in the induced
    product library.

    \item \emph{Coverage.}
    Each normalized smoother row receives non-vanishing effective horizon:
    \[
        T_{\mathrm{eff},j}
        =
        \frac{\Delta t}{\sum_{r,n}(\Pi_{j n}^{(r)})^2}
        \to\infty,
        \qquad
        \max_{r,n}|\Pi_{j n}^{(r)}|\to0 .
    \]

    \item \emph{Full rank.}
    The population weak design matrix
    \[
        \bar A_{jk}
        =
        \int \pi_j(x)\theta_k(x)\,\mathrm{d}\mu(x)
    \]
    has full column rank on the active library support, where $\pi_j$ denotes the
    population limit of the $j$th normalized local-polynomial intercept row.

    \item \emph{Tensor consistency.}
    The first-pass tensor estimator satisfies
    \[
        \|\hat\diff-\diff\|_{\infty}=o_p(1).
    \]
\end{enumerate}
\end{assumption}

These assumptions define the population identifiability, coverage, and library
envelope under which the weak system has a unique symbolic target. They do not,
by themselves, establish exact finite-sample unbiasedness or consistency for the
released data-dependent full-cloud smoother. The reported recovery evidence is
therefore empirical and restricted to the declared sampled-region benchmark
setting.

\subsection{Why the naive 1D-to-2D port fails}\label{sec:method_naive_fail}

The scalar estimator of \citet{eshwar2026weak} uses spatial Gaussian kernels to
build a weak design matrix and then solves two sparse systems: one for the drift
and one for the scalar diffusion. A direct 2D port would simply solve five
independent systems over a bivariate library:
\[
    \drift_1,\ \drift_2,\ \diff_{11},\ \diff_{12},\ \diff_{22}.
\]
This direct port fails for three structural reasons.

First, the raw bivariate design is ill-conditioned. If the two coordinates have
aspect ratio $\kappa$, then polynomial columns of degree $d$ can differ in scale
by $O(\kappa^d)$. In stochastic-volatility data, for example, the log-price
coordinate may be $O(1)$--$O(10)$ while the variance coordinate may be
$O(10^{-2})$--$O(10^{-1})$. The resulting normal equations are unstable.

Second, the drift target is heteroscedastic across kernels. For Euler-generated
benchmarks the sampled increment is
\begin{equation}
    \Delta X_n
    =
    \drift(X_{t_n})\Delta t
    +
    \sigma(X_{t_n})\Delta W_n,
    \label{eq:euler_increment}
\end{equation}
the drift signal scales as $\Delta t$ while the Brownian noise scales as
$\sqrt{\Delta t}$. Kernels located in high-diffusion regions therefore carry much
larger noise variance than kernels in low-diffusion regions. Ordinary least
squares treats these rows as equally reliable.

Third, an unconstrained regression for
$\diff_{11},\diff_{12},\diff_{22}$ does not guarantee
\[
    \hat\diff(x)\succeq0.
\]
A fitted tensor with negative eigenvalues cannot be interpreted as a local
covariance. This is not cosmetic: it breaks the leverage read-out
$\diff_{12}/\sqrt{\diff_{11}\diff_{22}}$ and invalidates the generator as a
diffusion generator.

On top of the base weak-form construction (coordinate standardization, spatial
Gaussian kernels, and the finite-step drift-square correction), the estimator
adds six mechanisms, each targeting one of the failure modes above:
(i) anisotropic covariance-shaped kernels, (ii) an order-2 local-polynomial
weak projection, (iii) adaptive-LASSO sparsification, (iv) feasible
generalized-least-squares drift reweighting, (v) a PSD Cholesky diffusion
read-out, and (vi) multi-trajectory pooling. These six ingredients are examined
through the cumulative-graft and leave-one-out analyses of
\Cref{sec:ablation}. Pooling, feasible diagonal GLS, and local-polynomial
smoothing show clear degradation when removed on the reported in-scope summary.
The PSD Cholesky read-out, adaptive sparsification, and anisotropic bandwidths
primarily provide structural validity or protection in stress regimes; the
leave-one-out matrix does not establish that each is individually necessary for
the in-scope median score.

\subsection{Weak-form moment equations}\label{sec:method_weak_moments}

For a smooth observable $\phi$, Dynkin's formula gives
\begin{equation}
    \E\!\left[
        \phi(X_{t_{n+1}})-\phi(X_{t_n})
        \,\middle|\,
        \mathcal{F}_{t_n}
    \right]
    =
    (\Lgen\phi)(X_{t_n})\Delta t
    +
    O(\Delta t^2).
    \label{eq:dynkin_method}
\end{equation}
For Euler--Maruyama trajectories the sampled increment is exactly
\begin{equation}
    \Delta X_n
    =
    \drift(X_{t_n})\,\Delta t
    +
    \sigma(X_{t_n})\,\Delta W_n,
    \qquad
    \Delta W_n\sim\mathcal{N}(0,\Delta t\,I_m)\perp\mathcal{F}_{t_n}.
    \label{eq:euler_exact_method}
\end{equation}
Because the drift and noise loading are frozen at $X_{t_n}$ over the step, the
first two conditional moments are then \emph{exact}. Taking $\phi(x)=x^{(p)}$,
\begin{equation}
    \E\!\left[
        \Delta X_n^{(p)}
        \,\middle|\,
        \mathcal{F}_{t_n}
    \right]
    =
    \drift_p(X_{t_n})\,\Delta t,
    \qquad p=1,2,
    \label{eq:drift_moment_method}
\end{equation}
and taking $\phi(x)=x^{(p)}x^{(q)}$, using
$\E[\Delta W_n^{(k)}\Delta W_n^{(\ell)}\mid\mathcal{F}_{t_n}]=\delta_{k\ell}\Delta t$
and the mean-zero cross term,
\begin{equation}
    \E\!\left[
        \Delta X_n^{(p)}\Delta X_n^{(q)}
        \,\middle|\,
        \mathcal{F}_{t_n}
    \right]
    =
    \diff_{pq}(X_{t_n})\,\Delta t
    +
    \drift_p(X_{t_n})\drift_q(X_{t_n})\,\Delta t^2 .
    \label{eq:diffusion_moment_method}
\end{equation}
Both identities are exact for Euler--Maruyama data. For a general continuous
It\^{o} diffusion sampled at spacing $\Delta t$, or for exact-transition simulators
such as the OU and GBM families used in the benchmark, they hold up to the
corresponding finite-step conditional-moment remainder; for clipped or projected
stress systems they describe the declared sampled process only on the retained
sampled region. The estimator treats the Euler increment as the working model.
The drift is therefore read from first increments, and the tensor from quadratic
increments once the drift-square term $\drift_p\drift_q\Delta t^2$ is removed
(\Cref{sec:method_correction}).

Let $\Pi\in\R^{J\times RN}$ denote the normalized spatial smoother matrix whose $j$th row
locally averages quantities around centre $c_j$. Let
$\Theta\in\R^{RN\times K}$ be the library matrix evaluated at all sampled states.
The shared weak design matrix is
\begin{equation}
    A=\Pi\Theta.
    \label{eq:shared_design_method}
\end{equation}
For drift component $p$, define
\begin{equation}
    B^{(p)}
    =
    \Pi\!\left(\frac{\Delta X^{(p)}}{\Delta t}\right).
    \label{eq:drift_target_method}
\end{equation}
For diffusion entry $(p,q)$, define the corrected pointwise target
\begin{equation}
    q_n^{(pq)}
    =
    \frac{1}{\Delta t}
    \left[
        \Delta X_n^{(p)}\Delta X_n^{(q)}
        -
        \hat\drift_p(X_{t_n})\hat\drift_q(X_{t_n})\Delta t^2
    \right],
    \label{eq:q_target_method}
\end{equation}
and smooth it:
\begin{equation}
    Q^{(pq)}=\Pi q^{(pq)}.
    \label{eq:tensor_target_method}
\end{equation}
The weak systems are then
\begin{equation}
    B^{(p)}\approx A c^{(p)},
    \qquad
    Q^{(pq)}\approx A d^{(pq)}.
    \label{eq:weak_systems_method}
\end{equation}
Both drift and diffusion use the same $A$. This is important: the method learns
a single symbolic generator, not separate unrelated regressions.

\subsection{Spatial weak rows and the dependence caveat}
\label{sec:method_unbiased_rows}

The spatial weak construction is motivated by a standard martingale identity.
For one Euler-generated trajectory and one component,
\[
    \Delta X_n^{(p)}
    =
    \drift_p(X_{t_n})\Delta t
    +
    \sum_{\ell}\sigma_{p\ell}(X_{t_n})\Delta W_n^{(\ell)}.
\]
Let $\pi_j$ be a fixed deterministic spatial weight function. The corresponding
martingale contribution is
\[
    M_j^{(p)}
    =
    \sum_n
    \pi_j(X_{t_n})
    \sum_{\ell}
    \sigma_{p\ell}(X_{t_n})\Delta W_n^{(\ell)}.
\]
Because $\pi_j(X_{t_n})\sigma_{p\ell}(X_{t_n})$ is
$\mathcal{F}_{t_n}$-measurable,
\[
    \E\!\left[
        \pi_j(X_{t_n})
        \sigma_{p\ell}(X_{t_n})
        \Delta W_n^{(\ell)}
        \,\middle|\,
        \mathcal{F}_{t_n}
    \right]
    =
    0.
\]
Thus a fixed or independently constructed spatial weight preserves the usual
martingale cancellation.

The released smoother is more data-dependent than this idealized calculation.
Coordinate standardization, $k$-means centres, covariance-shaped bandwidths, and
the local-polynomial Gram matrices are estimated from the full sampled state
cloud. Consequently, a finite-sample entry $\Pi_{jn}$ need not be
$\mathcal{F}_{t_n}$-measurable, because it can depend on states observed after
$t_n$. The preceding tower-property calculation therefore motivates the
population weak identity but does not establish exact finite-sample cancellation
for the released full-cloud smoother.

An exactly cross-fitted construction would estimate all nuisance
objects---standardization, centres, bandwidths, local-polynomial Gram matrices,
the resulting spatial weight functions, and the first-pass diffusion
field---using trajectories other than the trajectory on which the weak row is
evaluated. The resulting weight function would then be fixed relative to the
evaluated trajectory. This fully cross-fitted construction is not used in the
reported experiments, so no numerical result in this paper is presented as
evidence for such a theorem.

\subsection{Standardization, anisotropic kernels, and local-polynomial projection}
\label{sec:method_kernels}

\paragraph{Standardization.}
The states are standardized coordinate-wise before constructing kernels or
libraries:
\begin{equation}
    z_n
    =
    S_{\mathrm{std}}^{-1}(X_{t_n}-\bar X),
    \qquad
    S_{\mathrm{std}}=\operatorname{diag}(s_1,s_2),
    \label{eq:standardization_method}
\end{equation}
where $\bar X$ is the empirical mean and $s_1,s_2$ are empirical marginal
standard deviations. The library is evaluated in $z$-coordinates. After recovery,
coefficients may be mapped back to physical coordinates by the exact polynomial
change of variables induced by
\[
    z=S_{\mathrm{std}}^{-1}(x-\bar X).
\]
This removes the axis-scale disparity that causes the raw bivariate monomial
design to become ill-conditioned.

\paragraph{Kernel centres and bandwidth.}
Centres $\{c_j\}_{j=1}^{J}$ are placed by $k$-means on the standardized cloud
$\{z_n\}$. In the frozen implementation, $J=64$. The covariance-shaped bandwidth
is
\begin{equation}
    \Sigma_{\mathrm{bw}}
    =
    h_0^2
    \frac{\widehat{\operatorname{Cov}}(z)}
         {\operatorname{tr}\widehat{\operatorname{Cov}}(z)/2}
    +
    10^{-8}I,
    \qquad
    h_0
    =
    1.5\,
    \operatorname{median}_j
    \min_{j'\neq j}\|c_j-c_{j'}\|.
    \label{eq:bw_method}
\end{equation}
The unnormalized spatial kernel is
\begin{equation}
    w_{jn}
    =
    \exp\!\left[
        -\frac12
        (z_n-c_j)^\top
        \Sigma_{\mathrm{bw}}^{-1}
        (z_n-c_j)
    \right].
    \label{eq:kernel_weight_method}
\end{equation}
Unlike an isotropic kernel, this kernel adapts to residual covariance structure
in the standardized cloud.

\paragraph{Order-2 local-polynomial smoother.}
A local-constant Nadaraya--Watson estimate at centre $c_j$ is
\begin{equation}
    \hat g_{\mathrm{NW}}(c_j)
    =
    \frac{\sum_n w_{jn}g(z_n)}{\sum_n w_{jn}}.
    \label{eq:nw_method}
\end{equation}
Taylor expansion gives
\begin{equation}
    \E[\hat g_{\mathrm{NW}}(c_j)]
    =
    g(c_j)
    +
    \nabla g(c_j)^\top\nu_j
    +
    \frac12
    \operatorname{tr}
    \left(
        \nabla^2 g(c_j)\Sigma_{\mathrm{bw}}
    \right)
    +
    O(h_0^4),
    \label{eq:nw_bias_method}
\end{equation}
where
\[
    \nu_j=\E_{w_j}[z-c_j].
\]
In the interior $\nu_j=0$, but near the boundary the kernel is truncated and
$\nu_j\neq0$, producing a first-order boundary bias.

To remove this leading bias, we use an order-2 local-polynomial projection.
Let
\[
    \delta_{jn}=z_n-c_j,
\]
and define the quadratic local basis
\begin{equation}
    r(\delta)
    =
    \begin{bmatrix}
        1&
        \delta_1&
        \delta_2&
        \delta_1^2&
        \delta_1\delta_2&
        \delta_2^2
    \end{bmatrix}^{\top}.
    \label{eq:local_poly_basis_method}
\end{equation}
For each centre $c_j$, construct
\[
    R_j
    =
    \begin{bmatrix}
        r(\delta_{j1})^\top\\
        \vdots\\
        r(\delta_{j,RN})^\top
    \end{bmatrix},
    \qquad
    W_j=\operatorname{diag}(w_{j1},\ldots,w_{j,RN}).
\]
The local-polynomial fit for a target vector $g$ solves
\begin{equation}
    \hat\beta_j(g)
    =
    \arg\min_{\beta}
    \sum_n
    w_{jn}
    \left[
        g(z_n)-r(\delta_{jn})^\top\beta
    \right]^2
    +
    \varepsilon_j\|\beta\|_2^2,
    \label{eq:local_poly_fit_method}
\end{equation}
where $\varepsilon_j=10^{-8}\,\overline{\operatorname{diag}}(R_j^\top W_j R_j)$
is a small ridge scaled to the local Gram magnitude, present only for numerical
stability. The estimate at the centre is the intercept:
\[
    \hat g(c_j)=e_1^\top\hat\beta_j(g).
\]
Equivalently, the smoother row is
\begin{equation}
    \Pi_{j\cdot}
    =
    e_1^\top
    \left(R_j^\top W_j R_j+\varepsilon_j I\right)^{-1}
    R_j^\top W_j.
    \label{eq:smoother_row_method}
\end{equation}
Without the ridge term, the order-2 local-polynomial intercept reproduces all
quadratic polynomials exactly and removes the first-moment and curvature terms in
\eqref{eq:nw_bias_method}. The stabilized implementation uses the nonzero ridge
in \eqref{eq:local_poly_fit_method}; this adds a regularization bias controlled by
$\varepsilon_j$ that vanishes as the ridge scale tends to zero under stable local
conditioning. Thus the projection bias is $O(h_0^4)$ plus the ridge-induced term,
not a pure $O(h_0^4)$ bound at fixed ridge.

\subsection{First-pass drift and finite-step tensor correction}
\label{sec:method_correction}

The first-pass drift solves
\begin{equation}
    \hat c^{(p)}_{\mathrm{OLS}}
    =
    \arg\min_c
    \|B^{(p)}-Ac\|_2^2
    +
    \lambda\|c\|_2^2,
    \qquad p=1,2,
    \label{eq:first_pass_drift_method}
\end{equation}
or the corresponding sparse version described in
\Cref{sec:method_lasso_pooling}. This gives
\[
    \hat\drift_p(x)=\Theta(x)\hat c^{(p)}.
\]

The diffusion target must be corrected because quadratic increments contain a
drift-squared finite-step contribution. For Euler-generated data,
\eqref{eq:diffusion_moment_method} gives the exact conditional identity
\begin{equation}
    \E\!\left[
        \Delta X_n\Delta X_n^\top
        \,\middle|\,
        \mathcal{F}_{t_n}
    \right]
    =
    \diff(X_{t_n})\Delta t
    +
    \drift(X_{t_n})\drift(X_{t_n})^\top\Delta t^2.
    \label{eq:outer_expansion_method}
\end{equation}
For a continuously sampled It\^o diffusion rather than an Euler benchmark, the
same displayed expression carries the weak $O(\Delta t^2)$ conditional-moment
remainder from \eqref{eq:dynkin_method}.
Therefore, for each tensor entry,
\begin{equation}
    q_n^{(pq)}
    =
    \frac{
        \Delta X_n^{(p)}\Delta X_n^{(q)}
        -
        \hat\drift_p(X_{t_n})\hat\drift_q(X_{t_n})\Delta t^2
    }{\Delta t}.
    \label{eq:finite_step_method}
\end{equation}
The smoothed tensor target is
\[
    Q^{(pq)}=\Pi q^{(pq)}.
\]
Then
\begin{equation}
    \hat d^{(pq)}
    =
    \arg\min_d
    \|Q^{(pq)}-Ad\|_2^2
    +
    \lambda\|d\|_2^2.
    \label{eq:tensor_ridge_method}
\end{equation}

\paragraph{Optional observation-noise correction.}
If observations are contaminated by additive noise,
\[
    \widetilde X_n=X_n+\eta_n,
    \qquad
    \E[\eta_n]=0,
    \qquad
    \operatorname{Cov}(\eta_n)=\Sigma_{\eta},
\]
then lag-zero quadratic variation is inflated by $2\Sigma_\eta$, while
\[
    \E[
        \Delta\widetilde X_n
        \Delta\widetilde X_{n+1}^{\top}
    ]
    =
    -\Sigma_\eta
    +
    o(1).
\]
Thus $\Sigma_\eta$ can be estimated from the lag-one outer product and subtracted
from the raw quadratic increment before division by $\Delta t$:
\begin{equation}
    q_n
    =
    \frac{
        \Delta\widetilde X_n\Delta\widetilde X_n^\top
        -2\hat\Sigma_\eta
        -\hat\drift(X_{t_n})\hat\drift(X_{t_n})^\top\Delta t^2
    }{\Delta t}.
    \label{eq:obs_noise_corrected_target_method}
\end{equation}
In the reported clean synthetic benchmarks this correction is disabled to avoid
adding unnecessary variance. For pooled trajectories, the lag-one estimator must
be applied trajectory-aware so that artificial cross-trajectory pairs are not
used.

\subsection{PSD Cholesky diffusion read-out}
\label{sec:method_cholesky}

A direct linear fit for
$\diff_{11},\diff_{12},\diff_{22}$ does not guarantee
$\hat\diff(x)\succeq0$. We therefore recover the tensor through a Cholesky
read-out. The released implementation does not solve a nonlinear alternating
least-squares problem directly in the Cholesky coefficients. Instead, it uses a
three-stage procedure: first fit the unconstrained tensor entries, then project
the pointwise tensor estimates to the PSD cone, and finally regress the
corresponding Cholesky factor fields on the same library.

First, the unconstrained entrywise tensor estimate is evaluated at the data:
\[
    \hat\diff_{\mathrm{lin}}(X_{t_n})
    =
    \begin{pmatrix}
        \Theta(X_{t_n})\hat d^{(11)}
        &
        \Theta(X_{t_n})\hat d^{(12)}
        \\
        \Theta(X_{t_n})\hat d^{(12)}
        &
        \Theta(X_{t_n})\hat d^{(22)}
    \end{pmatrix}.
\]
Each pointwise matrix is projected to the PSD cone by eigenvalue clipping:
\begin{equation}
    \hat\diff_{\mathrm{PSD}}(X_{t_n})
    =
    \sum_{k=1}^{2}
    \max(\lambda_k,10^{-10})v_kv_k^\top,
    \label{eq:psd_projection_method}
\end{equation}
where $(\lambda_k,v_k)$ are the eigenpairs of
$\hat\diff_{\mathrm{lin}}(X_{t_n})$.

Next compute the pointwise Cholesky fields
\begin{align}
    \ell_{11}(X_{t_n})
    &=
    \sqrt{\hat\diff_{\mathrm{PSD},11}(X_{t_n})},
    \label{eq:chol_11_method}\\
    \ell_{21}(X_{t_n})
    &=
    \frac{
        \hat\diff_{\mathrm{PSD},12}(X_{t_n})
    }{
        \ell_{11}(X_{t_n})
    },
    \label{eq:chol_21_method}\\
    \ell_{22}(X_{t_n})
    &=
    \sqrt{
        \hat\diff_{\mathrm{PSD},22}(X_{t_n})
        -
        \ell_{21}(X_{t_n})^2
    },
    \label{eq:chol_22_method}
\end{align}
with denominators and radicands floored at $10^{-10}$.

Each Cholesky factor field is then weak-regressed against the same library:
\begin{equation}
    \ell_{ab}(x)=\Theta(x)\beta^{(ab)},
    \qquad
    (a,b)\in\{(1,1),(2,1),(2,2)\}.
    \label{eq:chol_library_method}
\end{equation}
The coefficient vectors are obtained from
\begin{equation}
    \hat\beta^{(ab)}
    =
    \arg\min_{\beta}
    \|\Pi\ell_{ab}-A\beta\|_2^2
    +
    \lambda\|\beta\|_2^2.
    \label{eq:chol_fit_method}
\end{equation}

At read-out, define
\[
    \hat L(x)
    =
    \begin{pmatrix}
        \hat\ell_{11}(x) & 0\\
        \hat\ell_{21}(x) & \hat\ell_{22}(x)
    \end{pmatrix}.
\]
The recovered tensor is
\begin{equation}
    \hat\diff(x)
    =
    \hat L(x)\hat L(x)^\top,
    \label{eq:chol_tensor_method}
\end{equation}
or explicitly
\begin{align}
    \hat\diff_{11}(x)
    &=
    \hat\ell_{11}(x)^2,
    \label{eq:readout_a11_method}\\
    \hat\diff_{12}(x)
    &=
    \hat\ell_{11}(x)\hat\ell_{21}(x),
    \label{eq:readout_a12_method}\\
    \hat\diff_{22}(x)
    &=
    \hat\ell_{21}(x)^2+\hat\ell_{22}(x)^2.
    \label{eq:readout_a22_method}
\end{align}
Thus $\hat\diff(x)\succeq0$ for every $x$ by construction. This stage is a PSD
Cholesky read-out/approximation unless the Cholesky factor fields themselves are
spanned by the chosen library. After expansion
\eqref{eq:readout_a11_method}--\eqref{eq:readout_a22_method}, the tensor entries
belong to the induced product library, not necessarily to the original linear
entrywise library.

Finally, the read-out tensor is mildly shrunk toward its isotropic part:
\begin{equation}
    \hat\diff(x)
    \leftarrow
    (1-\rho)\hat\diff(x)
    +
    \rho\,
    \frac12
    \operatorname{tr}\bigl(\hat\diff(x)\bigr)I,
    \qquad
    \rho=0.05.
    \label{eq:isotropic_shrink_method}
\end{equation}
This stabilizes near-singular and low-coverage regimes while leaving
well-identified tensor fields essentially unchanged.

\subsection{Feasible generalized-least-squares drift reweighting}
\label{sec:method_gls}

The first-pass drift weak system has the form
\begin{equation}
    B^{(p)}=Ac^{(p)}+M^{(p)},
    \label{eq:drift_noise_model_method}
\end{equation}
where $M^{(p)}$ is a martingale noise term. Although
$\E[M^{(p)}]=0$, its variance is not constant across kernel rows: rows centred in
high-diffusion regions are noisier. The frozen estimator corrects this with a
\emph{per-component} diagonal feasible GLS pass, using the first-pass diffusion
diagonal $\hat\diff_{pp}$ to build the row weights. For row $j$, the diagonal
proxy for the drift-noise variance is
\begin{equation}
    V_j^{(p)}
    =
    \frac{1}{\Delta t}
    \sum_n
    \Pi_{jn}^2
    \hat\diff_{pp}(X_{t_n}).
    \label{eq:noise_var_method}
\end{equation}
Rows in high-diffusion regions have larger $V_j^{(p)}$ and should be
down-weighted.

The feasible GLS row multiplier is
\begin{equation}
    s_j^{(p)}
    =
    \operatorname{clip}_{[0.05,\,20]}
    \left(
        \frac{(V_j^{(p)})^{-1/2}}
        {\operatorname{median}_{j'}(V_{j'}^{(p)})^{-1/2}}
    \right).
    \label{eq:gls_weight_method}
\end{equation}
Let
\[
    W_p=\operatorname{diag}\left((s_1^{(p)})^2,\ldots,(s_J^{(p)})^2\right).
\]
The drift is re-estimated by
\begin{equation}
    \hat c^{(p)}_{\mathrm{GLS}}
    =
    (A^\top W_p A)^{-1}A^\top W_pB^{(p)}.
    \label{eq:gls_estimator_method}
\end{equation}
Equivalently, rows of $A$ and $B^{(p)}$ are scaled by $s_j^{(p)}$ before solving.
In the frozen configuration this reweighting is run for a single pass, and the
clip in \eqref{eq:gls_weight_method} caps the influence of extreme rows. The normal-equation
weight is $W_p$, so the row multiplier $s_j^{(p)}$ corresponds to a squared weight in the GLS objective. The
weights use only the diagonal $\hat\diff_{pp}$, so they are diagonal GLS-style
weights for the per-component heteroscedastic model rather than a full
cross-component whitening; a full-tensor whitening variant, which also uses
$\hat\diff_{12}$, is available but is not part of the default stack.

\paragraph{Status of feasible diagonal GLS.}
The released estimator performs one in-sample per-component feasible diagonal
GLS pass. Its row weights are plug-in functions of the first-pass diffusion
estimate and the data-dependent spatial projection. The pass is intended as a
variance-stabilizing numerical reweighting and is evaluated through the ablation
study. We do not claim exact finite-sample unbiasedness, asymptotic equivalence
to oracle GLS, or attainment of oracle efficiency for this released
implementation. Establishing those properties would require a fully independent
or predictable construction of both the smoother and the GLS weights, together
with additional rate and stochastic-equicontinuity conditions.

\subsection{Adaptive LASSO, debiasing, and multi-trajectory pooling}
\label{sec:method_lasso_pooling}

Every weak system above is solved using the same sparse selection routine. First,
the design columns are $\ell_2$-normalized. An initial ridge estimate
$\hat c^{\mathrm{init}}$ is computed, and adaptive penalty weights are set as
\begin{equation}
    \omega_k=
    \max\!\bigl(|\hat c_k^{\mathrm{init}}|,\,\epsilon_0\bigr)^{-\gamma},
    \qquad
    \gamma=1,
    \qquad
    \epsilon_0=10^{-3},
    \label{eq:adaptive_weights_method}
\end{equation}
where the floor $\epsilon_0$ keeps the weight of a vanishing pilot coefficient
finite. The penalty parameter $\alpha$ is chosen by cross-validation. When
trajectory identifiers are supplied, this cross-validation is grouped by
trajectory; otherwise the implementation falls back to pseudo-block
cross-validation over the projected weak rows. The adaptive-LASSO problem is
\begin{equation}
    \hat c
    =
    \arg\min_c
    \left\{
        \frac12\|y-Xc\|_2^2
        +
        \alpha\sum_{k=1}^{K}\omega_k|c_k|
    \right\}.
    \label{eq:adaptive_lasso_method}
\end{equation}
Large pilot coefficients are penalized lightly; near-zero pilot coefficients are
penalized heavily. This reduces the common failure mode of uniform LASSO:
overshrinking true dominant terms while retaining small spurious terms.

After selection, the active support
\[
    \hat S=\{k:\hat c_k\neq0\}
\]
is refit by ordinary least squares:
\begin{equation}
    \hat c_{\hat S}^{\mathrm{debias}}
    =
    \arg\min_u
    \|y-X_{\hat S}u\|_2^2,
    \qquad
    \hat c_{\hat S^c}^{\mathrm{debias}}=0.
    \label{eq:debias_method}
\end{equation}
A sequentially-thresholded least-squares pruning step then removes terms whose
absolute coefficient is below a fixed relative threshold of the largest retained
coefficient for that target, refitting after each pruning step. In the frozen
estimator this internal threshold is
\[
    \tau_{\mathrm{STLSQ}}=0.12.
\]
This $12\%$ rule is the estimator's actual sparsification threshold. It should be
distinguished from the smaller $10^{-3}$ threshold used later only for reporting
whether a coefficient was nonzero in the coefficient-recovery table.

For multi-trajectory pooling, independent trajectories are pooled by stacking
their weak rows. When trajectory identifiers are supplied to the estimator, each
trajectory is smoothed separately, producing trajectory-specific rows
\[
    A^{(r)},\quad B^{(r,p)},\quad Q^{(r,pq)}.
\]
The pooled systems are stacked:
\begin{equation}
    A_{\mathrm{pool}}
    =
    \begin{bmatrix}
        A^{(1)}\\
        \vdots\\
        A^{(R)}
    \end{bmatrix},
    \qquad
    B_{\mathrm{pool}}^{(p)}
    =
    \begin{bmatrix}
        B^{(1,p)}\\
        \vdots\\
        B^{(R,p)}
    \end{bmatrix}.
    \label{eq:pooling_method}
\end{equation}
The same construction is used for $Q^{(pq)}$. In this trajectory-aware mode,
cross-validation is grouped by trajectory, so folds hold out whole trajectories
rather than individual kernel rows.

The main coefficient-recovery rerun uses the same statistical idea of
multi-trajectory pooling, but the released script stacks the $R=32$ trajectories
as a single pooled state-increment cloud before calling the estimator. Since that
script does not pass trajectory identifiers, the sparse-selection routine uses
its pseudo-block cross-validation fallback rather than one-fold-per-trajectory
cross-validation. Thus the coefficient table should be read as evidence for the
effect of pooled trajectory coverage and sample size, not as a
trajectory-held-out CV experiment.

\subsection{Symbolic generator read-out}\label{sec:method_readout}

The final recovered drift is
\[
    \hat\drift(x)
    =
    \begin{pmatrix}
        \Theta(x)\hat c^{(1)}\\
        \Theta(x)\hat c^{(2)}
    \end{pmatrix}.
\]
The final recovered diffusion tensor is the Cholesky read-out
\[
    \hat\diff(x)=\hat L(x)\hat L(x)^\top.
\]
The symbolic generator is therefore
\begin{equation}
    \hat\Lgen f(x)
    =
    \hat\drift_1(x)\partial_1 f(x)
    +
    \hat\drift_2(x)\partial_2 f(x)
    +
    \frac12
    \left[
        \hat\diff_{11}(x)\partial_{11}f(x)
        +
        2\hat\diff_{12}(x)\partial_{12}f(x)
        +
        \hat\diff_{22}(x)\partial_{22}f(x)
    \right].
    \label{eq:estimated_generator_method}
\end{equation}

Three physically meaningful quantities are read directly from this generator:
\begin{align}
    \text{leverage}(x)
    &=
    \frac{\hat\diff_{12}(x)}
         {\sqrt{\hat\diff_{11}(x)\hat\diff_{22}(x)}},
    \label{eq:leverage_method}\\
    \text{fluctuation fields}(x)
    &=
    \bigl(\hat\diff_{11}(x),\hat\diff_{22}(x)\bigr),
    \label{eq:fluctuation_method}\\
    \text{drift curl}(x)
    &=
    \frac12
    \left(
        \partial_1\hat\drift_2(x)
        -
        \partial_2\hat\drift_1(x)
    \right).
    \label{eq:rotation_method}
\end{align}
The implementation also reports the antisymmetric-drift field
$x\mapsto x\,\operatorname{anti}(\nabla\hat\drift(x))^\top$ as a directional
diagnostic. In the isotropic rotational OU case this diagnostic is analytically
aligned with the stationary current direction. For state-dependent diffusion or
general nonlinear systems it is not the full stationary Fokker--Planck current
$J=\drift\pi-\tfrac12\nabla\cdot(\diff\pi)$, which would require density
estimation and the diffusion-divergence term.

\subsection{Metrics and support rule}\label{sec:method_metrics}

All reported field errors are equal-weight relative $L^2$ errors on a Cartesian grid $\mathcal G$ spanning the
coordinatewise 2nd--98th percentiles of the trajectories used for fitting. They are in-sample, sampled-region
diagnostics rather than stationary-measure-weighted or held-out errors. For a scalar field $g$,
\begin{equation}
    \operatorname{err}_{\mathcal G}(\hat g,g)
    =
    \left[
    \frac{\sum_{x\in\mathcal G}(\hat g(x)-g(x))^2}{\sum_{x\in\mathcal G}g(x)^2}
    \right]^{1/2},
    \qquad
    \|h\|_{\mathcal G}^2
    =
    \sum_{x\in\mathcal G} h(x)^2.
    \label{eq:l2_metric_method}
\end{equation}
The joint drift metric implemented in the reported code is
\begin{equation}
    E_b
    =
    \left[
    \frac{
    \sum_{x\in\mathcal G}\|\hat \drift(x)-\drift(x)\|_2^2
    }{
    \sum_{x\in\mathcal G}\|\drift(x)\|_2^2
    }
    \right]^{1/2}.
    \label{eq:drift_metric_method}
\end{equation}
The tensor error is the Frobenius relative error
\begin{equation}
    E_a
    =
    \frac{
        \left(
        \sum_{x\in\mathcal G}
        \|\hat\diff(x)-\diff(x)\|_F^2
        \right)^{1/2}
    }{
        \left(
        \sum_{x\in\mathcal G}
        \|\diff(x)\|_F^2
        \right)^{1/2}
    }.
    \label{eq:tensor_metric_method}
\end{equation}
The off-diagonal cosine is
\begin{equation}
    \cos_{12}
    =
    \frac{
        \sum_{x\in\mathcal G}\hat\diff_{12}(x)\diff_{12}(x)
    }{
        \left(\sum_{x\in\mathcal G}\hat\diff_{12}(x)^2\right)^{1/2}
        \left(\sum_{x\in\mathcal G}\diff_{12}(x)^2\right)^{1/2}
    }.
    \label{eq:offdiag_cosine_method}
\end{equation}
The PSD-valid fraction is
\begin{equation}
    \operatorname{PSD}
    =
    \frac1{|\mathcal G|}\sum_{x\in\mathcal G}\mathbb{1}\{\hat\diff(x)\succeq0\}.
    \label{eq:psd_metric_method}
\end{equation}
Because the final tensor is constructed through the Cholesky read-out and
isotropic shrinkage, PSD validity is a structural validity check rather than
independent evidence of tensor-entry accuracy. The generator-action error
compares $\hat\Lgen f$ and $\Lgen f$ on the monomial
test set
\[
    f\in\{x_1,x_2,x_1^2,x_1x_2,x_2^2\}.
\]

Support is scored as stable projected support. The estimator performs sparse
selection internally through adaptive LASSO, OLS debiasing, and relative STLSQ
pruning with $\tau_{\mathrm{STLSQ}}=0.12$, while the final PSD tensor is the
expanded and isotropically shrunk Cholesky read-out. For reporting, the final
drift fields and expanded tensor entries are projected back into the declared
library for the two drift components $\drift_1,\drift_2$ and the three tensor
entries $\diff_{11},\diff_{12},\diff_{22}$. The resulting count is therefore a
stable projected-support false-positive count, not a direct comparison of sparse
Cholesky-factor coefficients.

The paper-level stable-support filter is deliberately conservative. A truth term is active only if its magnitude is
at least 2\% of the target scale. A seed-level recovered term is selected only if its magnitude is at least 20\% of the
recovered target scale. A recovered inactive term is counted as a stable projected false positive only if it recurs in
at least 80\% of seeds, the corresponding truth-active recurrence is at most 20\%, and the recovered median magnitude is
at least 5\% of the system-wide maximum recovered magnitude. Using the corresponding 80\% recurrence threshold for true
active terms, the reported ledger recovers 104 of 134 stable active terms. Thus the support result is a statement about
no large recurring projected false positives under this post-processing rule, not perfect symbolic recall.

For the coefficient-recovery table, we use a separate reporting convention across
seeds. A term is counted as selected in seed $s$ when its recovered coefficient is
numerically nonzero at the reporting threshold
\begin{equation}
    |\hat c_{k,s}|>10^{-3}.
    \label{eq:reporting_selection_threshold_method}
\end{equation}
The selection rate of term $k$ is then
\begin{equation}
    \rho_k
    =
    \frac1{S}
    \sum_{s=1}^{S}
    \mathbb{1}\{|\hat c_{k,s}|>10^{-3}\},
    \label{eq:selection_rate_method}
\end{equation}
and the reported coefficient is the conditional median over selecting seeds:
\begin{equation}
    \hat c_k
    =
    \operatorname{median}
    \{\hat c_{k,s}:|\hat c_{k,s}|>10^{-3}\}.
    \label{eq:conditional_median_method}
\end{equation}
This reporting threshold is intentionally much smaller than the internal
$\tau_{\mathrm{STLSQ}}=0.12$ pruning threshold. It is used only to summarize
whether a term survived in a fitted seed, not to choose the model during fitting.
A false positive is a reported selected projected-support term whose
corresponding true coefficient is zero under the declared library support.

For descriptive cross-method summaries we use the composite score
\begin{equation}
S=0.38\left[1-\frac{\min(E_b,2)}{2}\right]_+
+0.34\left[1-\frac{\min(E_a,1.5)}{1.5}\right]_+
+0.16\,\operatorname{PSD}
+0.12\,C_{12},
\label{eq:composite_score_method}
\end{equation}
where $C_{12}=1$ when $a_{12}$ is not applicable and otherwise
$C_{12}=\operatorname{clip}((\cos_{12}+1)/2,0,1)$. Raw field errors, PSD validity, and support counts are the primary
evidence; the composite score is only a compact descriptive summary.

\subsection{Algorithmic summary}\label{sec:method_algorithm}

Given trajectories $\{X^{(r)}_{t_n}\}$, step size $\Delta t$, library
$\Theta$, and kernel count $J$, the estimator proceeds as follows.

\begin{enumerate}
    \item Standardize all states using \eqref{eq:standardization_method}.

    \item Place $J=64$ centres by $k$-means in standardized coordinates.

    \item Build anisotropic spatial kernels using
    \eqref{eq:bw_method}--\eqref{eq:kernel_weight_method}.

    \item Construct the order-2 local-polynomial smoother $\Pi$ using
    \eqref{eq:local_poly_fit_method}--\eqref{eq:smoother_row_method}.

    \item Form the shared weak design $A=\Pi\Theta$.

    \item Estimate the first-pass drift from
    $B^{(p)}=\Pi(\Delta X^{(p)}/\Delta t)$.

    \item Form the finite-step-corrected tensor targets
    $q^{(pq)}$ using \eqref{eq:finite_step_method}.

    \item Estimate the unconstrained tensor entries, project pointwise to the
    PSD cone, and refit through the Cholesky factor fields using
    \eqref{eq:psd_projection_method}--\eqref{eq:chol_tensor_method}.

    \item Compute kernel-local drift-noise variances
    $V_j^{(p)}$ using \eqref{eq:noise_var_method}.

    \item Re-solve the drift by feasible diagonal GLS using
    \eqref{eq:gls_weight_method}--\eqref{eq:gls_estimator_method}.

    \item Apply adaptive LASSO, debiasing, and sequential thresholding using
    \eqref{eq:adaptive_lasso_method}--\eqref{eq:debias_method}; the frozen
    internal STLSQ threshold is $\tau_{\mathrm{STLSQ}}=0.12$.

    \item Pool independent trajectories by stacking weak rows. When trajectory
    identifiers are supplied, sparse-penalty selection can use trajectory-grouped
    cross-validation; when they are not supplied, the implementation uses
    pseudo-block cross-validation on the pooled weak rows.

    \item Return the symbolic generator $\hat\Lgen$ in
    \eqref{eq:estimated_generator_method}, together with leverage, fluctuation,
    rotational-drift, PSD, field-error, and support metrics.
\end{enumerate}

\section{Results}\label{sec:results}
\subsection{Generator recovery across the benchmark zoo}\label{sec:showcase}
\paragraph{Experimental protocol.}
Table~\ref{tab:master} combines two frozen synthetic campaigns. The main v6 campaign contains 13 systems with
$R=16$ trajectories, seeds 9101--9110, 1600 steps for most systems and 2600 steps for financial/Heston cases. The v6.2
additions contain 16 systems with $R=16$ trajectories and seeds 9601--9610; their system-specific step counts and
$\Delta t$ values are recorded in \texttt{v6\_2\_extra\_summary\_raw.csv}. Field errors use the fitting-trajectory central grid
defined in \Cref{sec:method_metrics}; there is no held-out system or held-out evaluation trajectory. Hyperparameters
were frozen after development on an overlapping synthetic zoo, so the results are descriptive benchmark diagnostics, not
independent generalization estimates. A row is eligible for PASS only when it belongs to the predeclared in-scope class
and its declared recovery contract satisfies drift $<0.80$, tensor $<0.45$, PSD $\ge0.99$, and off-diagonal cosine
$>0.85$ when applicable. Predeclared stress and limit rows retain their limit designation even when their realized
numerical metrics happen to satisfy these gates. In particular, the non-polynomial-drift and too-large-$\Delta t$ rows
remain named stress tests rather than being reclassified after observing their results. The two Heston PASS rows are
scoped-target rows outside the stationary-theory envelope for the price/log-price coordinate: their declared drift metric
is the variance-component $b_2$ error, while the low-SNR price/log-price drift $b_1$ is excluded from the PASS contract
and remains a reported null. The Heston and SABR benchmark definitions follow
\citet{heston1993closed} and \citet{hagan2002sabr}, respectively.

The experiments ask whether the lifted weak-form estimator recovers the full
two-dimensional generator, not just separate scalar moment fits. The object being
judged is the five-field tuple $(\drift_1,\drift_2,\diff_{11},\diff_{12},
\diff_{22})$. \Cref{tab:master} is the navigation layer: each row reports drift
error, diffusion-tensor error, off-diagonal cosine when $\diff_{12}$ is present,
PSD validity, stable projected-support false-positive count, and a verdict.

We evaluate the frozen estimator on 29 synthetic two-dimensional systems: 19 meet
their predeclared in-scope recovery contracts, eight are retained as predeclared
stress or named-limit rows, and two
remain scoped reviews. Across the 19 PASS rows, the median central-grid drift
metric is $0.204$, the median tensor error is $0.0397$, and the median finite,
non-degenerate off-diagonal cosine is $0.997$. The recovered tensor is PSD at all
reported grid points by construction of the Cholesky read-out. Under the declared
stable projected-support filter, no large recurring false positives are counted;
using the corresponding recurrence threshold, 104 of 134 stable active terms are
recovered. The scoped-review and limit rows are retained to mark
cases where data or library information is missing: near-singular tensors,
boundary concentration, non-polynomial drift, poor coverage, or coarse sampling
that stresses plug-in drift correction and recovery.

\begin{table}[!htbp]\centering\small
\caption{Master system index (navigation layer). Recovery metrics are medians over $n=10$ seeds at the frozen
WG-SINDy configuration. ``Drift'' is the central-grid relative $L^2$ drift metric; ``Tensor'' is the central-grid
relative diffusion-tensor error; ``off-diag. cosine'' is the off-diagonal cosine (``--'' where diffusion is diagonal,
so $a_{12}\equiv0$); ``PSD'' is the fraction of central-grid points with a positive-semidefinite recovered tensor; and
``FP'' is the stable projected-support false-positive count under the declared post-processing rule. $\checkmark$ PASS,
$\circ$ scoped review, $\times$ named physical limit. Each system links to its
datasheet in \Cref{app:datasheets}.}\label{tab:master}
\begin{tabular}{l l c c c c c c}
\toprule
System & Family/Tier & Drift & Tensor & off-diag. cosine & PSD & FP & Verdict \\
\midrule
\multicolumn{8}{l}{\textit{Linear Ornstein--Uhlenbeck}}\\
\hyperref[sec:v62-indep-ou]{Independent OU} & linear/1 & 0.098 & 0.033 & -- & 1.00 & 0 & $\checkmark$ \\
\hyperref[sec:v62-correlated-ou]{Correlated OU} & linear/1 & 0.136 & 0.051 & 1.000 & 1.00 & 0 & $\checkmark$ \\
\hyperref[sec:v62-coupled-ou]{Coupled OU} & linear/1 & 0.214 & 0.028 & -- & 1.00 & 0 & $\checkmark$ \\
\hyperref[sec:v62-two-factor-vasicek]{Two-factor Vasicek} & linear/9 & 0.350 & --$^\dagger$ & -- & 1.00 & 0 & $\circ$ \\
\midrule
\multicolumn{8}{l}{\textit{Rotational / non-reversible}}\\
\hyperref[sec:v62-rotational-ou]{Rotational OU} & rotational/2 & 0.058 & 0.038 & -- & 1.00 & 0 & $\checkmark$ \\
\hyperref[sec:v62-spiral-sink-corr]{Spiral sink + corr.\ noise} & rotational/2 & 0.078 & 0.043 & 1.000 & 1.00 & 0 & $\checkmark$ \\
\hyperref[sec:v62-nongradient-circulation]{Non-gradient drift-curl} & rotational/3 & 0.239 & 0.024 & -- & 1.00 & 0 & $\checkmark$ \\
\midrule
\multicolumn{8}{l}{\textit{Bistable / gradient}}\\
\hyperref[sec:v62-double-well-transverse]{Double well + transverse} & bistable/3 & 0.226 & 0.024 & -- & 1.00 & 0 & $\checkmark$ \\
\hyperref[sec:v62-gradient-potential]{Gradient potential} & bistable/3 & 0.246 & 0.024 & -- & 1.00 & 0 & $\checkmark$ \\
\hyperref[sec:v62-maier-stein]{Maier--Stein} & bistable/8 & 0.235 & 0.038 & -- & 1.00 & 0 & $\checkmark$ \\
\hyperref[sec:v62-duffing]{Duffing oscillator} & bistable/8 & 0.240 & 0.031 & -- & 1.00 & 0 & $\checkmark$ \\
\hyperref[sec:v62-mueller-brown]{M\"uller--Brown} & bistable/8 & 1.826 & 0.075 & -- & 1.00 & 0 & $\times$ \\
\midrule
\multicolumn{8}{l}{\textit{Multiplicative / state-dependent diffusion}}\\
\hyperref[sec:v62-diag-multiplicative]{Diagonal multiplicative} & multipl./4 & 0.200 & 0.112 & -- & 1.00 & 0 & $\checkmark$ \\
\hyperref[sec:v62-nondiag-cholesky]{Non-diagonal Cholesky} & multipl./4 & 0.204 & 0.079 & 0.988 & 1.00 & 0 & $\checkmark$ \\
\hyperref[sec:v62-near-singular]{Near-singular tensor} & multipl./4 & 0.870 & 1.505 & 0.724 & 1.00 & 0 & $\times$ \\
\midrule
\multicolumn{8}{l}{\textit{Financial / stochastic volatility}}\\
\hyperref[sec:v62-heston-logsv]{Log-Heston} & financial/5 & 0.143 & 0.054 & 0.999 & 1.00 & 0 & $\checkmark$ \\
\hyperref[sec:v62-heston-sv]{Heston $(S,V)$} & financial/5 & 0.273 & 0.089 & 0.993 & 1.00 & 0 & $\checkmark$ \\
\hyperref[sec:v62-cir-pair]{CIR pair} & financial/5 & 0.393 & 0.116 & 0.995 & 1.00 & 0 & $\checkmark$ \\
\hyperref[sec:v62-sabr]{SABR} & financial/9 & --$^\ddagger$ & 0.195 & 0.990 & 1.00 & 0 & $\times$ \\
\hyperref[sec:v62-gbm-2d]{Correlated 2D GBM} & financial/9 & 2.49 & 0.096 & 0.997 & 1.00 & 0 & $\circ$ \\
\midrule
\multicolumn{8}{l}{\textit{Stochastic limit cycles}}\\
\hyperref[sec:v62-van-der-pol]{Van der Pol} & limit-cycle/7 & 0.066 & 0.040 & -- & 1.00 & 0 & $\checkmark$ \\
\hyperref[sec:v62-fitzhugh-nagumo]{FitzHugh--Nagumo} & limit-cycle/7 & 0.205 & 0.040 & -- & 1.00 & 0 & $\checkmark$ \\
\hyperref[sec:v62-stuart-landau]{Stuart--Landau} & limit-cycle/7 & 0.126 & 0.018 & -- & 1.00 & 0 & $\checkmark$ \\
\hyperref[sec:v62-brusselator]{Brusselator} & limit-cycle/7 & 0.013 & 0.067 & -- & 1.00 & 0 & $\checkmark$ \\
\midrule
\multicolumn{8}{l}{\textit{Honest limits (named, reported)}}\\
\hyperref[sec:v62-underdamped-langevin]{Underdamped Langevin} & limit/6 & 0.157 & 0.563 & -- & 1.00 & 0 & $\times$ \\
\hyperref[sec:v62-near-boundary-heston]{Near-boundary Heston} & limit/6 & 1.413 & 0.267 & 0.982 & 1.00 & 0 & $\times$ \\
\hyperref[sec:v62-nonpoly-drift]{Non-polynomial drift} & limit/6 & 0.210 & 0.037 & -- & 1.00 & 0 & $\times$ \\
\hyperref[sec:v62-bad-coverage]{Bad coverage} & limit/6 & 0.815 & 0.784 & -- & 1.00 & 0 & $\times$ \\
\hyperref[sec:v62-too-large-dt]{Too-large $\Delta t$} & limit/6 & 0.098 & 0.119 & -- & 1.00 & 0 & $\times$ \\
\bottomrule
\end{tabular}
\\[2pt]
\footnotesize $^\dagger$ Constant tiny tensor: relative-$L^2$ is degenerate; absolute error is small (see datasheet).
$^\ddagger$ SABR is driftless ($b\equiv0$); the central-grid drift metric is undefined while tensor and leverage are recovered.
For the two Heston rows, the declared drift metric is the variance-component $b_2$ error; the low-SNR price/log-price
drift $b_1$ is excluded from the PASS contract and remains a reported null.
\end{table}

\subsection{The 1D method fails in 2D}\label{sec:naive}
\begin{figure}[!htbp]\centering
\includegraphics[width=0.95\linewidth]{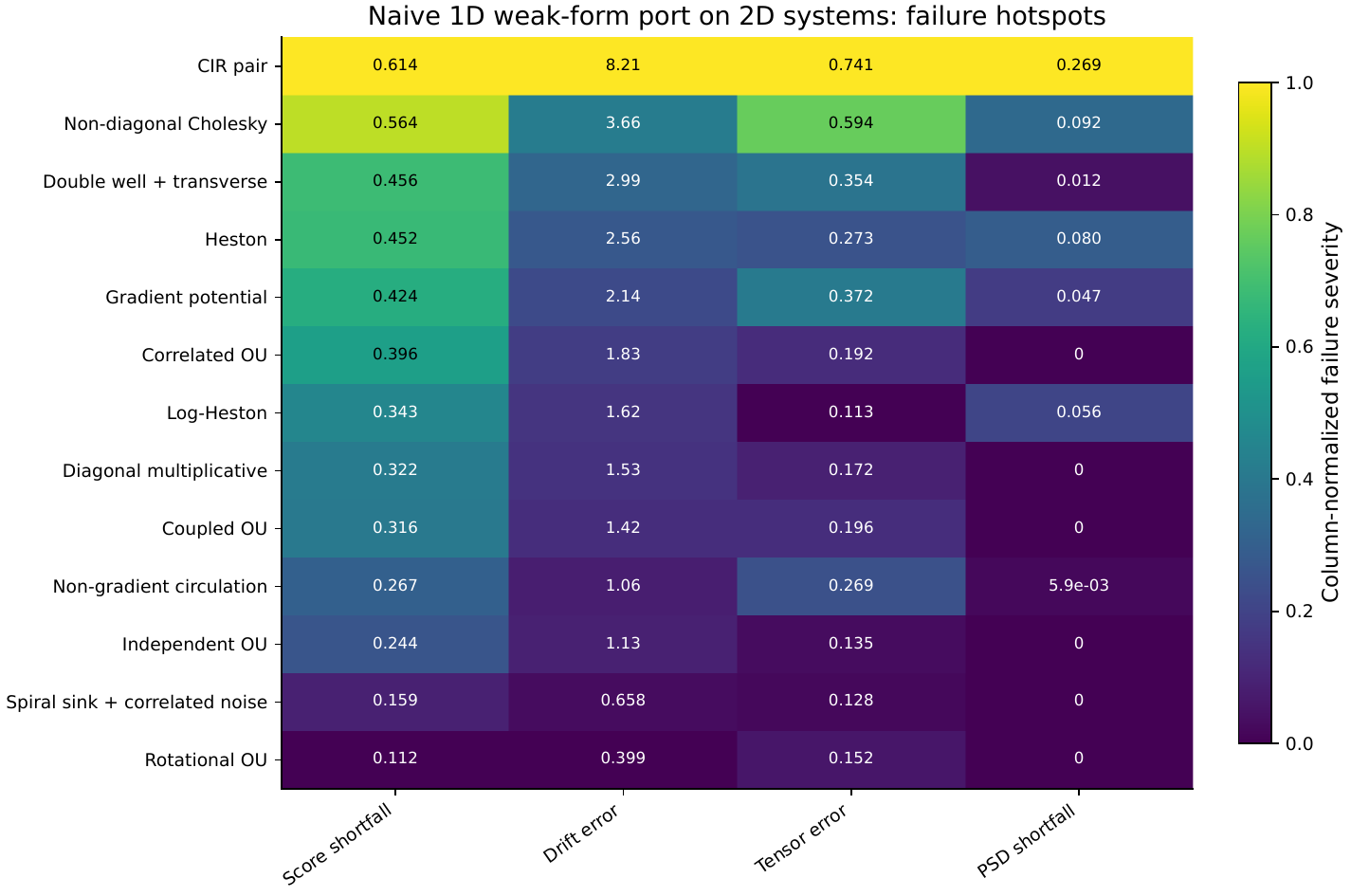}
\caption{Naive 1D-on-2D baseline. Each column is normalized to show where that failure mode concentrates; the
baseline lacks conditioning, drift whitening, and a covariance-constrained tensor read-out.}\label{fig:naive}
\end{figure}

The natural first baseline is to run the one-dimensional estimator component by
component on $\drift_1$, $\drift_2$, $\diff_{11}$, $\diff_{12}$, and
$\diff_{22}$.  This keeps the adapted spatial-kernel identity but not the
geometry needed in two dimensions: raw bivariate monomials are ill-conditioned on
anisotropic state clouds, drift rows have unequal noise levels, and independently
fitted tensor entries need not form covariance matrices.  \Cref{fig:naive}
shows that the dominant failure is drift in low-SNR or scale-disparate systems,
with tensor and PSD failures appearing where the scalar fits ignore covariance
structure.

\subsection{Ablation: mechanism contributions}\label{sec:ablation}
\begin{figure}[!htbp]\centering
\includegraphics[width=0.78\linewidth]{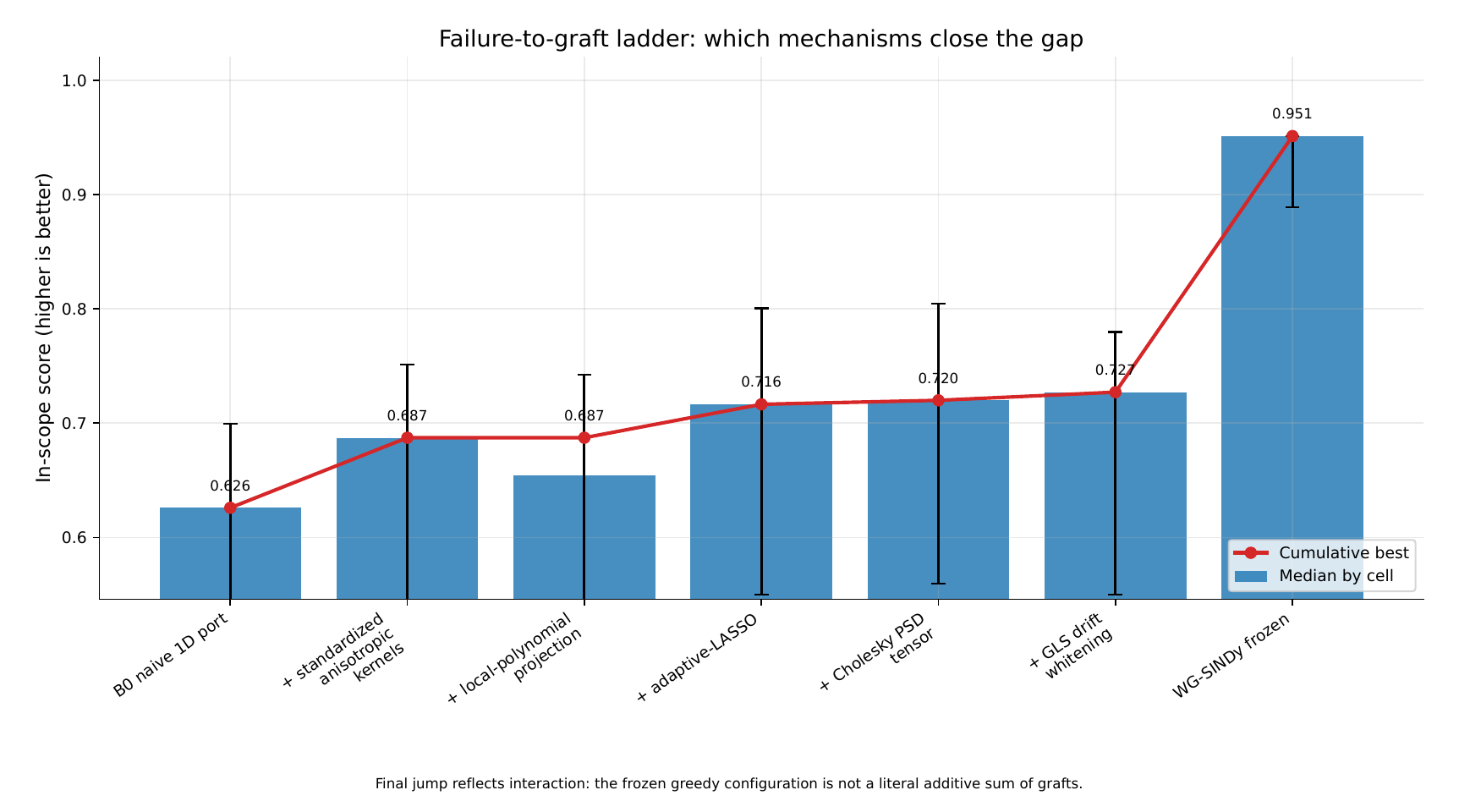}
\caption{Cumulative graft ladder. The in-scope score improves from the naive port to the frozen WG-SINDy
configuration; the last jump reflects interaction among fitted mechanisms.}\label{fig:ladder}
\end{figure}

\begin{figure}[!htbp]\centering
\includegraphics[width=0.85\linewidth]{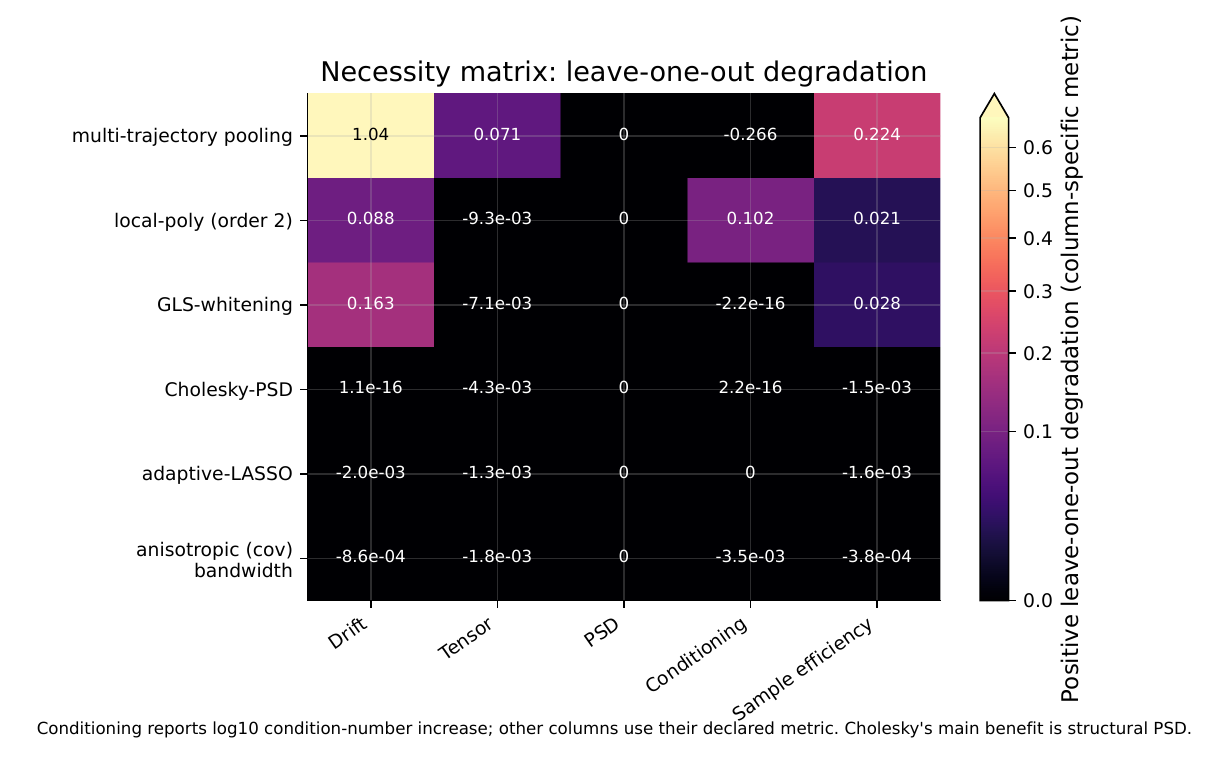}
\caption{Leave-one-out analysis of the six estimator components. Conditioning reports the increase in the design
condition number, while the other columns use their declared capability-specific metrics. Pooling, feasible diagonal
GLS, and local-polynomial smoothing show the clearest in-scope degradation when removed. The remaining components are
retained principally for structural validity or stress-regime protection; this panel does not establish their individual
necessity for the median in-scope score.}\label{fig:necessity}
\end{figure}

\Cref{fig:ladder} adds the mechanisms in the order used by the estimator:
standardized anisotropic kernels, local-polynomial projection, adaptive
sparsification, Cholesky tensor recovery, GLS drift whitening, and pooled
trajectory selection.  The score rises from the naive port to the frozen method,
with the final jump reflecting interaction among jointly tuned components.
\Cref{fig:necessity} gives the leave-one-out view. Pooling, feasible GLS, and
local-polynomial smoothing have clear in-scope effects in the checked-in
necessity matrix. Cholesky recovery, adaptive LASSO, and anisotropic bandwidths
are important structural or stress-regime safeguards, but this matrix does not by
itself establish that each is empirically necessary on the in-scope median score.


\subsection{Descriptive comparison with in-repository baseline proxies}\label{sec:headtohead}
\begin{figure}[!htbp]\centering
\includegraphics[width=0.9\linewidth]{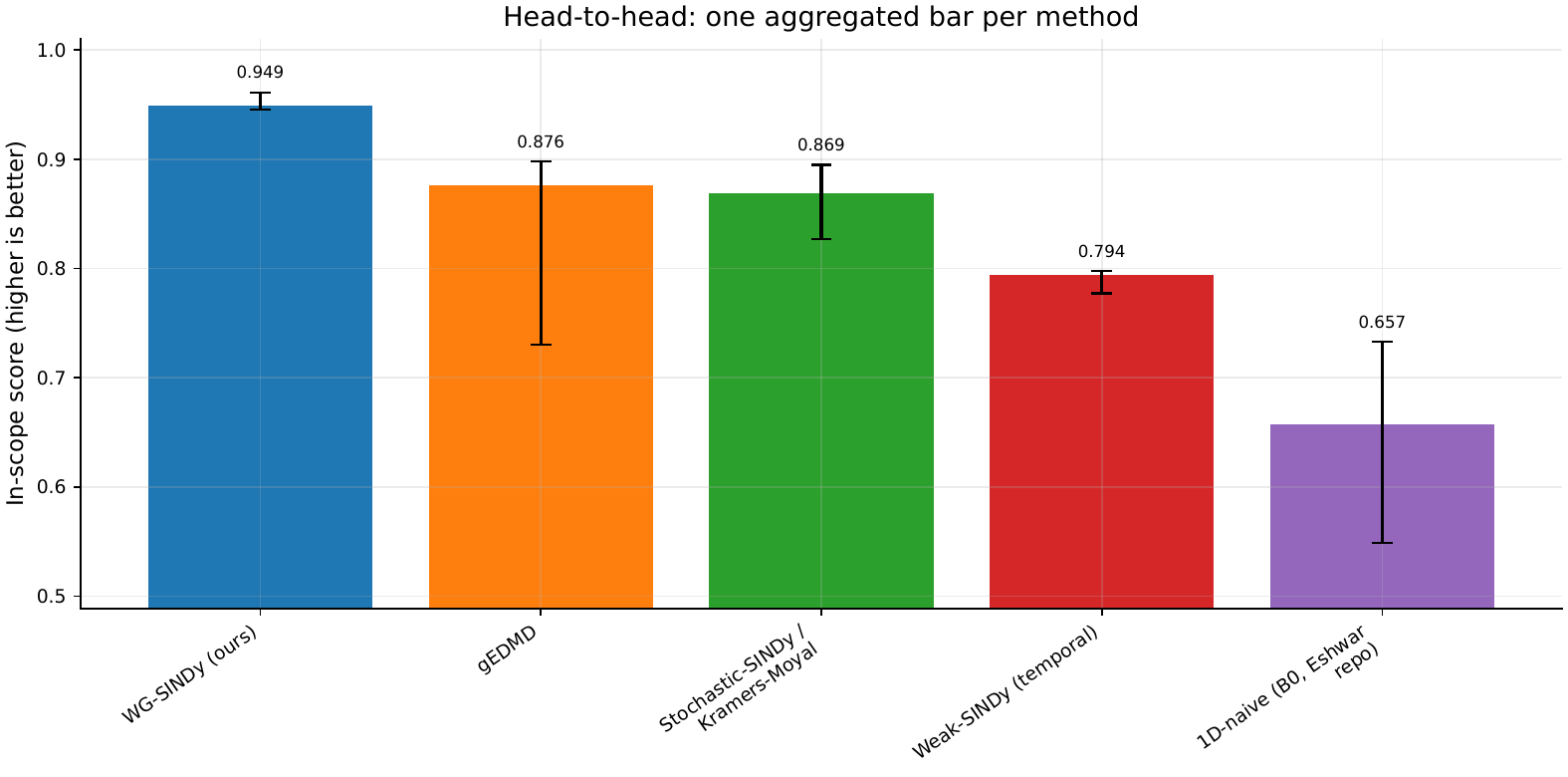}
\caption{Descriptive median in-scope scores for WG-SINDy and four in-repository baseline proxies. Data budgets differ
across configurations, so the panel is a diagnostic comparison rather than a controlled method-level
ranking.}\label{fig:h2h}
\end{figure}

\Cref{fig:h2h} compares WG-SINDy with four lightweight in-repository baseline
proxies: a local Kramers--Moyal/stochastic-SINDy moment estimator, a temporal
Weak-SINDy proxy, a dense generator-EDMD proxy, and the naive one-dimensional
weak-form port. These are method-family diagnostics rather than certified
reproductions of the original published implementations. The data budgets are
also unequal: WG-SINDy and the generator-EDMD proxy use $R=16$, the local-moment
and temporal-weak proxies use $R=4$, and the naive port uses $R=1$. The resulting
scores are therefore descriptive and should not be interpreted as a controlled
ranking of the underlying published methods.

\subsection{Per-system baseline-proxy comparison}\label{sec:per-system-baselines}
\begin{figure}[!htbp]\centering
\includegraphics[width=0.98\linewidth]{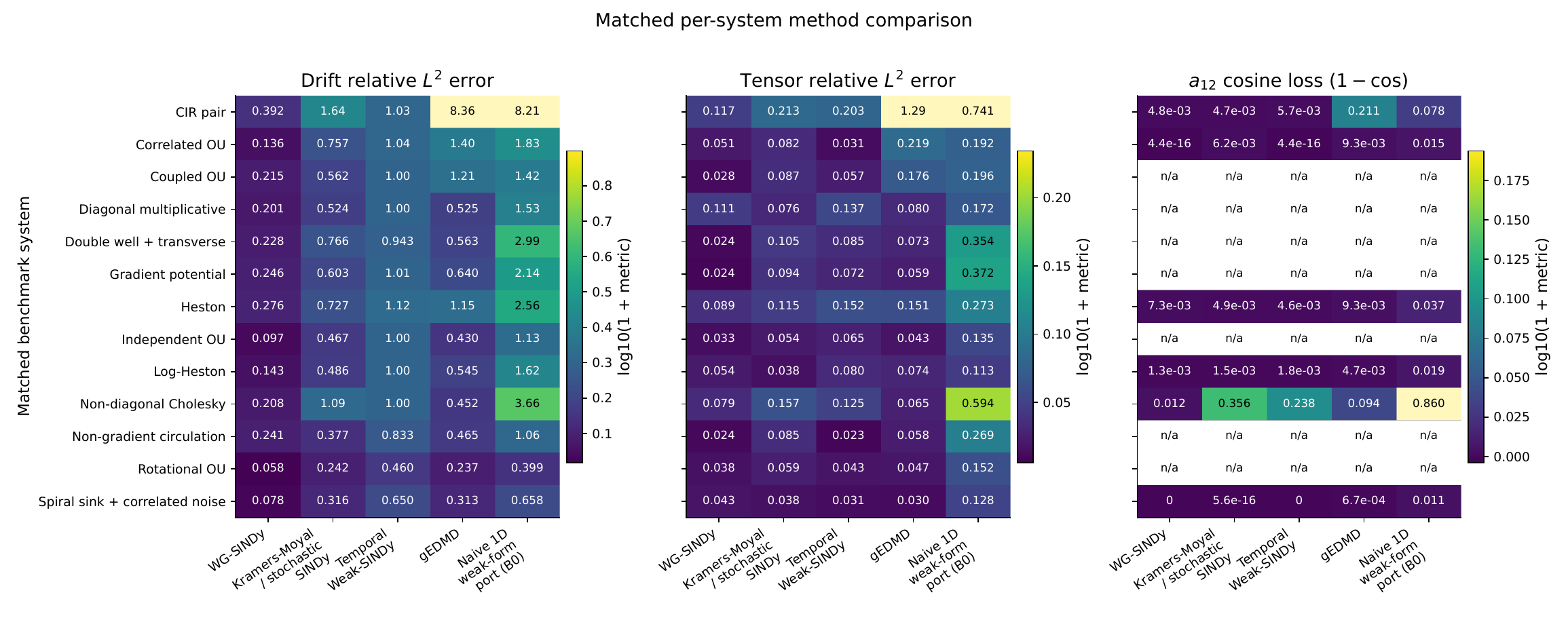}
\caption{Per-system comparison using the shipped in-repository baseline proxies. Cell values are raw central-grid drift
relative errors, tensor relative errors, and $1-\cos(\hat\diff_{12},\diff_{12})$. Lower is better; colour uses
$\log_{10}(1+\mathrm{metric})$. Proxy implementations and data budgets differ across
columns.}\label{fig:per-system-method-heatmap}
\end{figure}

\Cref{fig:per-system-method-heatmap} reports the same in-repository proxy
comparison system by system. Cell values are the raw common metrics available
from the shipped outputs. The colour transformation is used only for
readability. Because these are proxy implementations with unequal trajectory
budgets, the panel should be read as a failure-mode diagnostic rather than as
evidence that one published method universally outperforms another.

\subsection{The three read-outs}\label{sec:readouts}
\begin{figure}[!htbp]\centering
\includegraphics[width=0.7\linewidth]{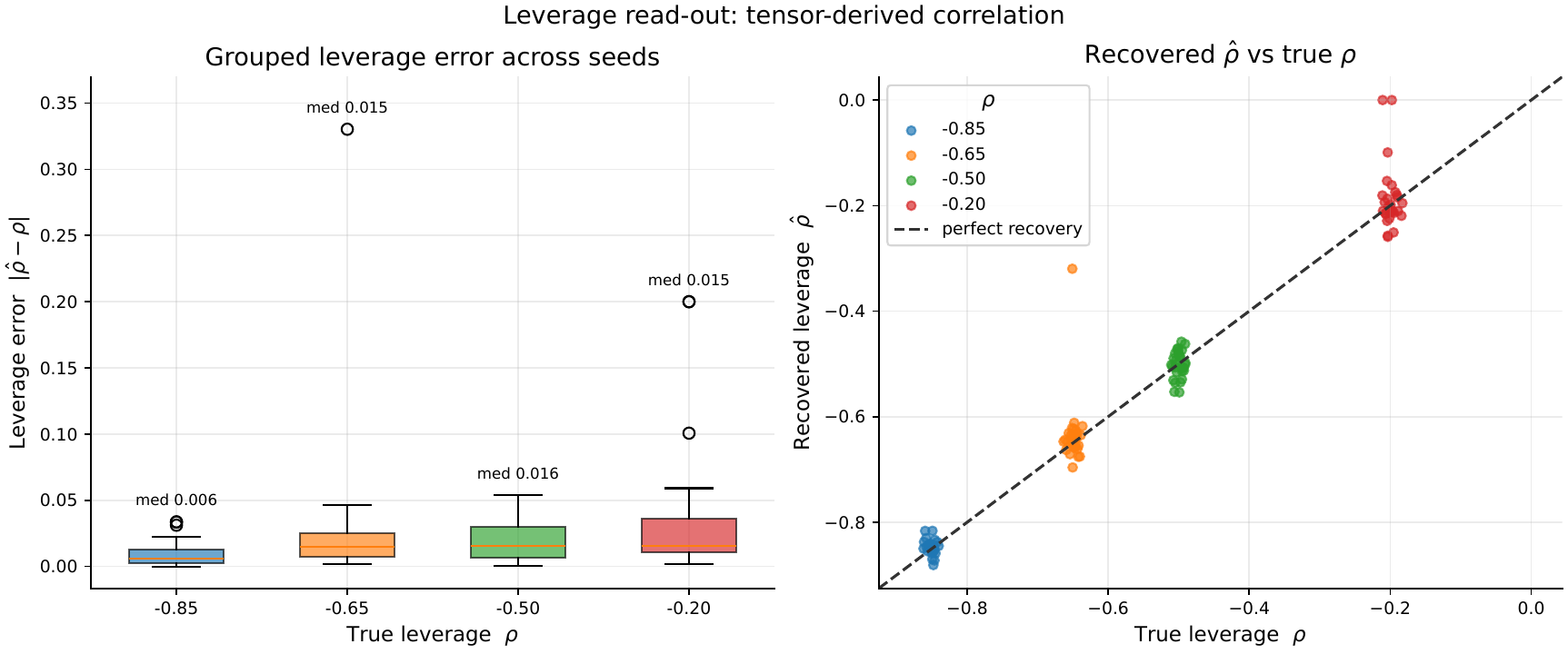}
\caption{Leverage read-out from the recovered tensor. The estimator preserves the sign and magnitude of
$\rho=\diff_{12}/\sqrt{\diff_{11}\diff_{22}}$.}\label{fig:leverage}
\end{figure}

Once the generator is recovered, physical summaries become read-outs rather than
separate fits. Normalizing $\hat\diff_{12}$ by the diagonal tensor entries gives
the instantaneous shock correlation, and \Cref{fig:leverage} shows that its sign
and magnitude are preserved across leverage regimes. The antisymmetric part of
the recovered drift gives a drift-curl diagnostic; in the isotropic rotational OU
case this diagnostic agrees with the stationary-current direction, while for
nonlinear systems it is reported only as an antisymmetric-drift direction. The
full tensor field provides the third read-out: local fluctuation geometry, shown
in the datasheets.

\begin{figure}[!htbp]\centering
\includegraphics[width=0.7\linewidth]{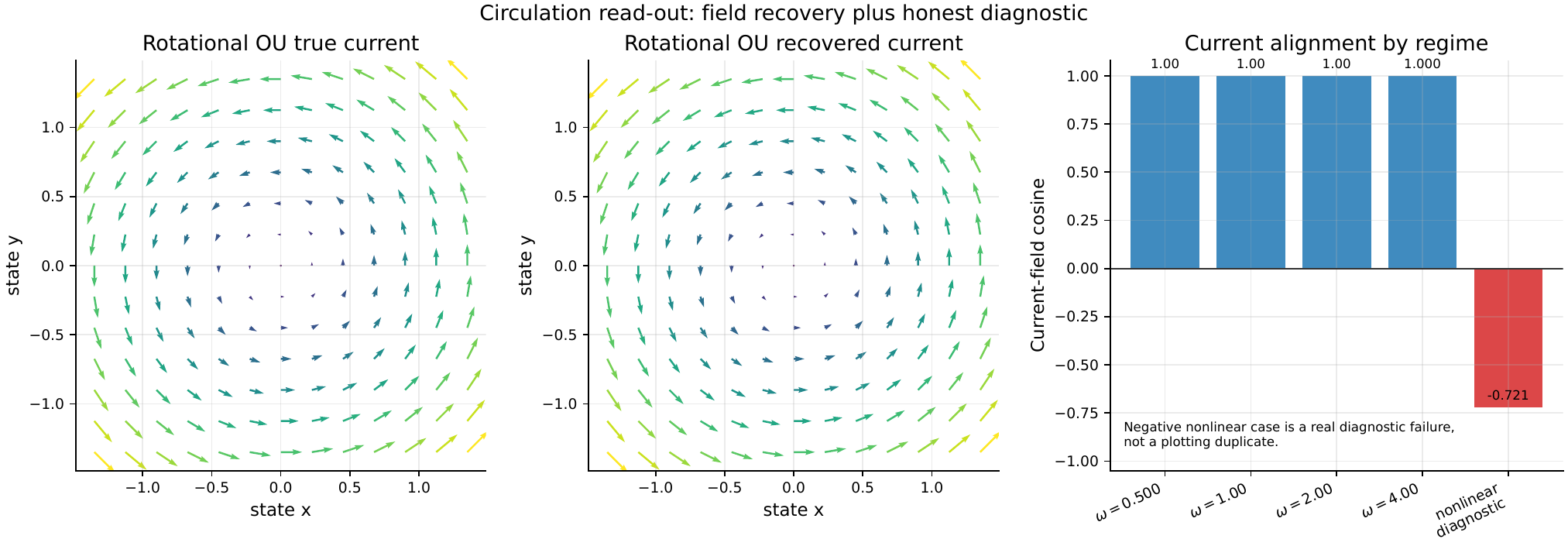}
\caption{Recovered antisymmetric-drift direction in the rotational OU benchmark. The stationary-current
interpretation is restricted to this isotropic linear case; nonlinear rows are drift-curl diagnostics only.}\label{fig:circulation}
\end{figure}

\subsection{Convergence and honest nulls}\label{sec:nulls}
\begin{figure}[!htbp]\centering
\includegraphics[width=0.7\linewidth]{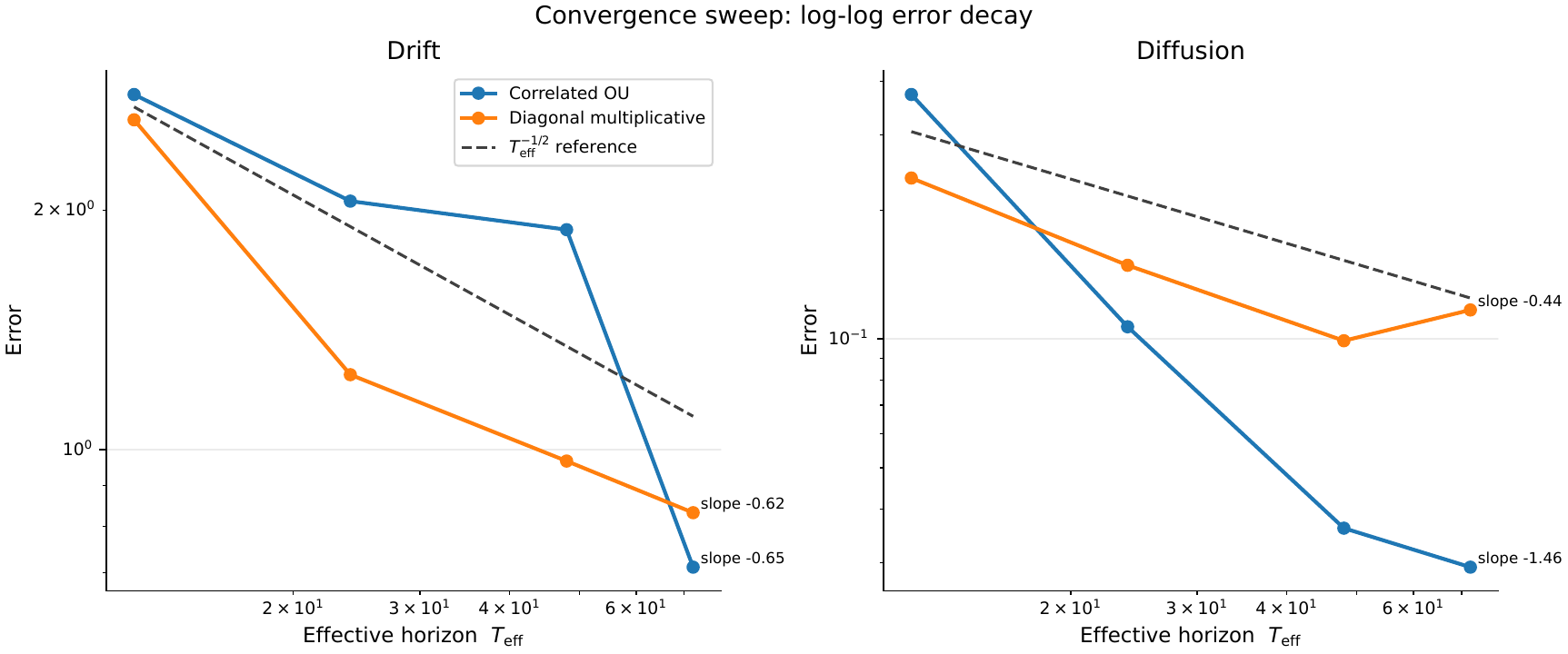}
\caption{Convergence with effective horizon. Errors decrease until finite-step, library, or conditioning floors dominate.}\label{fig:conv}
\end{figure}

\begin{figure}[!htbp]\centering
\includegraphics[width=0.85\linewidth]{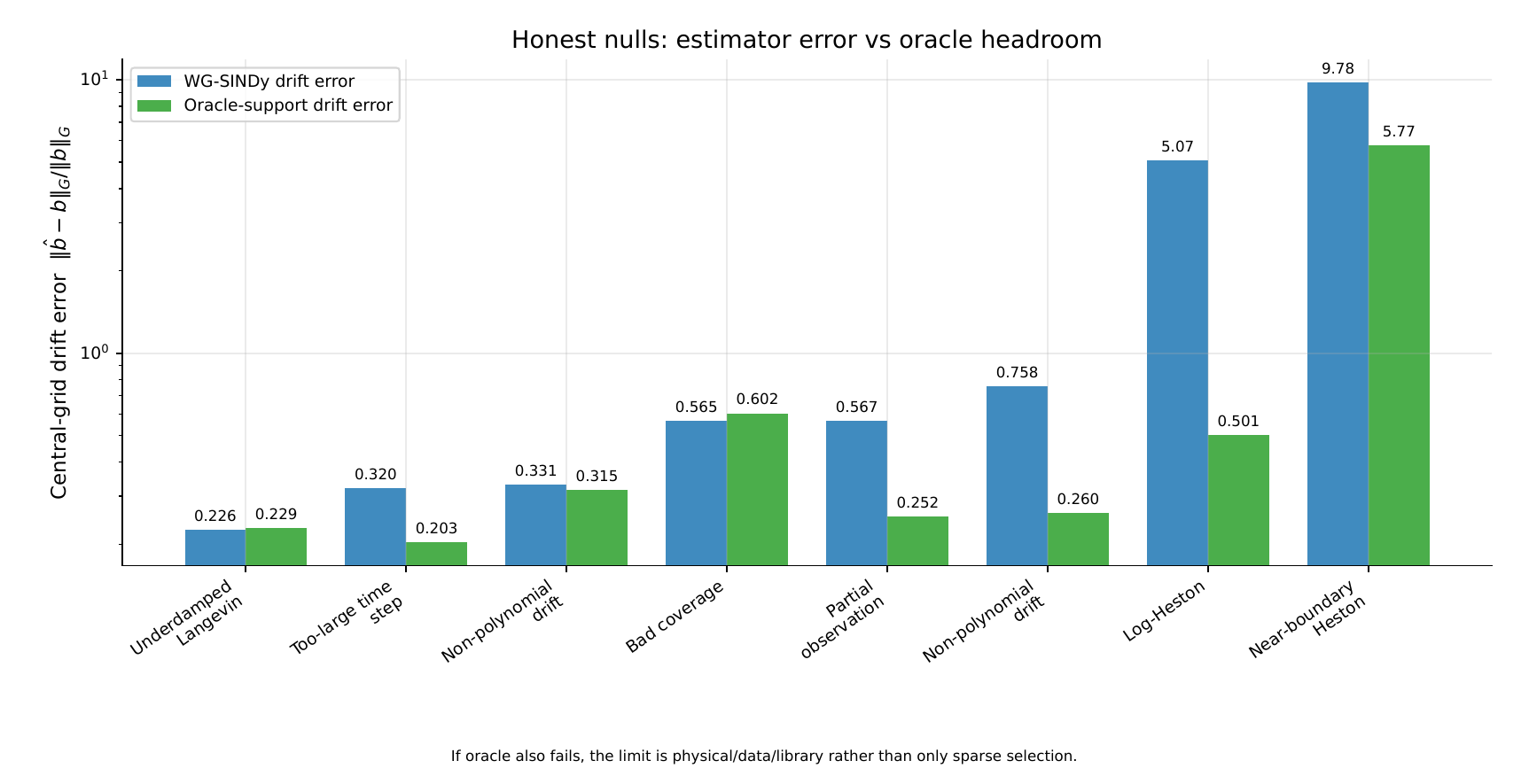}
\caption{Honest nulls. WG-SINDy is compared with an oracle-support solve to separate sparse-selection error from
physical, data-geometric, or library limits.}\label{fig:nulls}
\end{figure}

\Cref{fig:conv} checks that errors decrease with effective horizon when the
problem is identifiable, until finite-step, library, or conditioning floors
dominate.  \Cref{fig:nulls} gives the complementary check by comparing WG-SINDy
with an oracle-support solve on systems that violate recovery conditions.  When
the oracle also has large error, the failure is not sparse selection alone.  The
panel names the limiting axis---low drift SNR, near-singular tensors, boundary
concentration, missing library terms, partial observation, insufficient coverage,
or too-large $\Delta t$---so the method reports when the data do not identify the
claimed symbolic generator.


\section{Discussion and Limitations}\label{sec:discussion}
\paragraph{How to read the relative errors.}
The errors in \Cref{tab:master} are generator-recovery diagnostics, not claims of
sub-percent parametric precision.  A useful recovery must select the right terms,
suppress spurious ones, match the fields on the sampled region, recover the
off-diagonal tensor direction when present, and return a valid covariance matrix.
This is why field errors, support counts, off-diagonal cosines, and PSD fractions
are reported separately.

The main limitation is the information content of the paths.  The weak form
removes derivative estimation and reduces martingale noise, but it cannot excite
unvisited state regions, detect terms below the weak-row noise floor, or span a
generator outside the library.  Tensor and leverage terms in stochastic-volatility
models are recoverable because they enter quadratic variation, while log-price
drift is an $O(\Delta t)$ signal hidden under $O(\sqrt{\Delta t})$ fluctuations.
Thus the method should be used with coverage, conditioning, support, and PSD
diagnostics: when those checks fail, the sampling regime or library must change
before a symbolic generator claim is defensible.

\section{Conclusion}\label{sec:conclusion}
We extended weak-form spatial-kernel generator recovery from scalar diffusions to
two-dimensional It\^o systems. The extension is motivated by the adapted
fixed-weight spatial identity used in the one-dimensional construction, while
adding the machinery needed in two dimensions: standardized anisotropic kernels, local-polynomial weak
projection, sparse adaptive selection, feasible GLS drift reweighting, a PSD
Cholesky read-out, and trajectory pooling. Because the released smoother is
estimated from the full sampled state cloud, the present paper treats this
identity as population motivation rather than as an exact finite-sample
unbiasedness theorem for the implemented estimator. The resulting frozen estimator
recovers both drift components and the full symmetric diffusion tensor inside the
declared identifiability envelope, including off-diagonal noise correlation,
while returning a positive-semidefinite tensor by construction. The benchmark
shows useful recovery across 19 in-scope systems and also identifies 10 systems
where the data, library, or sampling regime do not support a stronger symbolic
recovery claim.

\bibliographystyle{plainnat}
\bibliography{references}
\appendix
\clearpage

\section*{Supplementary Material: per-system datasheets}\label{app:datasheets}
\addcontentsline{toc}{section}{Supplementary Material}
The following per-system datasheets are supplementary. The main-text claims rest on the flagship examples
(\Cref{sec:readouts}) and the navigation index \Cref{tab:master}; each datasheet below gives the SDE, its
analytic generator, the recovered symbolic coefficients, the recovered-vs-true fields, the quantitative
verdict (with the failing axis named for limit cases, per \Cref{sec:method_metrics}), and a discussion. Systems are
grouped by family.
\IfFileExists{ds_linear.tex}{
\paragraph{Linear Ornstein--Uhlenbeck systems}\;

\subsection{Independent OU}\label{sec:v62-indep-ou}
\paragraph{Context.} The independent two-dimensional Ornstein--Uhlenbeck process is the cleanest linear control in the benchmark. Its coordinates have unequal relaxation rates and unequal marginal noise amplitudes, but no cross-drift coupling and no instantaneous noise correlation. It therefore checks that the estimator recovers two distinct mean-reversion scales without inventing either drift coupling or an off-diagonal diffusion term.
\paragraph{System.} $$\mathrm{d}X=-X\,\mathrm{d}t+\mathrm{d}W_1,\qquad \mathrm{d}Y=-2Y\,\mathrm{d}t+0.7\,\mathrm{d}W_2,\qquad \mathrm{d}W_1\mathrm{d}W_2=0.$$
\paragraph{Generator.} The drift is $\drift=(-x,-2y)$ and the constant diffusion tensor is $\diff=\operatorname{diag}(1,0.49)$. The true off-diagonal entry is zero, so an $a_{12}$ cosine is not defined; recovery is judged by drift and tensor errors, PSD validity, and the stable projected-support rule.
\paragraph{Recovery.} At the frozen pooled WG-SINDy configuration, the 10-seed median drift central-grid relative $L^2$ error is $0.098$ and the tensor relative $L^2$ error is $0.033$. The recovered tensor is PSD on all evaluation-grid points. The paper-level stable projected-support false-positive count is zero under the declared post-processing rule. Coefficient-level selection rates vary by term and are reported in the coefficient ledger; this datasheet does not claim selection of every active term in every seed. This supports the stated in-scope linear-control claim; it does not by itself establish universal recovery from arbitrary sampling regimes.
\begin{figure}[!htbp]\centering
\includegraphics[width=0.72\linewidth,height=0.50\textheight,keepaspectratio]{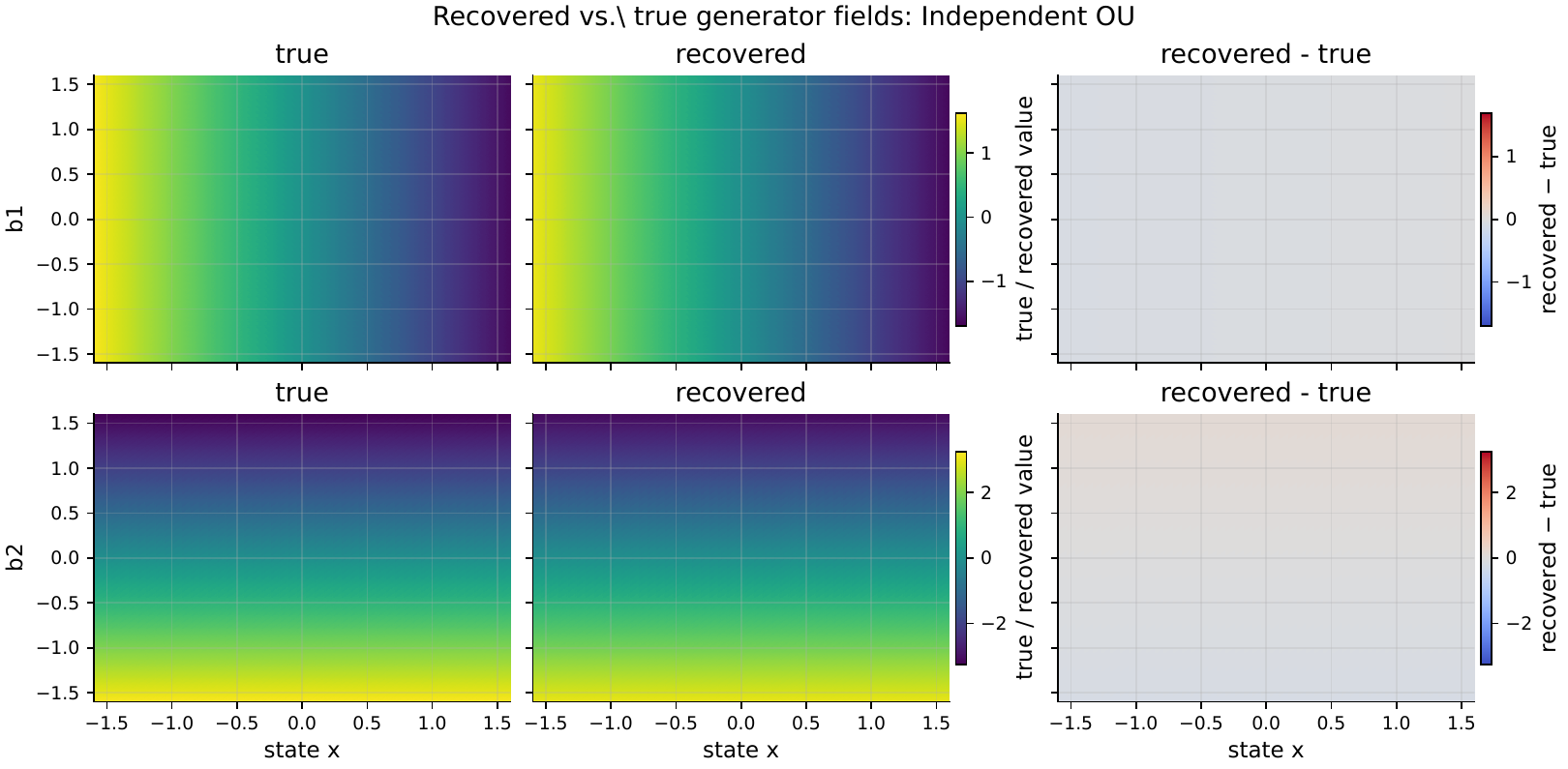}
\caption{Independent OU: recovered versus true generator fields (shared per-field colour scale; error column centred at zero).}
\end{figure}
\paragraph{Verdict.} Drift central-grid rel-$L^2=0.098$, tensor rel-$L^2=0.033$, $a_{12}$ cosine $\mathrm{n/a}$, PSD $1.00$, stable projected-support FP $0$ ($n=10$ seeds). \textbf{PASS}.
\medskip

\subsection{Correlated OU}\label{sec:v62-correlated-ou}
\paragraph{Context.} The Ornstein--Uhlenbeck process is the canonical mean-reverting diffusion and the natural two-dimensional starting point. Here the two coordinates relax independently but are driven by \emph{correlated} Brownian motions, so the coupling lives entirely in the diffusion tensor rather than the drift. It is the simplest system in which a constant off-diagonal $\diff_{12}$ must be recovered, and it isolates the cross-variation channel that the one-dimensional theory never exercises: any spurious off-diagonal here would be a false leverage signal, so it is also a strict false-positive control.
\paragraph{System.} $$\mathrm{d}X=-X\,\mathrm{d}t+\sigma_1\mathrm{d}W_1,\quad \mathrm{d}Y=-1.5Y\,\mathrm{d}t+\sigma_2\mathrm{d}W_2,\ \mathrm{d}W_1\mathrm{d}W_2=\rho\,\mathrm{d}t.$$
\paragraph{Generator.} Diagonal mean reversion with correlated noise ($\rho=-0.6$). $\drift$ pulls each coordinate to zero; the constant off-diagonal $\diff_{12}=\rho\sigma_1\sigma_2=-0.48$ is the leverage channel.
\paragraph{Recovery.} WG-SINDy recovers the drift at central-grid relative $L^2=0.136$ and the diffusion tensor at $0.051$ relative error, with off-diagonal (leverage) cosine $1.000$; the recovered generator gives the relaxation rates and cross-coupling (and, where present, the leverage correlation). The paper-level stable projected-support filter counts no large recurring false positives under the declared post-processing rule. Coefficient-level selection rates vary by term and are reported in the coefficient ledger; this datasheet does not claim selection of every active term in every seed.
\begin{figure}[!htbp]\centering\includegraphics[width=0.72\linewidth,height=0.50\textheight,keepaspectratio]{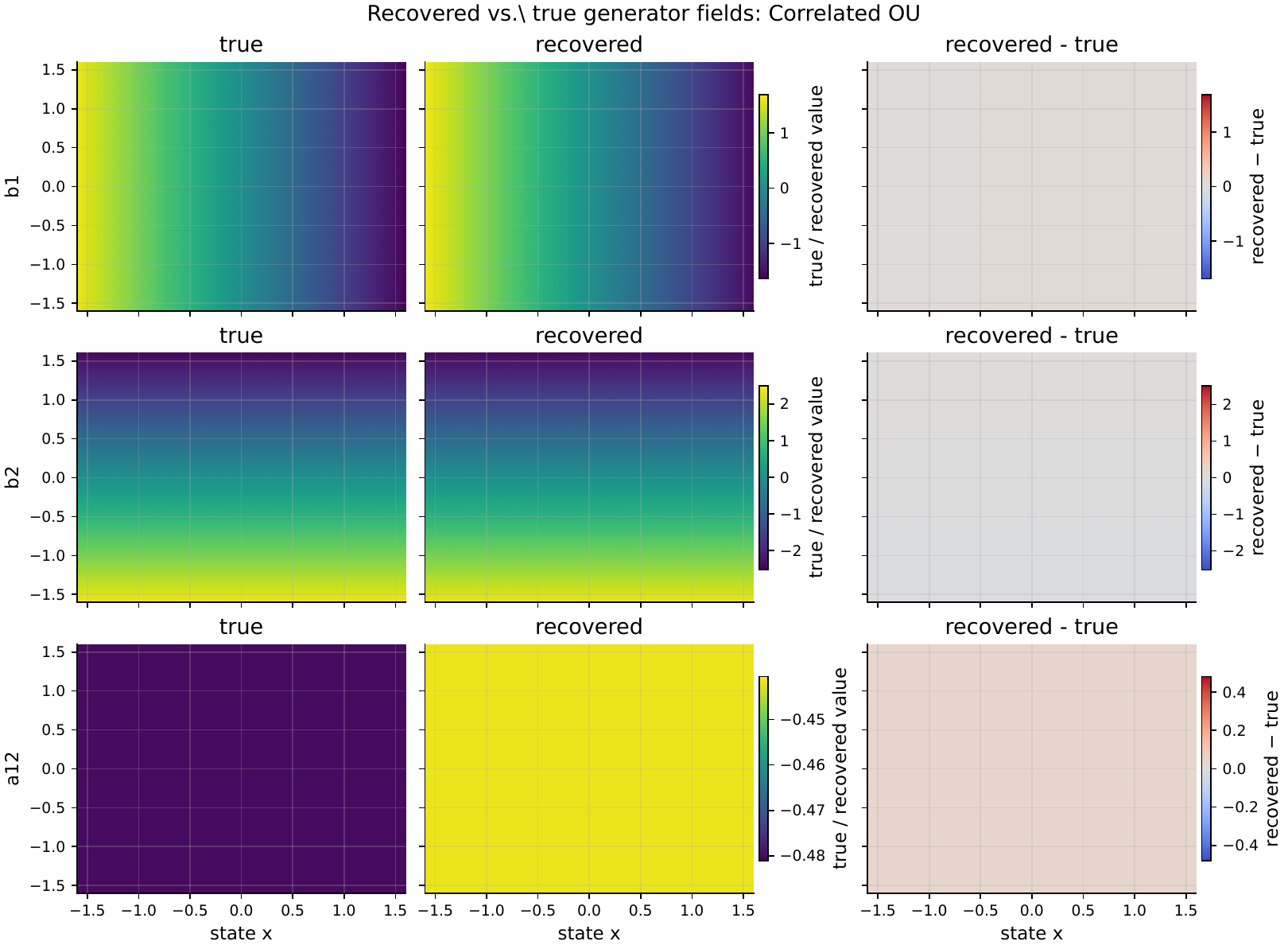}
\caption{Correlated OU: recovered versus true generator fields (shared per-field colour scale; error column centred at zero).}\end{figure}
\paragraph{Verdict.} Drift central-grid rel-$L^2=0.136$, tensor rel-$L^2=0.051$, $a_{12}$ cosine $1.000$, PSD $1.00$, stable projected-support FP $0$ ($n=10$ seeds). \textbf{PASS}.
\medskip

\subsection{Coupled OU}\label{sec:v62-coupled-ou}
\paragraph{Context.} A linear system in which the two coordinates are coupled through the \emph{drift} rather than the noise, the analogue of a two-body linear relaxation. Because the drift matrix is symmetric the dynamics remain reversible, so it serves as a negative control for the antisymmetric-drift diagnostic: the recovered drift Jacobian should be numerically symmetric, distinguishing genuine coupling from non-equilibrium rotation.
\paragraph{System.} $$\mathrm{d}X=(-X+0.5Y)\mathrm{d}t+\mathrm{d}W_1,\quad \mathrm{d}Y=(0.5X-Y)\mathrm{d}t+\mathrm{d}W_2.$$
\paragraph{Generator.} Cross-state linear drift, isotropic noise. The symmetric drift Jacobian makes it reversible (no antisymmetric drift); $\diff=I$.
\paragraph{Recovery.} WG-SINDy recovers the drift at central-grid relative $L^2=0.214$ and the diffusion tensor at $0.028$ relative error; the recovered generator gives the relaxation rates and cross-coupling (and, where present, the leverage correlation). The paper-level stable projected-support filter counts no large recurring false positives under the declared post-processing rule. Coefficient-level selection rates vary by term and are reported in the coefficient ledger; this datasheet does not claim selection of every active term in every seed.
\begin{figure}[!htbp]\centering\includegraphics[width=0.72\linewidth,height=0.50\textheight,keepaspectratio]{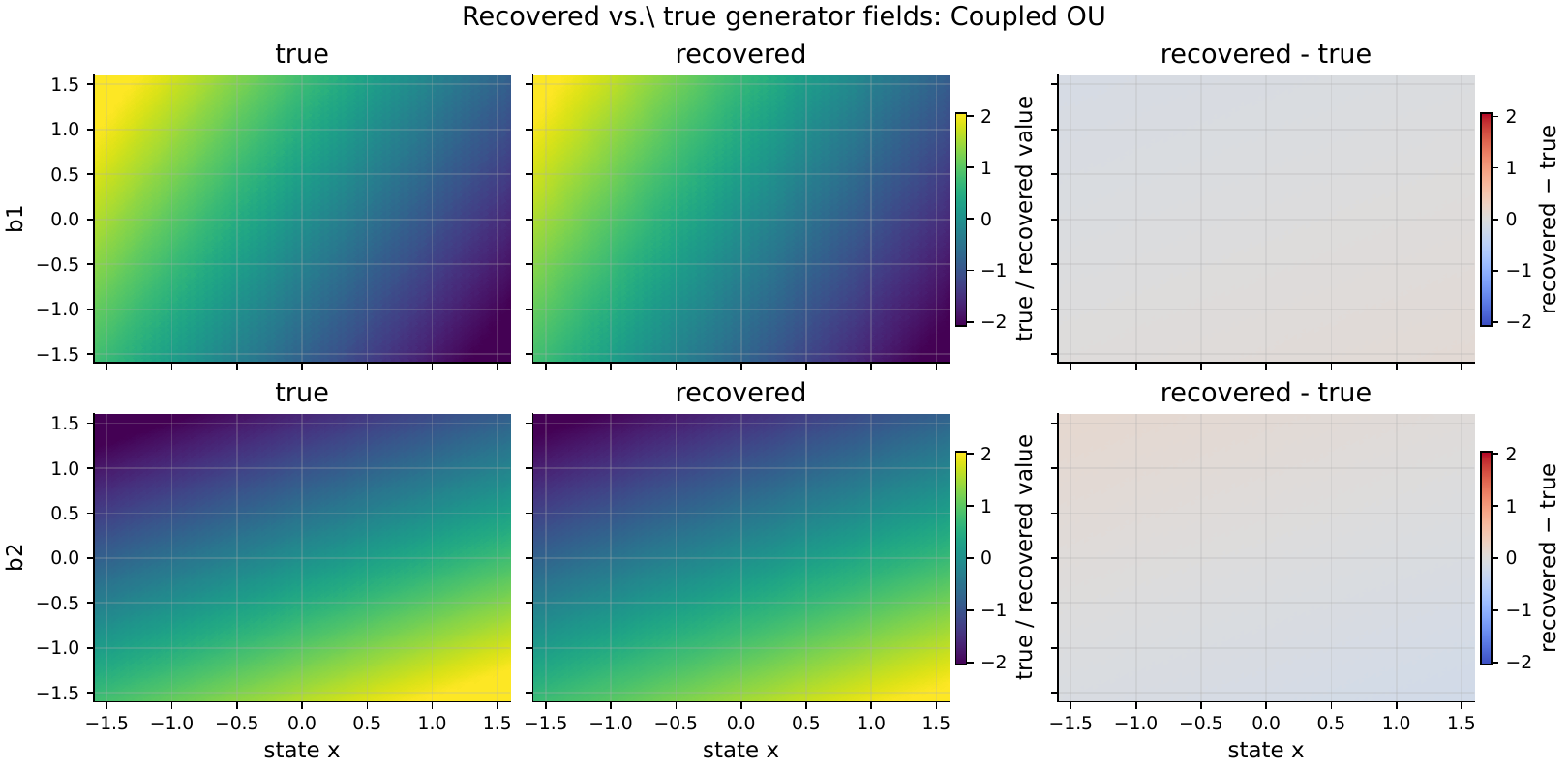}
\caption{Coupled OU: recovered versus true generator fields (shared per-field colour scale; error column centred at zero).}\end{figure}
\paragraph{Verdict.} Drift central-grid rel-$L^2=0.214$, tensor rel-$L^2=0.028$, $a_{12}$ cosine $\mathrm{n/a}$, PSD $1.00$, stable projected-support FP $0$ ($n=10$ seeds). \textbf{PASS}.
\medskip}{}
\IfFileExists{ds_rotational.tex}{
\paragraph{Rotational and non-reversible systems}\;

\subsection{Rotational OU}\label{sec:v62-rotational-ou}
\paragraph{Context.} The prototypical \emph{non-reversible} linear diffusion: a damped rotation about the origin. It is the linear template for the antisymmetric-drift diagnostic. The drift Jacobian carries a non-zero antisymmetric part $\omega J$, which in this isotropic linear case aligns with the stationary-current direction; the radial damping $\alpha$ fixes the spectral gap and $\omega$ the rotation frequency, both of which must be read off the recovered generator.
\paragraph{System.} $$\mathrm{d}(X,Y)^\top=\begin{psmallmatrix}-1&-2\\2&-1\end{psmallmatrix}(X,Y)^\top\mathrm{d}t+\mathrm{d}W.$$
\paragraph{Generator.} Damped rotation: $\alpha=1$ sets the radial decay (spectral gap), $\omega=2$ the rotation; antisymmetric drift gives the rotational-drift diagnostic. $\diff=I$.
\paragraph{Recovery.} WG-SINDy recovers the drift at central-grid relative $L^2=0.058$ and the diffusion tensor at $0.038$ relative error; the antisymmetric part of the recovered drift gives the antisymmetric-drift diagnostic, while the symmetric part is the relaxation. The paper-level stable projected-support filter counts no large recurring false positives under the declared post-processing rule. Coefficient-level selection rates vary by term and are reported in the coefficient ledger; this datasheet does not claim selection of every active term in every seed.
\begin{figure}[!htbp]\centering\includegraphics[width=0.72\linewidth,height=0.50\textheight,keepaspectratio]{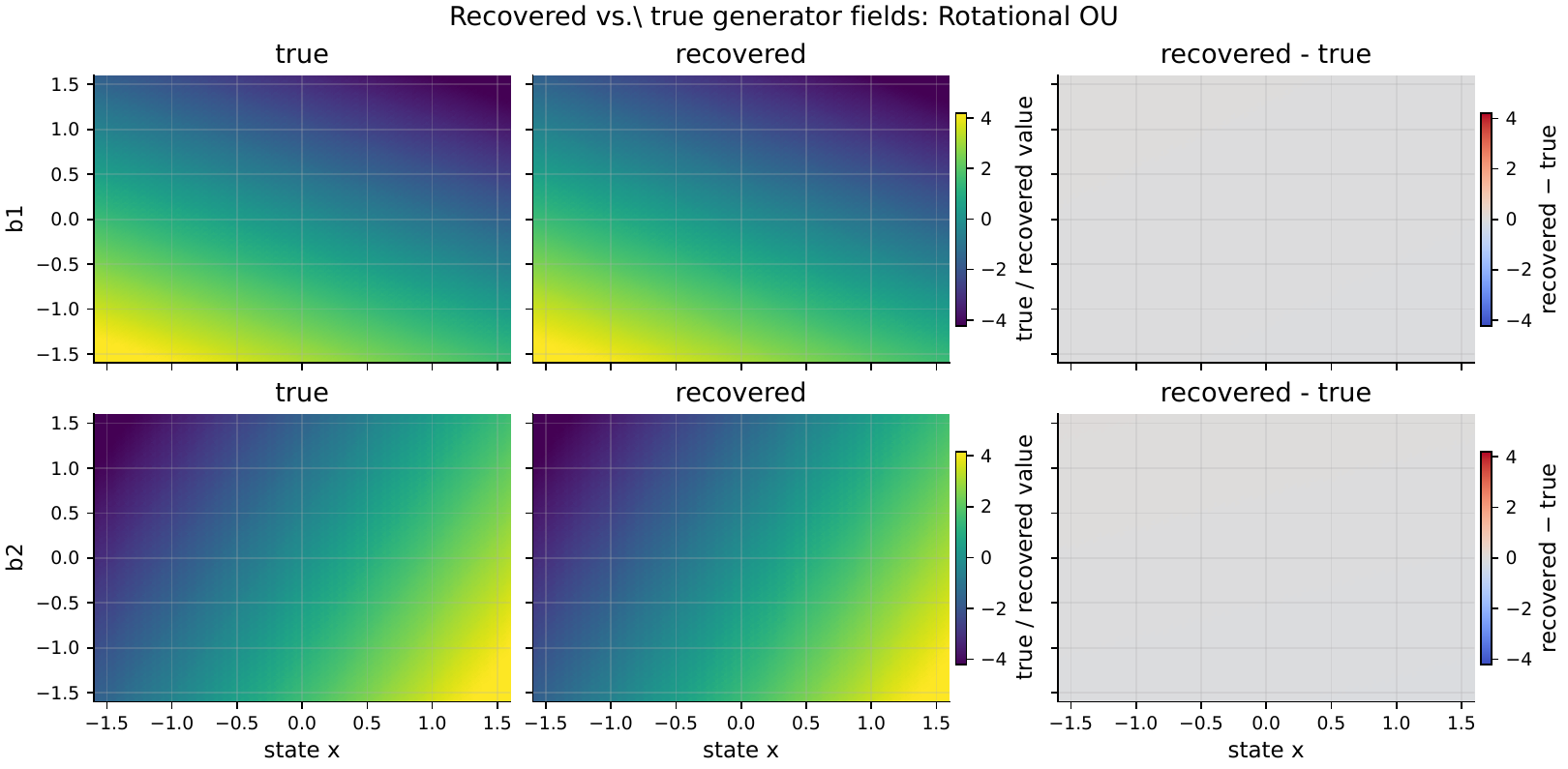}
\caption{Rotational OU: recovered versus true generator fields (shared per-field colour scale; error column centred at zero).}\end{figure}
\paragraph{Verdict.} Drift central-grid rel-$L^2=0.058$, tensor rel-$L^2=0.038$, $a_{12}$ cosine $\mathrm{n/a}$, PSD $1.00$, stable projected-support FP $0$ ($n=10$ seeds). \textbf{PASS}.
\medskip

\subsection{Spiral sink + correlated noise}\label{sec:v62-spiral-sink-corr}
\paragraph{Context.} A stress test that switches on both two-dimensional channels at once: a non-reversible rotational drift \emph{and} a correlated (off-diagonal) diffusion. It probes whether the shared design matrix entangles the drift and tensor estimates or keeps them separable, since recovering the rotation and the noise correlation simultaneously is exactly the regime where a naive estimator confounds the two.
\paragraph{System.} $$\drift=(-X-1.5Y,\,1.5X-Y),\quad \diff_{12}=\rho\sigma_1\sigma_2=-0.40.$$
\paragraph{Generator.} Non-reversible rotation and correlated noise together: exercises the rotational-drift diagnostic and the leverage channel at once.
\paragraph{Recovery.} WG-SINDy recovers the drift at central-grid relative $L^2=0.078$ and the diffusion tensor at $0.043$ relative error, with off-diagonal (leverage) cosine $1.000$; the antisymmetric part of the recovered drift gives the antisymmetric-drift diagnostic, while the symmetric part is the relaxation. The paper-level stable projected-support filter counts no large recurring false positives under the declared post-processing rule. Coefficient-level selection rates vary by term and are reported in the coefficient ledger; this datasheet does not claim selection of every active term in every seed.
\begin{figure}[!htbp]\centering\includegraphics[width=0.72\linewidth,height=0.50\textheight,keepaspectratio]{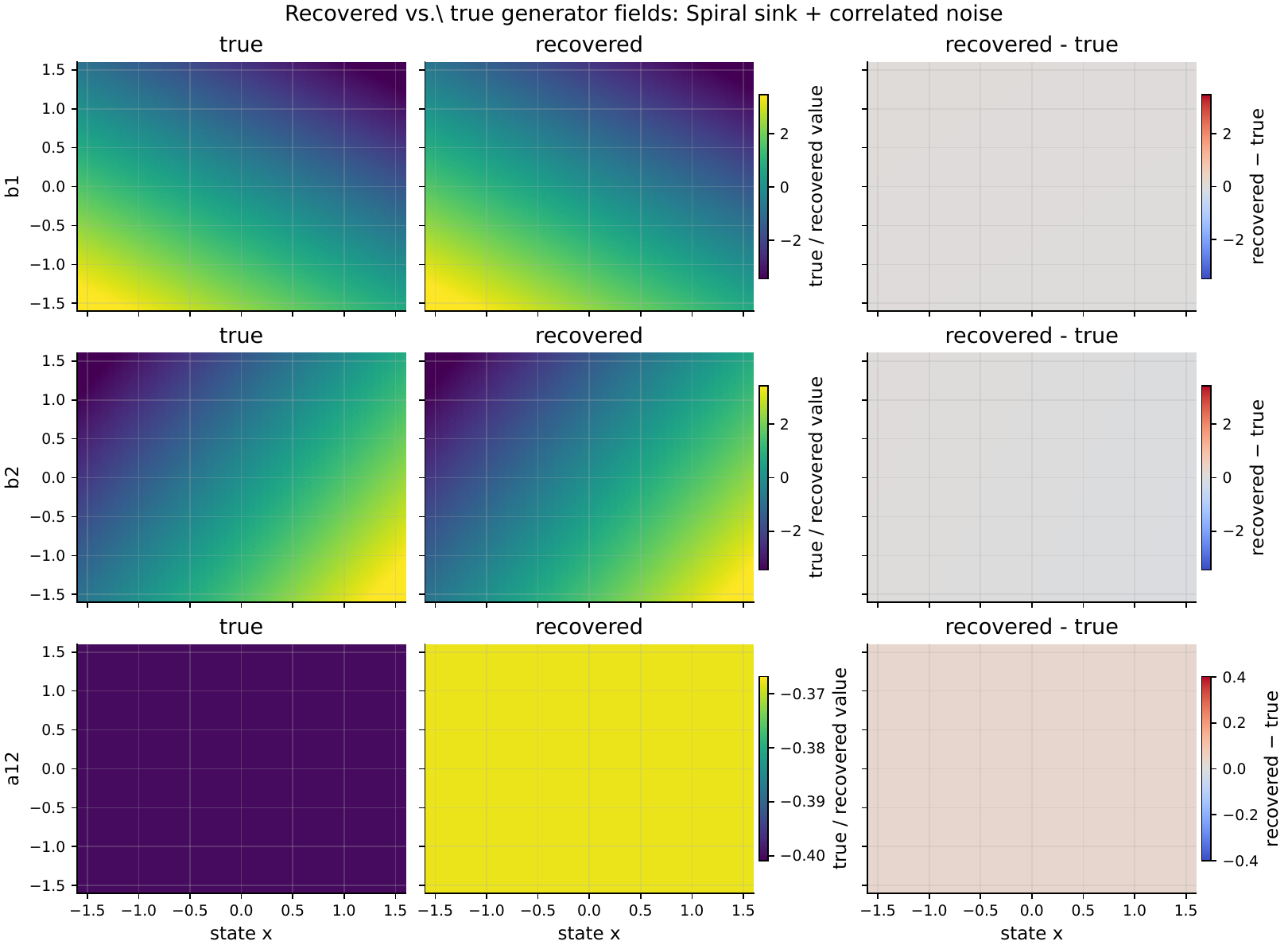}
\caption{Spiral sink + correlated noise: recovered versus true generator fields (shared per-field colour scale; error column centred at zero).}\end{figure}
\paragraph{Verdict.} Drift central-grid rel-$L^2=0.078$, tensor rel-$L^2=0.043$, $a_{12}$ cosine $1.000$, PSD $1.00$, stable projected-support FP $0$ ($n=10$ seeds). \textbf{PASS}.
\medskip

\subsection{Non-gradient drift-curl}\label{sec:v62-nongradient-circulation}
\paragraph{Context.} A bistable energy landscape with an added non-conservative rotational drift, $\drift=-\nabla V+\omega J\nabla V$. It is the nonlinear counterpart of rotational OU and a nonlinear test of the drift decomposition: the method must split the recovered drift into a conservative part that rebuilds the double-well potential and a rotational part that diagnoses broken detailed balance in the sampled region.
\paragraph{System.} $$\drift=-\nabla V+\omega J\nabla V,\ V=\tfrac14(x^2-1)^2+\tfrac12 y^2+\tfrac14 x^2y^2.$$
\paragraph{Generator.} Same bistable potential as the gradient case but with an added curl $\omega J\nabla V$; the conservative and rotational drift parts are recovered separately.
\paragraph{Recovery.} WG-SINDy recovers the drift at central-grid relative $L^2=0.239$ and the diffusion tensor at $0.024$ relative error; the antisymmetric part of the recovered drift gives the antisymmetric-drift diagnostic, while the symmetric part is the relaxation. The paper-level stable projected-support filter counts no large recurring false positives under the declared post-processing rule. Coefficient-level selection rates vary by term and are reported in the coefficient ledger; this datasheet does not claim selection of every active term in every seed.
\begin{figure}[!htbp]\centering\includegraphics[width=0.72\linewidth,height=0.50\textheight,keepaspectratio]{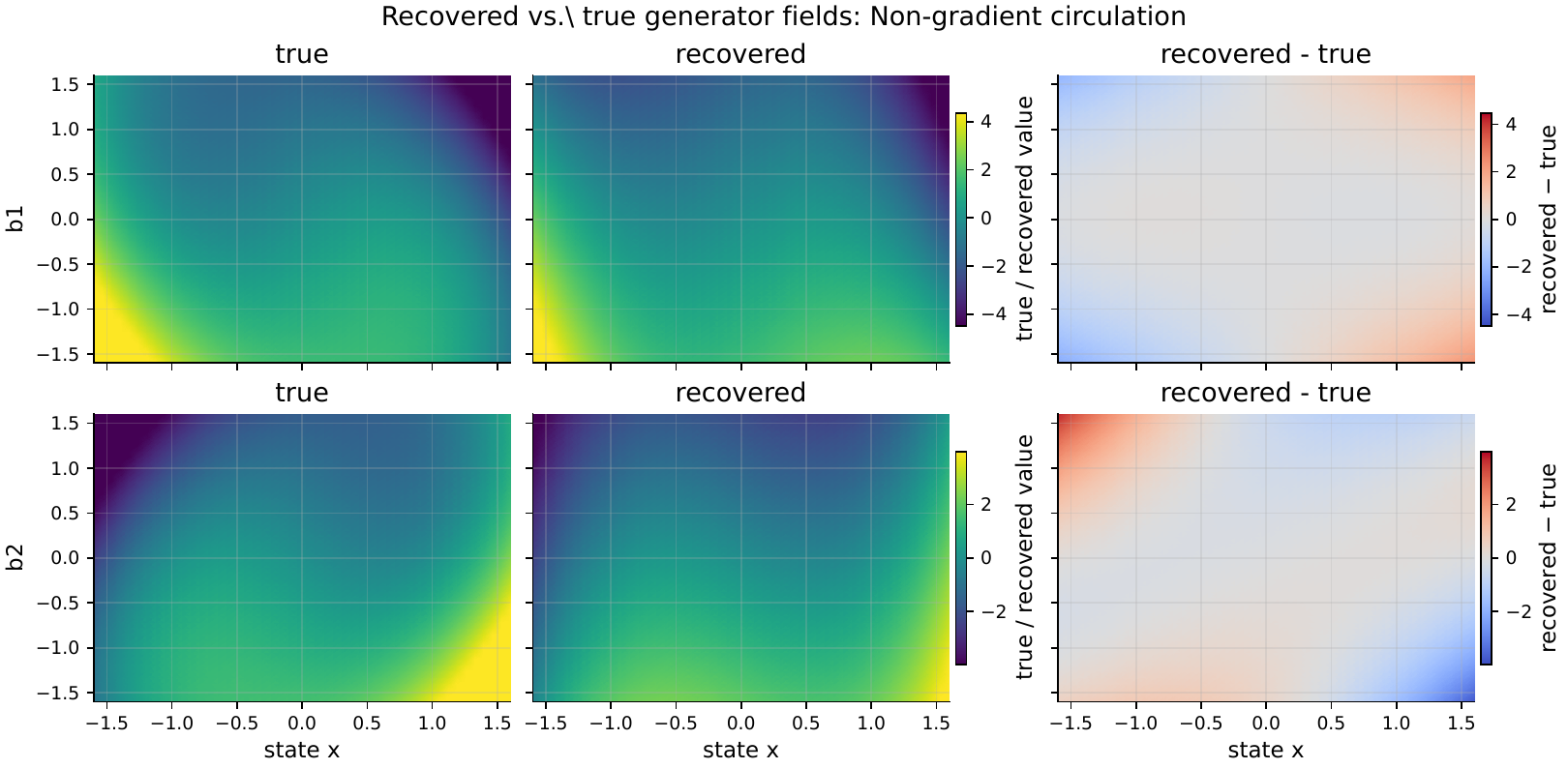}
\caption{Non-gradient drift-curl: recovered versus true generator fields (shared per-field colour scale; error column centred at zero).}\end{figure}
\paragraph{Verdict.} Drift central-grid rel-$L^2=0.239$, tensor rel-$L^2=0.024$, $a_{12}$ cosine $\mathrm{n/a}$, PSD $1.00$, stable projected-support FP $0$ ($n=10$ seeds). \textbf{PASS}.
\medskip}{}
\IfFileExists{ds_bistable.tex}{
\paragraph{Bistable and gradient systems}\;

\subsection{Double well + transverse}\label{sec:v62-double-well-transverse}
\paragraph{Context.} A one-dimensional double well coupled to a stable transverse mode, the simplest metastable two-dimensional system. The cubic restoring force encodes two wells separated by a barrier, while the transverse direction relaxes linearly; recovering the cubic accurately is what lets the identified generator reproduce the metastable two-state structure and the escape geometry.
\paragraph{System.} $$\drift=(X-X^3-0.5Y,\,-Y+0.5X),\ \diff=0.49 I.$$
\paragraph{Generator.} Cubic bistable $x$ coupled to a stable transverse mode; the cubic encodes the two wells and barrier.
\paragraph{Recovery.} WG-SINDy recovers the drift at central-grid relative $L^2=0.226$ and the diffusion tensor at $0.024$ relative error; the recovered drift reconstructs the potential landscape, its metastable wells, and the barrier between them. The paper-level stable projected-support filter counts no large recurring false positives under the declared post-processing rule. Coefficient-level selection rates vary by term and are reported in the coefficient ledger; this datasheet does not claim selection of every active term in every seed.
\begin{figure}[!htbp]\centering\includegraphics[width=0.72\linewidth,height=0.50\textheight,keepaspectratio]{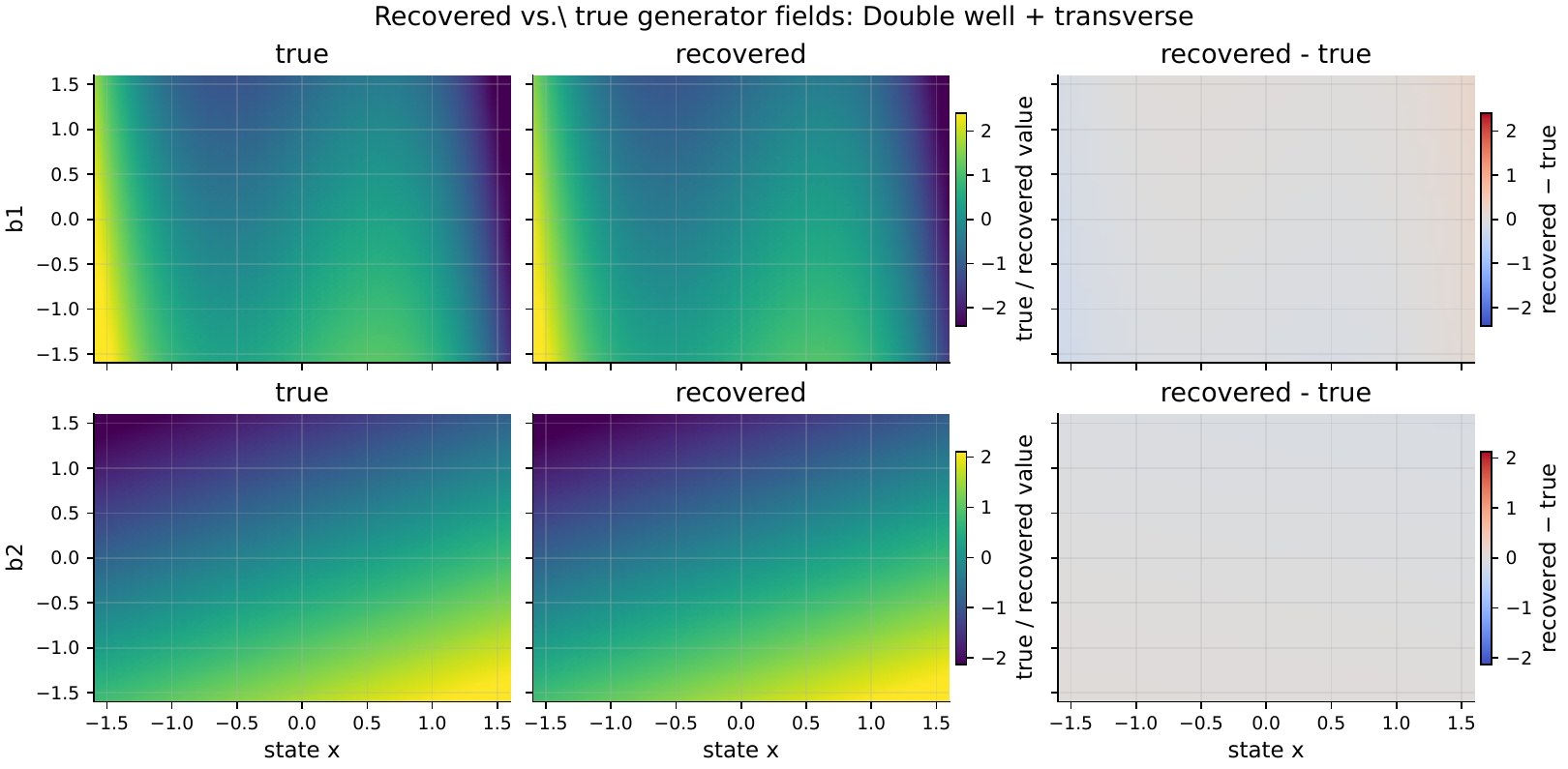}
\caption{Double well + transverse: recovered versus true generator fields (shared per-field colour scale; error column centred at zero).}\end{figure}
\paragraph{Verdict.} Drift central-grid rel-$L^2=0.226$, tensor rel-$L^2=0.024$, $a_{12}$ cosine $\mathrm{n/a}$, PSD $1.00$, stable projected-support FP $0$ ($n=10$ seeds). \textbf{PASS}.
\medskip

\subsection{Gradient potential}\label{sec:v62-gradient-potential}
\paragraph{Context.} A genuine two-dimensional gradient flow on a quartic potential with an $x^2y^2$ coupling. As the reversible twin of the non-gradient drift-curl system, it tests recovery of a full 2D potential (not a separable one) and of the resulting Boltzmann stationary density $\pi\propto e^{-2V/\sigma^2}$; the drift-curl diagnostic must return essentially zero, the discriminating contrast against its non-gradient counterpart.
\paragraph{System.} $$\drift=-\nabla V,\ V=\tfrac14(x^2-1)^2+\tfrac12 y^2+\tfrac14 x^2y^2,\ \diff=0.49 I.$$
\paragraph{Generator.} Reversible gradient flow with an $x^2y^2$ coupling; stationary density $\propto e^{-2V/\sigma^2}$; drift-curl diagnostic correctly near zero.
\paragraph{Recovery.} WG-SINDy recovers the drift at central-grid relative $L^2=0.246$ and the diffusion tensor at $0.024$ relative error; the recovered drift reconstructs the potential landscape, its metastable wells, and the barrier between them. The paper-level stable projected-support filter counts no large recurring false positives under the declared post-processing rule. Coefficient-level selection rates vary by term and are reported in the coefficient ledger; this datasheet does not claim selection of every active term in every seed.
\begin{figure}[!htbp]\centering\includegraphics[width=0.72\linewidth,height=0.50\textheight,keepaspectratio]{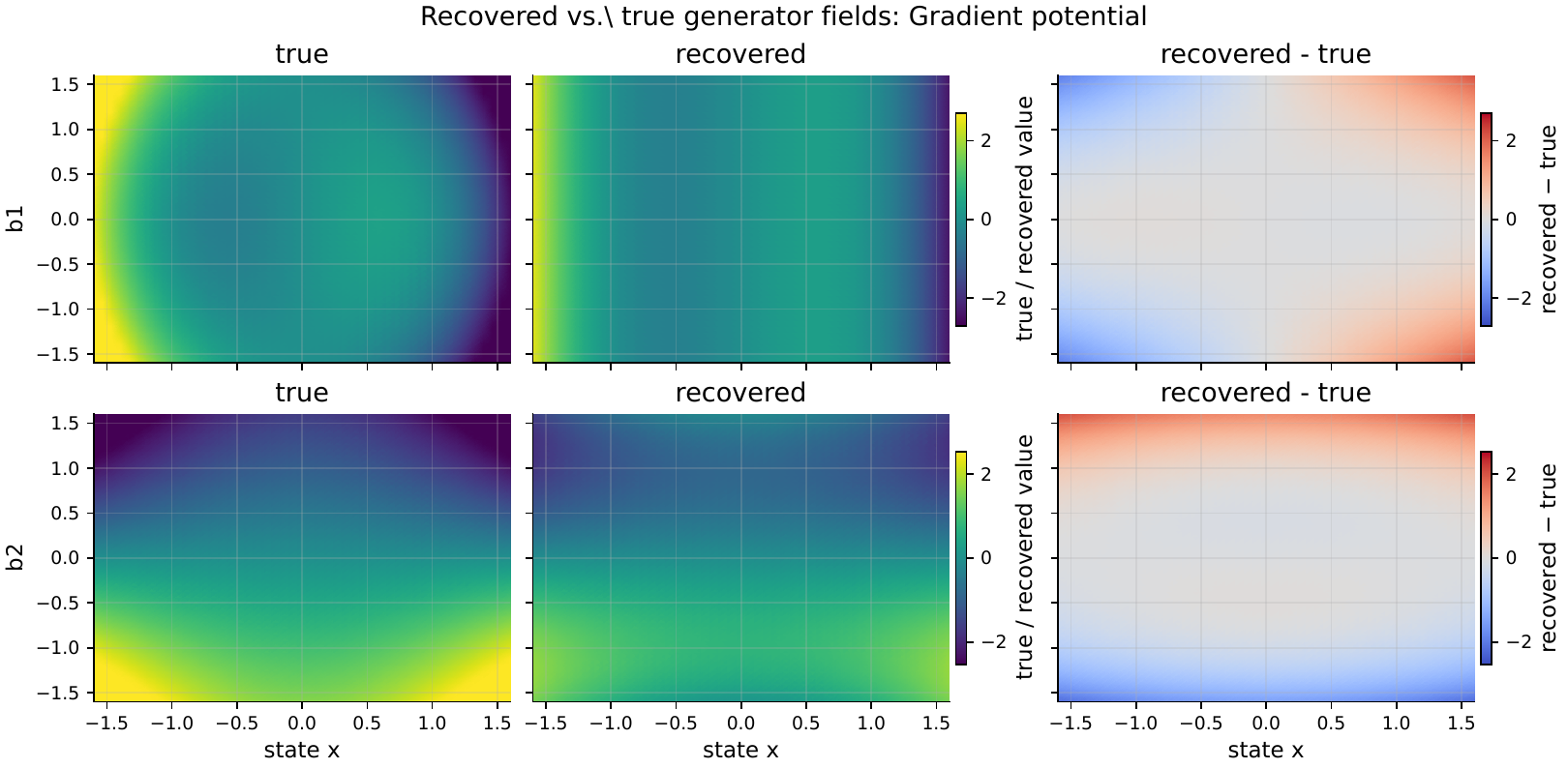}
\caption{Gradient potential: recovered versus true generator fields (shared per-field colour scale; error column centred at zero).}\end{figure}
\paragraph{Verdict.} Drift central-grid rel-$L^2=0.246$, tensor rel-$L^2=0.024$, $a_{12}$ cosine $\mathrm{n/a}$, PSD $1.00$, stable projected-support FP $0$ ($n=10$ seeds). \textbf{PASS}.
\medskip

\subsection{Maier--Stein}\label{sec:v62-maier-stein}
\paragraph{Context.} The Maier--Stein system is a standard benchmark in large-deviation and transition-path theory, modelling noise-activated escape over a non-gradient barrier. Its drift mixes a cubic bistability with a state-dependent transverse term $-(1+x^2)y$, whose $y$ and $x^2y$ pieces are strongly collinear on the sampled region; recovering the symbolic drift means the quasipotential and most-probable escape path become computable from data, which is why it is included despite being one of the harder fits.
\paragraph{System.} $$\drift=(X-X^3-\beta XY^2,\,-(1+X^2)Y),\ \beta=0.35,\ \diff=0.1225 I.$$
\paragraph{Generator.} Canonical non-gradient escape model. The $-(1+x^2)y$ term ($y$ and $x^2y$) is collinear on the sampled region; the small $-\beta xy^2$ term is the hardest to identify.
\paragraph{Recovery.} WG-SINDy recovers the drift at central-grid relative $L^2=0.235$ and the diffusion tensor at $0.038$ relative error; the recovered drift reconstructs the potential landscape, its metastable wells, and the barrier between them. The paper-level stable projected-support filter counts no large recurring false positives under the declared post-processing rule. Coefficient-level selection rates vary by term and are reported in the coefficient ledger; this datasheet does not claim selection of every active term in every seed.
\begin{figure}[!htbp]\centering\includegraphics[width=0.72\linewidth,height=0.50\textheight,keepaspectratio]{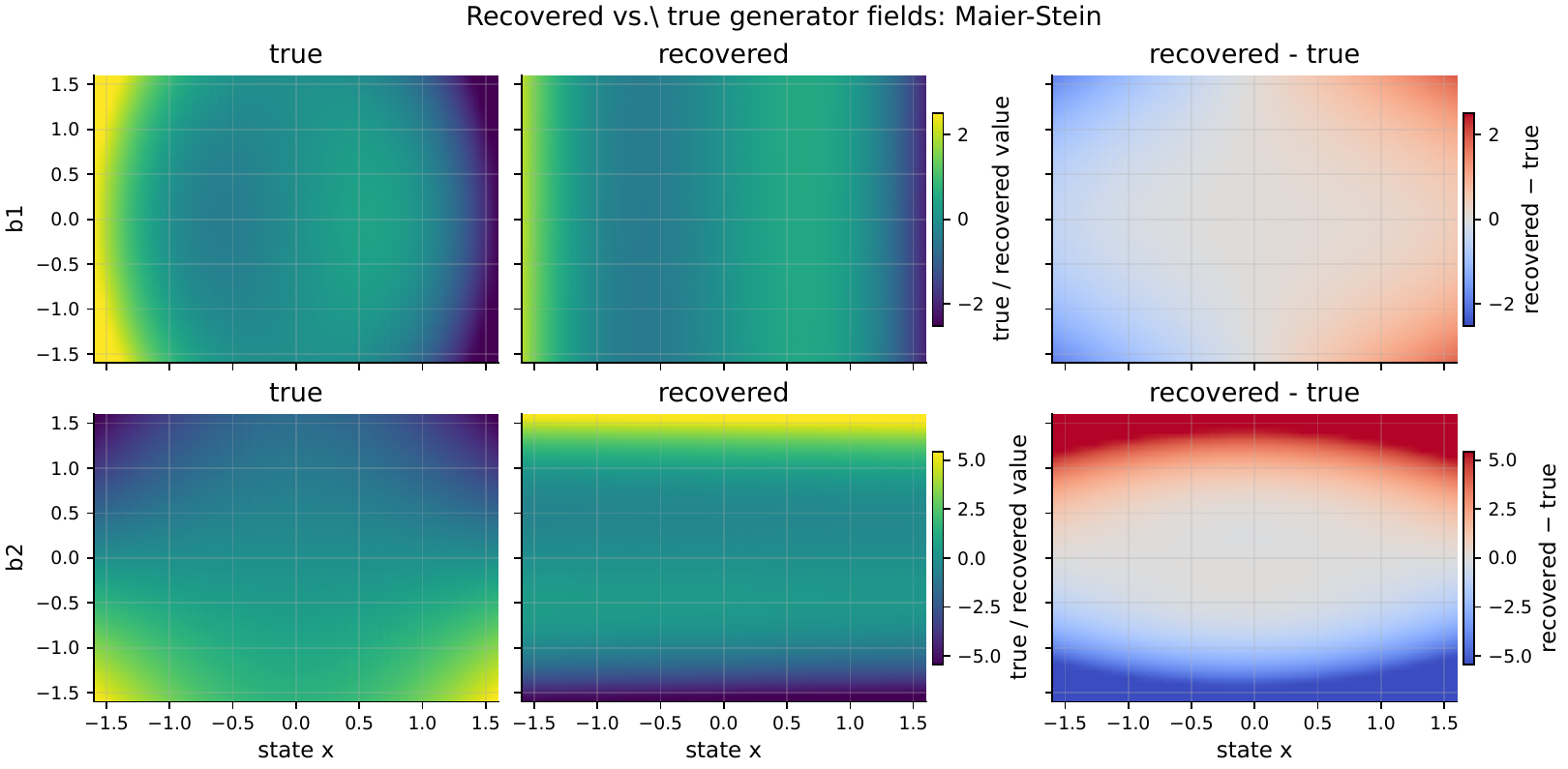}
\caption{Maier--Stein: recovered versus true generator fields (shared per-field colour scale; error column centred at zero).}\end{figure}
\paragraph{Verdict.} Drift central-grid rel-$L^2=0.235$, tensor rel-$L^2=0.038$, $a_{12}$ cosine $\mathrm{n/a}$, PSD $1.00$, stable projected-support FP $0$ ($n=10$ seeds). \textbf{PASS}.
\medskip

\subsection{Duffing oscillator}\label{sec:v62-duffing}
\paragraph{Context.} The Duffing oscillator is a textbook nonlinear mechanical resonator written in position--velocity phase space. It tests recovery of a deterministic skeleton with a cubic restoring force and weak linear damping; because the damping coefficient is small relative to the restoring force and the noise, it is a controlled probe of how the estimator handles a genuine but low-amplitude term.
\paragraph{System.} $$\mathrm{d}X=Y\,\mathrm{d}t,\ \mathrm{d}Y=(-0.35Y+X-X^3)\mathrm{d}t+\sigma\mathrm{d}W,\ \diff_{22}=0.09.$$
\paragraph{Generator.} Noisy bistable oscillator in position--velocity form; the small damping $-0.35y$ is low-SNR and intermittently selected.
\paragraph{Recovery.} WG-SINDy recovers the drift at central-grid relative $L^2=0.240$ and the diffusion tensor at $0.031$ relative error; the recovered drift reconstructs the potential landscape, its metastable wells, and the barrier between them. The paper-level stable projected-support filter counts no large recurring false positives under the declared post-processing rule. Coefficient-level selection rates vary by term and are reported in the coefficient ledger; this datasheet does not claim selection of every active term in every seed.
\begin{figure}[!htbp]\centering\includegraphics[width=0.72\linewidth,height=0.50\textheight,keepaspectratio]{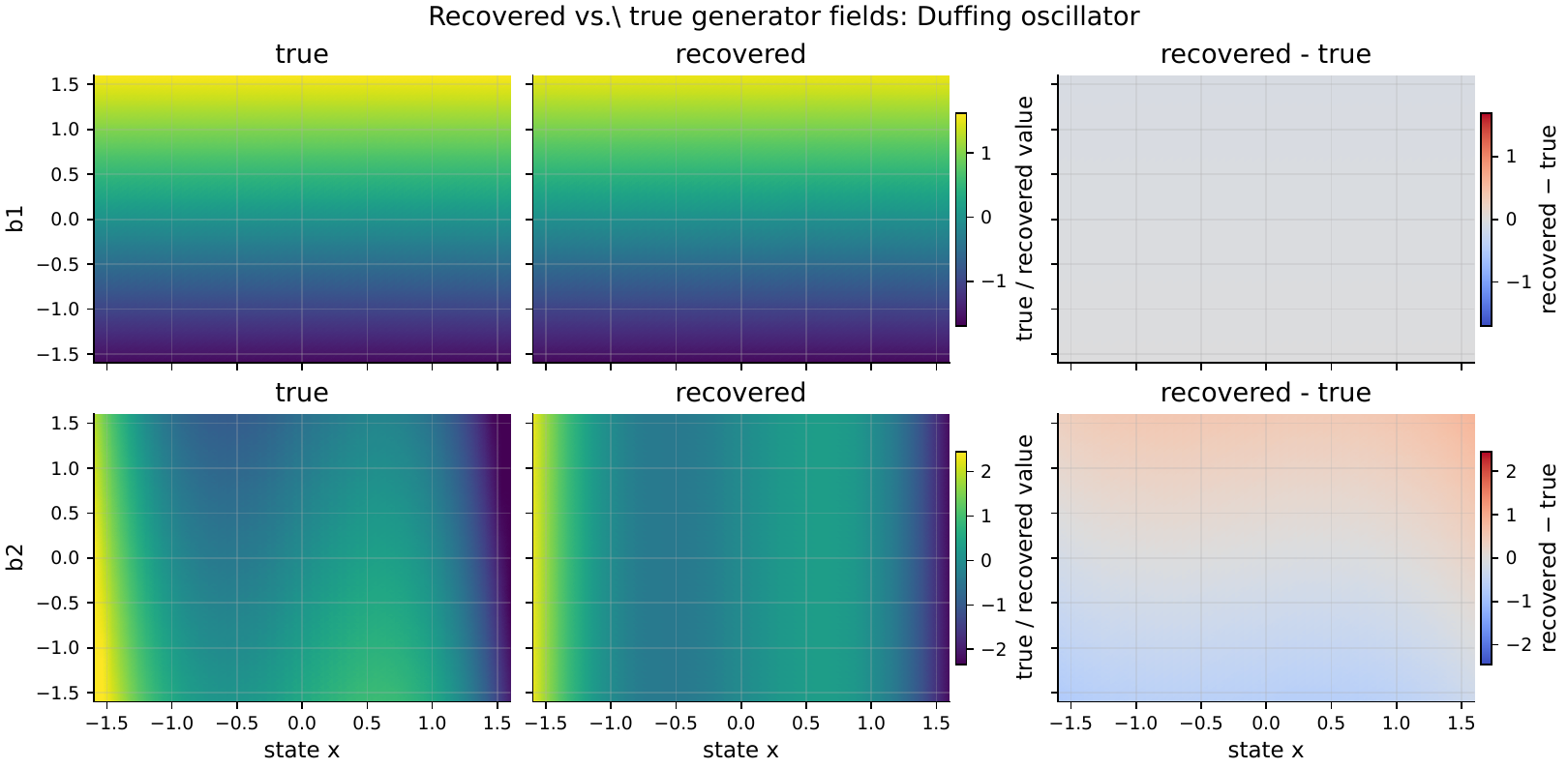}
\caption{Duffing oscillator: recovered versus true generator fields (shared per-field colour scale; error column centred at zero).}\end{figure}
\paragraph{Verdict.} Drift central-grid rel-$L^2=0.240$, tensor rel-$L^2=0.031$, $a_{12}$ cosine $\mathrm{n/a}$, PSD $1.00$, stable projected-support FP $0$ ($n=10$ seeds). \textbf{PASS}.
\medskip

\subsection{M\"uller--Brown}\label{sec:v62-mueller-brown}
\paragraph{Context.} The M\"uller--Brown surface is a standard multi-well molecular-dynamics benchmark. Its gradient drift is a sum of anisotropic Gaussians, strongly non-polynomial and stiff, and the small mobility depresses the drift signal; the diffusion is recovered while the drift is a representability null, an honest boundary that a radial-basis library, not a polynomial one, would be needed to cross.
\paragraph{System.} $$\drift=-m\nabla V_{\mathrm{MB}}\ (\text{four-Gaussian potential}),\ m=0.004.$$
\paragraph{Generator.} Stiff multi-well molecular potential; the strongly non-polynomial gradient is not spanned by a polynomial library. Named limit.
\paragraph{Recovery.} WG-SINDy recovers the drift at central-grid relative $L^2=1.826$ and the diffusion tensor at $0.075$ relative error; the recovered drift reconstructs the potential landscape, its metastable wells, and the barrier between them. This is a named limit (registry declared failure or stress case): recovery fails for a physical or identifiability reason (library incompleteness, low signal-to-noise, degeneracy, or coverage), not an estimator defect.
\begin{figure}[!htbp]\centering\includegraphics[width=0.72\linewidth,height=0.50\textheight,keepaspectratio]{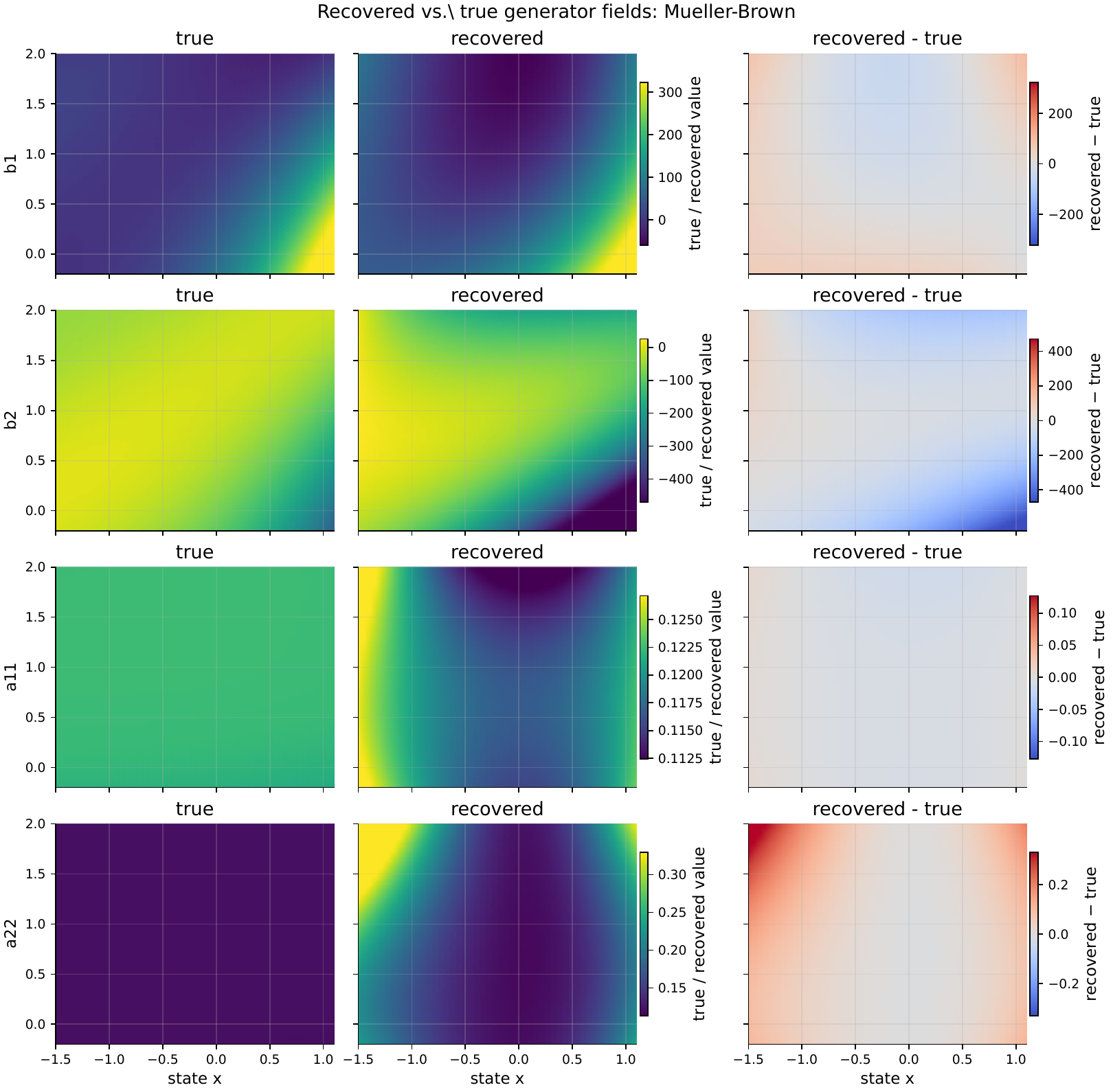}
\caption{M\"uller--Brown: recovered versus true generator fields (shared per-field colour scale; error column centred at zero).}\end{figure}
\paragraph{Verdict.} Drift central-grid rel-$L^2=1.826$, tensor rel-$L^2=0.075$, $a_{12}$ cosine $\mathrm{n/a}$, PSD $1.00$, stable projected-support FP $0$ ($n=10$ seeds). \textbf{NAMED\_NULL}.
\medskip}{}
\IfFileExists{ds_multiplicative.tex}{
\paragraph{State-dependent (multiplicative) diffusion}\;

\subsection{Diagonal multiplicative}\label{sec:v62-diag-multiplicative}
\paragraph{Context.} A linear-drift system with \emph{state-dependent} (multiplicative) diffusion whose amplitude grows quadratically with position, the situation in population dynamics and fluctuating-environment models. It is the first test of recovering a diffusion \emph{field} rather than a constant, and of correctly returning a zero off-diagonal while both diagonal variances grow, i.e. not hallucinating leverage where none exists.
\paragraph{System.} $$\drift=-X,\ \diff=\mathrm{diag}(0.5+0.1x^2+0.1y^2,\,0.4+0.1x^2+0.1y^2).$$
\paragraph{Generator.} State-dependent diagonal diffusion (a field, not a constant); off-diagonal correctly zero. Quadratic diffusion terms are low-SNR.
\paragraph{Recovery.} WG-SINDy recovers the drift at central-grid relative $L^2=0.200$ and the diffusion tensor at $0.112$ relative error; the recovered diffusion tensor is a position-dependent field giving the local fluctuation amplitude and relaxation. The paper-level stable projected-support filter counts no large recurring false positives under the declared post-processing rule. Coefficient-level selection rates vary by term and are reported in the coefficient ledger; this datasheet does not claim selection of every active term in every seed.
\begin{figure}[!htbp]\centering\includegraphics[width=0.72\linewidth,height=0.50\textheight,keepaspectratio]{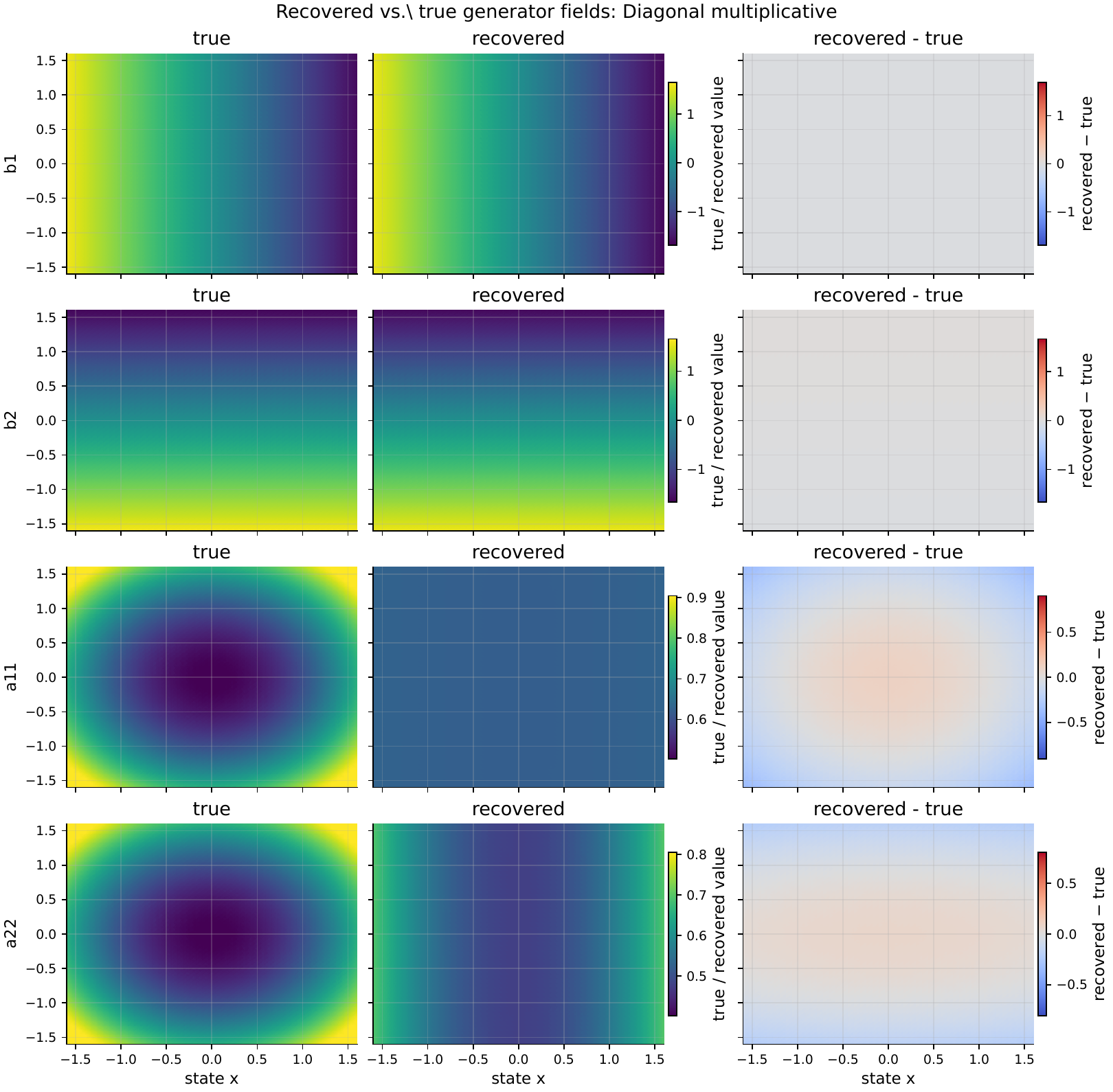}
\caption{Diagonal multiplicative: recovered versus true generator fields (shared per-field colour scale; error column centred at zero).}\end{figure}
\paragraph{Verdict.} Drift central-grid rel-$L^2=0.200$, tensor rel-$L^2=0.112$, $a_{12}$ cosine $\mathrm{n/a}$, PSD $1.00$, stable projected-support FP $0$ ($n=10$ seeds). \textbf{PASS}.
\medskip

\subsection{Non-diagonal Cholesky}\label{sec:v62-nondiag-cholesky}
\paragraph{Context.} The hardest clean synthetic test of a fully \emph{state-dependent off-diagonal} diffusion, constructed through a Cholesky factor so the target is positive semidefinite everywhere by design. The off-diagonal varies in space and changes sign across the axes, so it probes both the recovery of a spatially varying leverage field and the structural PSD guarantee that a naive entrywise regression of cross-increments would violate.
\paragraph{System.} $$\diff=LL^\top,\ L=\begin{psmallmatrix}0.5+0.1x^2&0\\0.2xy&0.4+0.1y^2\end{psmallmatrix},\ \drift=-X.$$
\paragraph{Generator.} State-dependent off-diagonal diffusion, PSD by construction; the leading $a_{12}\!\approx\!0.1xy$ term is recovered, higher-order factor terms are weak.
\paragraph{Recovery.} WG-SINDy recovers the drift at central-grid relative $L^2=0.204$ and the diffusion tensor at $0.079$ relative error, with off-diagonal (leverage) cosine $0.988$; the recovered diffusion tensor is a position-dependent field giving the local fluctuation amplitude and relaxation. The paper-level stable projected-support filter counts no large recurring false positives under the declared post-processing rule. Coefficient-level selection rates vary by term and are reported in the coefficient ledger; this datasheet does not claim selection of every active term in every seed.
\begin{figure}[!htbp]\centering\includegraphics[width=0.72\linewidth,height=0.50\textheight,keepaspectratio]{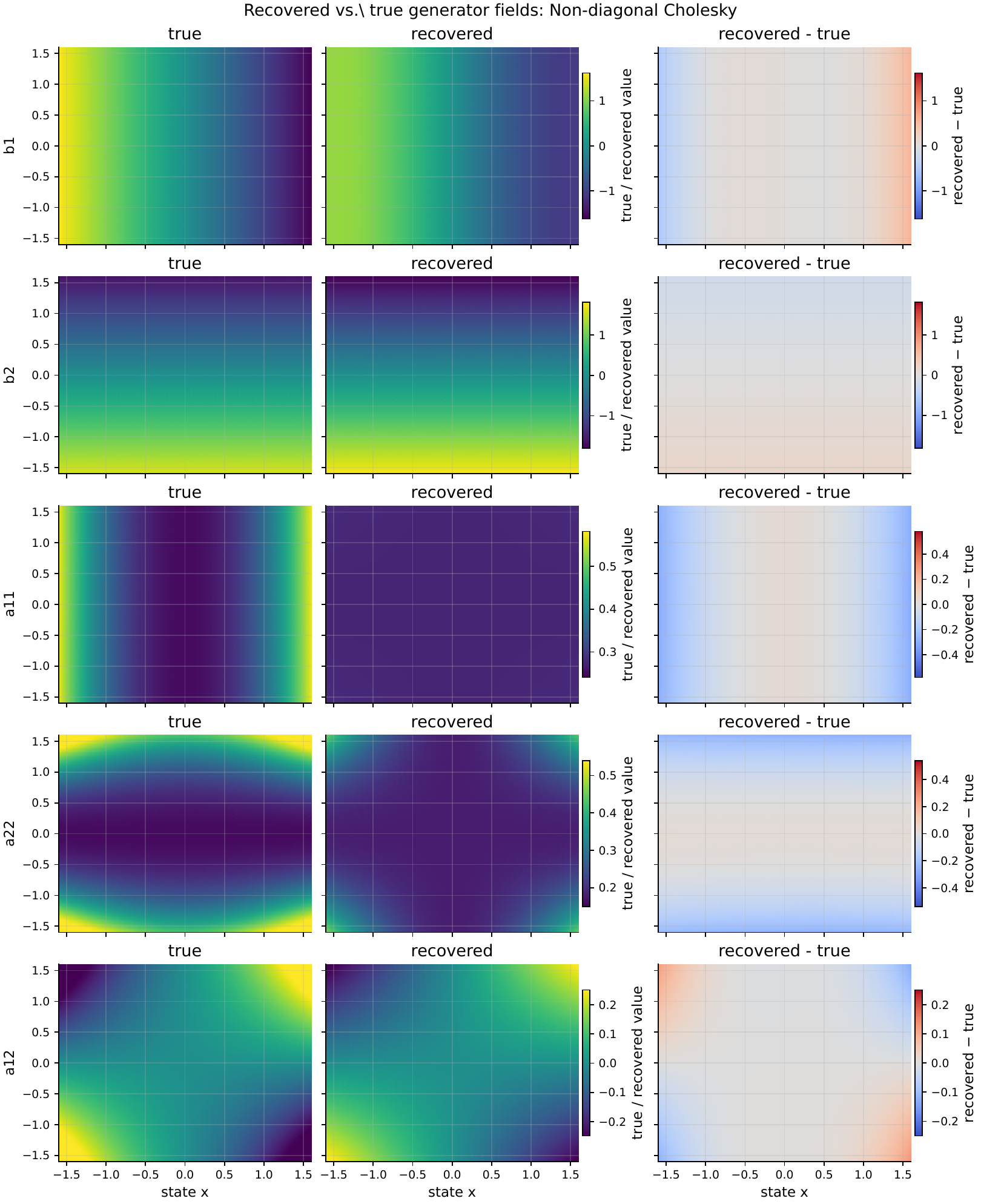}
\caption{Non-diagonal Cholesky: recovered versus true generator fields (shared per-field colour scale; error column centred at zero).}\end{figure}
\paragraph{Verdict.} Drift central-grid rel-$L^2=0.204$, tensor rel-$L^2=0.079$, $a_{12}$ cosine $0.988$, PSD $1.00$, stable projected-support FP $0$ ($n=10$ seeds). \textbf{PASS}.
\medskip}{}
\IfFileExists{ds_financial.tex}{
\paragraph{Financial / stochastic-volatility systems}\;

\subsection{Log-Heston}\label{sec:v62-heston-logsv}
\paragraph{Context.} The Heston model is a cornerstone of mathematical finance, with stochastic instantaneous variance driving the asset. In the numerically stable log-price coordinate it is the flagship leverage system: the negative price--variance correlation $\rho$ enters only through the off-diagonal $\diff_{12}=\rho\xi v$, the single parameter that shapes the implied-volatility skew used in option pricing. It also exposes the central honest limit, the risk-neutral log-price drift $\mu-\tfrac12 v$, whose per-step magnitude signal is about two orders below the diffusion scale in the nominal regime (with squared signal-to-noise near four orders smaller).
\paragraph{System.} $$\mathrm{d}X=(\mu-\tfrac12 v)\mathrm{d}t+\sqrt{v}\mathrm{d}W_1,\ \mathrm{d}v=\kappa(\theta-v)\mathrm{d}t+\xi\sqrt{v}\mathrm{d}W_2,\ \mathrm{d}W_1\mathrm{d}W_2=\rho\,\mathrm{d}t.$$
\paragraph{Generator.} Flagship leverage system. $\drift_v=\kappa(\theta-v)$ variance mean reversion; $\diff_{11}=v$, $\diff_{22}=\xi^2 v$, $\diff_{12}=\rho\xi v$ leverage. Log-price drift $\mu-\tfrac12 v$ is the low-SNR null.
\paragraph{Recovery.} WG-SINDy recovers the drift at central-grid relative $L^2=0.143$ and the diffusion tensor at $0.054$ relative error, with off-diagonal (leverage) cosine $0.999$; the off-diagonal recovers the price--variance leverage correlation while the variance drift and vol-of-vol terms remain identifiable. The paper-level stable projected-support filter counts no large recurring false positives under the declared post-processing rule. Coefficient-level selection rates vary by term and are reported in the coefficient ledger; this datasheet does not claim selection of every active term in every seed.
\begin{figure}[!htbp]\centering\includegraphics[width=0.72\linewidth,height=0.50\textheight,keepaspectratio]{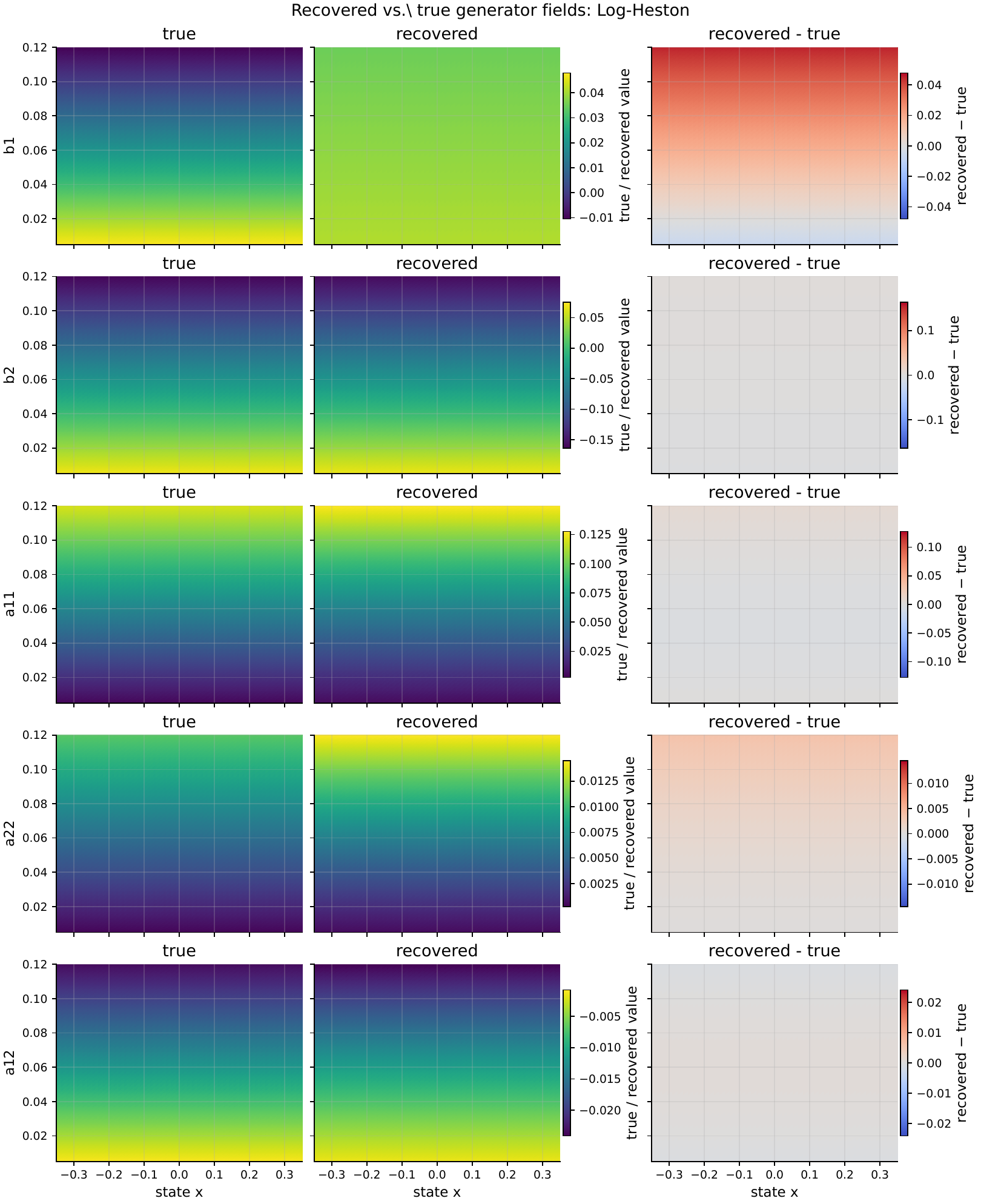}
\caption{Log-Heston: recovered versus true generator fields (shared per-field colour scale; error column centred at zero).}\end{figure}
\paragraph{Verdict.} Drift central-grid rel-$L^2=0.143$, tensor rel-$L^2=0.054$, $a_{12}$ cosine $0.999$, PSD $1.00$, stable projected-support FP $0$ ($n=10$ seeds). \textbf{PASS}.
\medskip

\subsection{\texorpdfstring{Heston $(S,V)$}{Heston (S,V)}}\label{sec:v62-heston-sv}
\paragraph{Context.} The raw-price formulation of the same Heston dynamics, retained to show the leverage result is not an artefact of the log transform. Its diagonal variance $S^2v$ spans an enormous dynamic range, the very ill-conditioning that motivates practitioners to take logs; recovering the leverage and variance generator here demonstrates that the anisotropic kernels and whitening absorb extreme coordinate scaling without manual transformation.
\paragraph{System.} $$\mathrm{d}S=\mu S\,\mathrm{d}t+S\sqrt{v}\mathrm{d}W_1,\ \mathrm{d}v=\kappa(\theta-v)\mathrm{d}t+\xi\sqrt{v}\mathrm{d}W_2,\ \mathrm{d}W_1\mathrm{d}W_2=\rho\,\mathrm{d}t.$$
\paragraph{Generator.} Raw-price Heston: $\diff_{11}=S^2v$ (large dynamic range), $\diff_{12}=\rho\xi Sv$ leverage. Recovered despite the scale; price drift $\mu S$ is the low-SNR null.
\paragraph{Recovery.} WG-SINDy recovers the drift at central-grid relative $L^2=0.273$ and the diffusion tensor at $0.089$ relative error, with off-diagonal (leverage) cosine $0.993$; the off-diagonal recovers raw price--variance leverage despite the large $S^2v$ dynamic range. The paper-level stable projected-support filter counts no large recurring false positives under the declared post-processing rule. Coefficient-level selection rates vary by term and are reported in the coefficient ledger; this datasheet does not claim selection of every active term in every seed.
\begin{figure}[!htbp]\centering\includegraphics[width=0.72\linewidth,height=0.50\textheight,keepaspectratio]{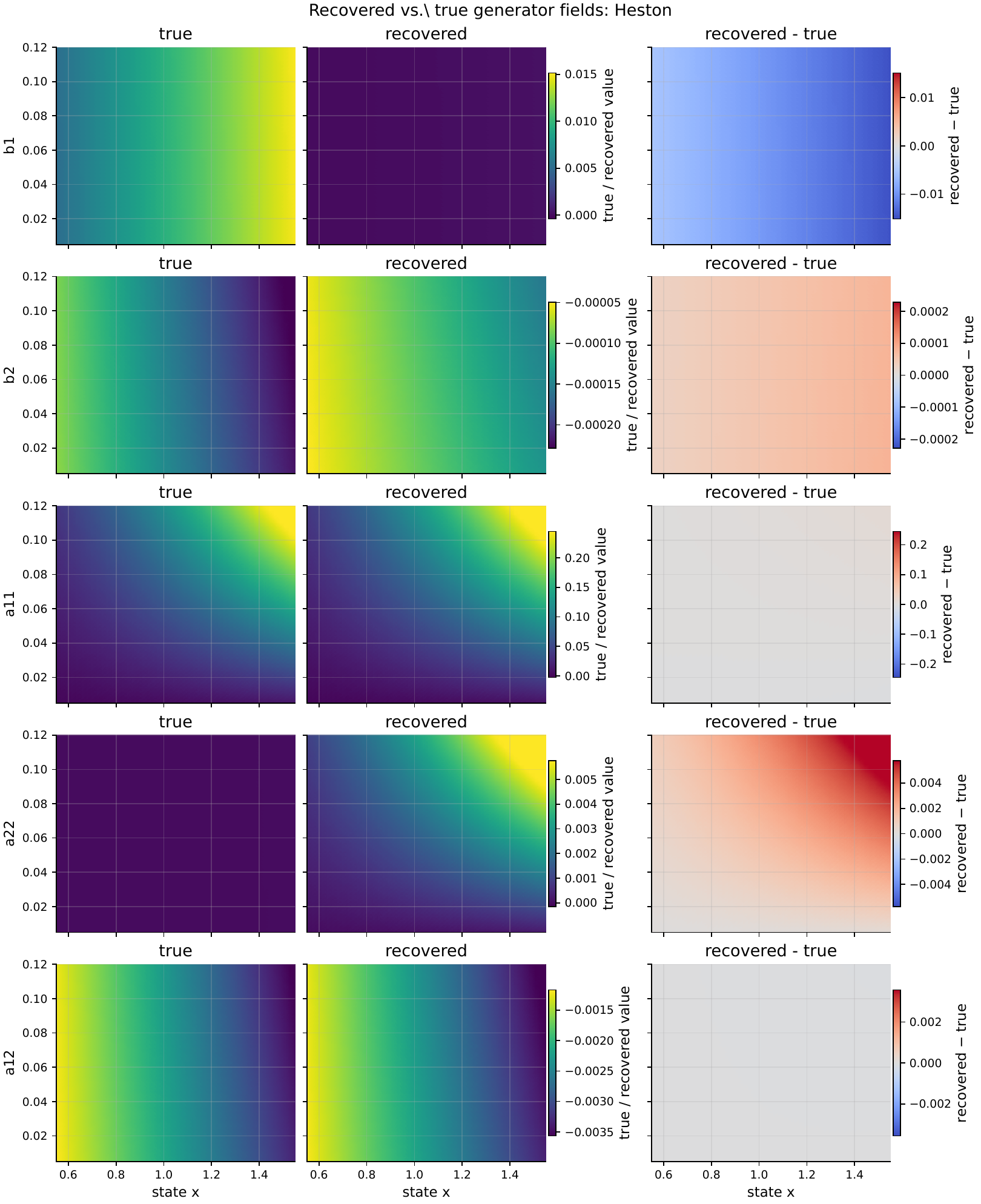}
\caption{Heston $(S,V)$: recovered versus true generator fields (shared per-field colour scale; error column centred at zero).}\end{figure}
\paragraph{Verdict.} Drift central-grid rel-$L^2=0.273$, tensor rel-$L^2=0.089$, $a_{12}$ cosine $0.993$, PSD $1.00$, stable projected-support FP $0$ ($n=10$ seeds). \textbf{PASS}.
\medskip

\subsection{CIR pair}\label{sec:v62-cir-pair}
\paragraph{Context.} A pair of correlated Cox--Ingersoll--Ross processes, the building block of multi-factor interest-rate and competing-population models. Its off-diagonal diffusion is genuinely non-polynomial ($\propto\sqrt{xy}$), so it tests whether the estimator accommodates a square-root feature library; both square-root mean-reverting drifts are recovered, unlike the stochastic-volatility log-price drift, because mean reversion carries adequate signal.
\paragraph{System.} $$\mathrm{d}X=\kappa(\theta-X)\mathrm{d}t+\xi\sqrt{X}\mathrm{d}W_1,\ \mathrm{d}Y=\kappa(\theta-Y)\mathrm{d}t+\xi\sqrt{Y}\mathrm{d}W_2,\ \mathrm{d}W_1\mathrm{d}W_2=\rho\,\mathrm{d}t.$$
\paragraph{Generator.} Correlated square-root processes; $\diff_{12}=\rho\xi^2\sqrt{xy}$ needs the square-root library F. Both mean-reverting drifts recover.
\paragraph{Recovery.} WG-SINDy recovers the drift at central-grid relative $L^2=0.393$ and the diffusion tensor at $0.116$ relative error, with off-diagonal (leverage) cosine $0.995$; the off-diagonal recovers the correlation between the two square-root factors, while both mean-reverting drifts remain visible. The paper-level stable projected-support filter counts no large recurring false positives under the declared post-processing rule. Coefficient-level selection rates vary by term and are reported in the coefficient ledger; this datasheet does not claim selection of every active term in every seed.
\begin{figure}[!htbp]\centering\includegraphics[width=0.72\linewidth,height=0.50\textheight,keepaspectratio]{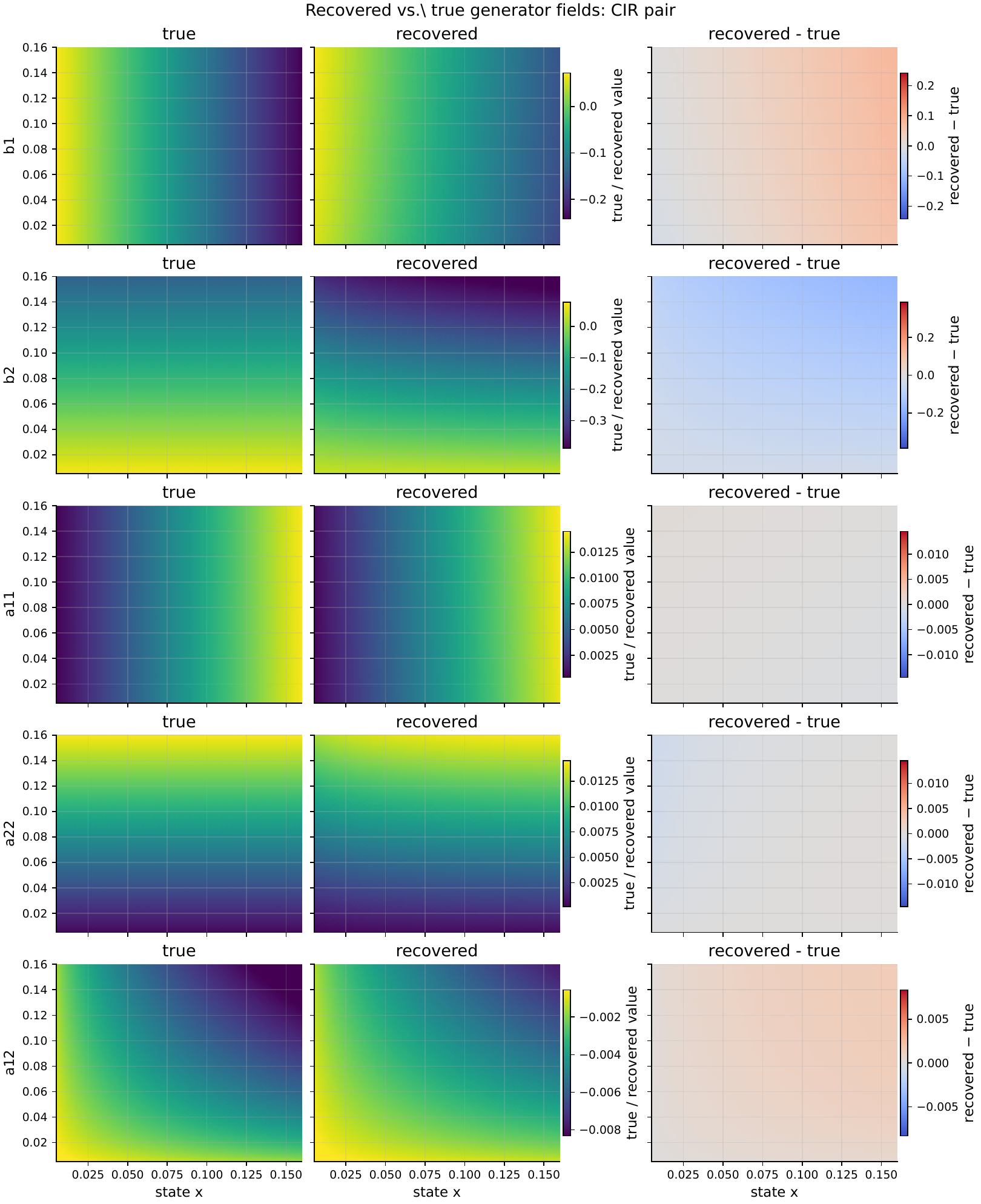}
\caption{CIR pair: recovered versus true generator fields (shared per-field colour scale; error column centred at zero).}\end{figure}
\paragraph{Verdict.} Drift central-grid rel-$L^2=0.393$, tensor rel-$L^2=0.116$, $a_{12}$ cosine $0.995$, PSD $1.00$, stable projected-support FP $0$ ($n=10$ seeds). \textbf{PASS}.
\medskip

\subsection{SABR}\label{sec:v62-sabr}
\paragraph{Context.} The SABR model is an industry-standard stochastic-volatility specification for interest-rate and FX smiles. It is a driftless martingale with power-law ($F^\beta$) diffusion, so it tests two edges at once: a generator with no drift to recover (which must come back as exactly zero) and a fractional-power diffusion that a polynomial library can only approximate, making it a named limit on the tensor while the leverage sign is still recovered.
\paragraph{System.} $$\mathrm{d}F=\sigma F^\beta\mathrm{d}W_1,\ \mathrm{d}\sigma=\nu\sigma\mathrm{d}W_2,\ \mathrm{d}W_1\mathrm{d}W_2=\rho\,\mathrm{d}t,\ \beta=0.5.$$
\paragraph{Generator.} Driftless martingale with power-law diffusion; leverage cosine recovers, the $F^{2\beta}$ power-law is only approximately spanned. Named limit.
\paragraph{Recovery.} WG-SINDy recovers the drift at central-grid relative $L^2=--$ and the diffusion tensor at $0.195$ relative error, with off-diagonal (leverage) cosine $0.990$; the off-diagonal recovers the forward--volatility correlation and the tensor captures the power-law volatility geometry only approximately in the declared library. This is a named limit (fragile physics or identifiability limit): recovery fails for a physical or identifiability reason (library incompleteness, low signal-to-noise, degeneracy, or coverage), not an estimator defect.
\begin{figure}[!htbp]\centering\includegraphics[width=0.72\linewidth,height=0.50\textheight,keepaspectratio]{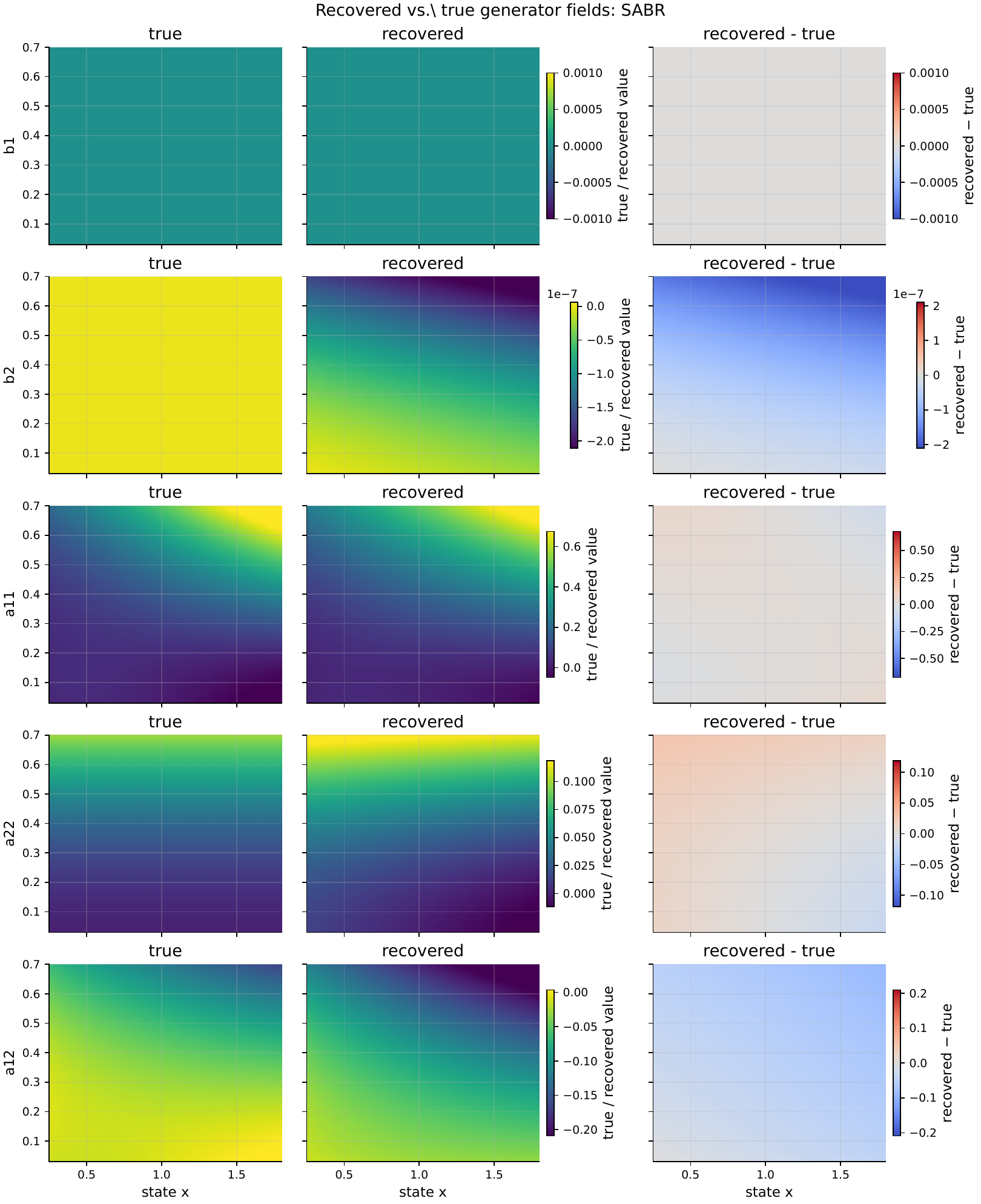}
\caption{SABR: recovered versus true generator fields (shared per-field colour scale; error column centred at zero).}\end{figure}
\paragraph{Verdict.} Drift central-grid rel-$L^2=--$, tensor rel-$L^2=0.195$, $a_{12}$ cosine $0.990$, PSD $1.00$, stable projected-support FP $0$ ($n=10$ seeds). \textbf{NAMED\_NULL}.
\medskip

\subsection{Correlated 2D GBM}\label{sec:v62-gbm-2d}
\paragraph{Context.} Two correlated geometric Brownian motions, the canonical multi-asset price model underlying portfolio and basket-option risk. The asset--asset correlation lives in the off-diagonal $\diff_{12}=\rho\sigma_1\sigma_2 S_1S_2$, which is recovered cleanly; the per-asset drifts $\mu_iS_i$ are dominated by the diffusion and share the low signal-to-noise character of the Heston log-price drift, so the system is reported as a scoped review.
\paragraph{System.} $$\mathrm{d}S_i=\mu_i S_i\,\mathrm{d}t+\sigma_i S_i\mathrm{d}W_i,\ \mathrm{d}W_1\mathrm{d}W_2=\rho\,\mathrm{d}t.$$
\paragraph{Generator.} Two correlated lognormal assets; tensor and $S_1S_2$ leverage recover (cosine $0.997$); the small $\mu_i S_i$ drift is low-SNR. Scoped review.
\paragraph{Recovery.} WG-SINDy recovers the drift at central-grid relative $L^2=2.492$ and the diffusion tensor at $0.096$ relative error, with off-diagonal (leverage) cosine $0.997$; the off-diagonal recovers asset--asset correlation, while the per-asset drift remains low-SNR. Scoped review (metric gate not met but truth is representable): the truth lies in the library but the metric gate is not met, typically a low-SNR drift component.
\begin{figure}[!htbp]\centering\includegraphics[width=0.72\linewidth,height=0.50\textheight,keepaspectratio]{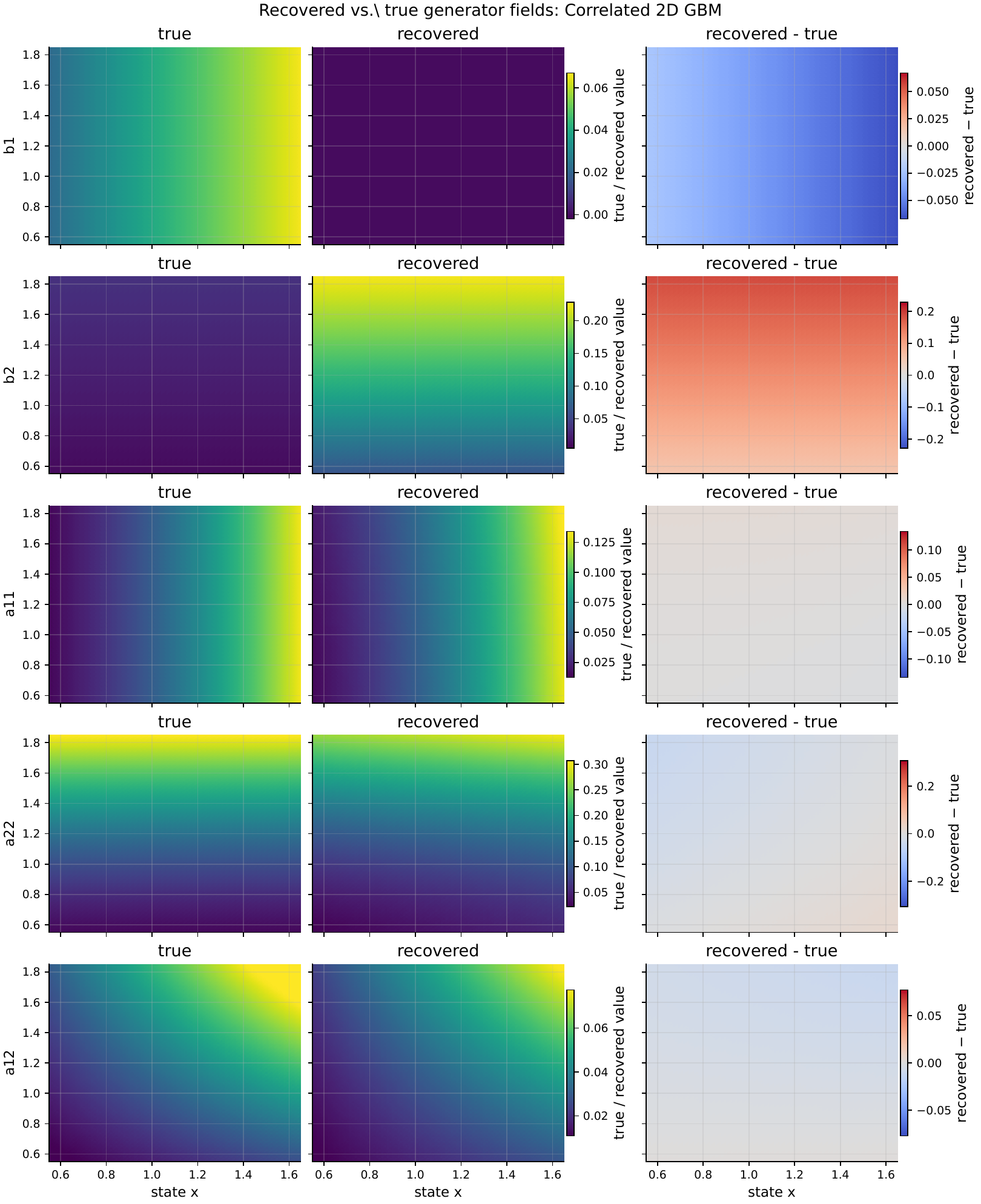}
\caption{Correlated 2D GBM: recovered versus true generator fields (shared per-field colour scale; error column centred at zero).}\end{figure}
\paragraph{Verdict.} Drift central-grid rel-$L^2=2.492$, tensor rel-$L^2=0.096$, $a_{12}$ cosine $0.997$, PSD $1.00$, stable projected-support FP $0$ ($n=10$ seeds). \textbf{SCOPED\_REVIEW}.
\medskip

\subsection{Two-factor Vasicek}\label{sec:v62-two-factor-vasicek}
\paragraph{Context.} A coupled two-factor Vasicek short-rate model, the affine-rates analogue of coupled OU at realistic (micro) scale. Its constant diffusion is so small that relative tensor error becomes a degenerate metric; the affine coupled drift, which is the financially relevant object for term-structure work, is recovered, so it is flagged as a scoped review rather than a pass.
\paragraph{System.} $$\drift=(\kappa_1(\theta_1-X)+c_{12}(Y-\theta_2),\ \kappa_2(\theta_2-Y)+c_{21}(X-\theta_1)).$$
\paragraph{Generator.} Affine coupled short-rate model; constant micro-scale tensor (relative metric degenerate, absolute error small). Scoped review.
\paragraph{Recovery.} WG-SINDy recovers the drift at central-grid relative $L^2=0.350$ and the diffusion tensor at $1.000$ relative error; the recovered generator gives the affine relaxation rates and cross-coupling, while the tiny constant tensor makes relative tensor error a degenerate metric. Scoped review (metric gate not met but truth is representable): the truth lies in the library but the metric gate is not met, typically a low-SNR drift component.
\begin{figure}[!htbp]\centering\includegraphics[width=0.72\linewidth,height=0.50\textheight,keepaspectratio]{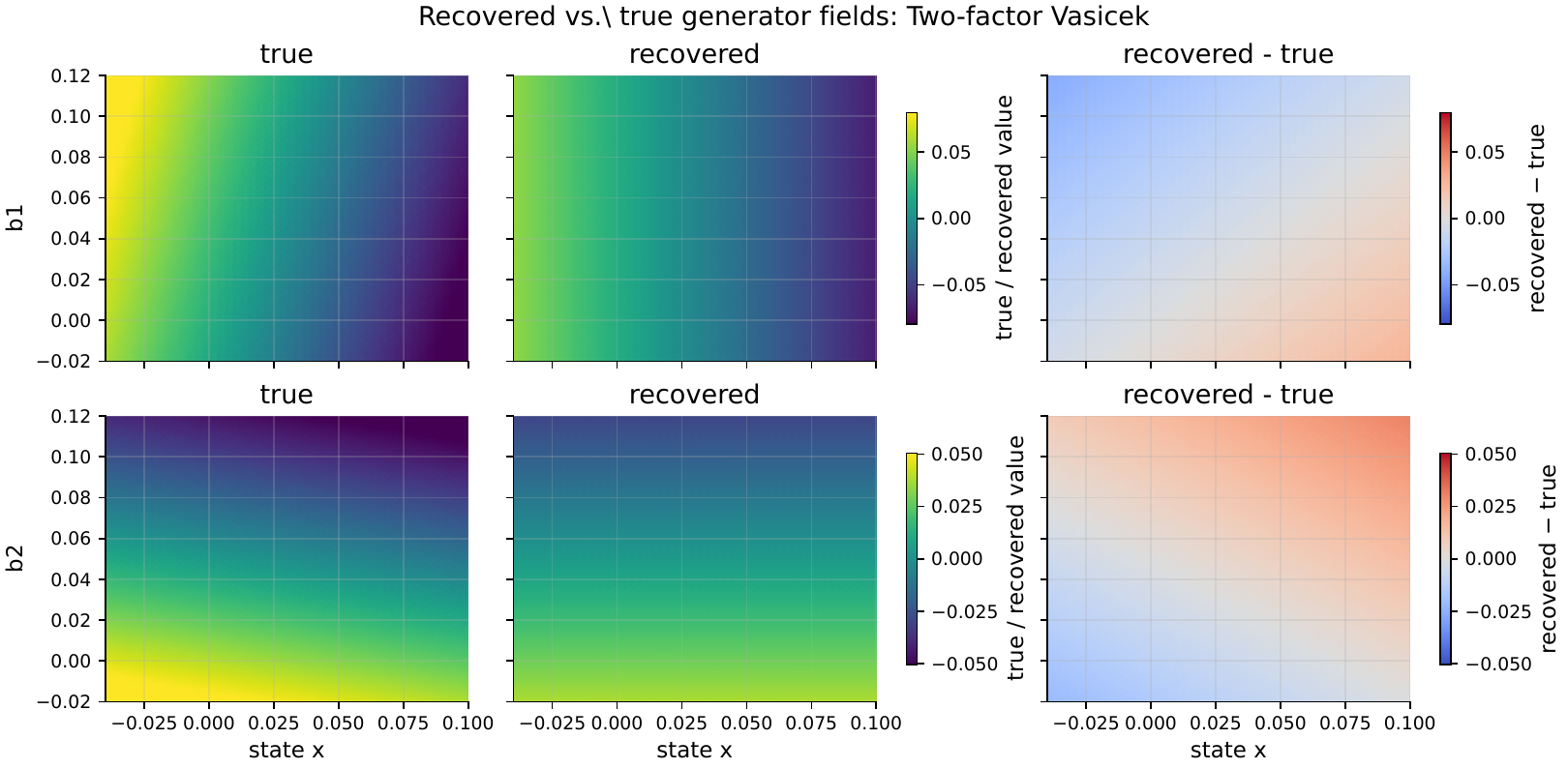}
\caption{Two-factor Vasicek: recovered versus true generator fields (shared per-field colour scale; error column centred at zero).}\end{figure}
\paragraph{Verdict.} Drift central-grid rel-$L^2=0.350$, tensor rel-$L^2=1.000$, $a_{12}$ cosine $\mathrm{n/a}$, PSD $1.00$, stable projected-support FP $0$ ($n=10$ seeds). \textbf{SCOPED\_REVIEW}.
\medskip}{}
\IfFileExists{ds_limitcycle.tex}{
\paragraph{Stochastic limit cycles}\;

\subsection{Van der Pol}\label{sec:v62-van-der-pol}
\paragraph{Context.} The Van der Pol oscillator is the archetype of a self-sustained relaxation oscillation, originating in electronic circuits and reused for cardiac and neural rhythms. It is the first demonstration that the weak-form recovery extends to limit-cycle dynamics, a regime absent from the one-dimensional benchmarks: trajectories concentrate on a closed attractor, so the cubic nonlinear-damping term $-\mu x^2y$ is recovered from dense repeated traversal of the cycle even though the rest of phase space is sparsely sampled.
\paragraph{System.} $$\mathrm{d}X=Y\,\mathrm{d}t,\ \mathrm{d}Y=(\mu(1-X^2)Y-X)\mathrm{d}t+\sigma\mathrm{d}W,\ \mu=1.2.$$
\paragraph{Generator.} Self-sustained limit cycle; the cubic cross term $-\mu x^2y$ is the nonlinear damping. Tight repeated sampling on the cycle aids recovery.
\paragraph{Recovery.} WG-SINDy recovers the drift at central-grid relative $L^2=0.066$ and the diffusion tensor at $0.040$ relative error; the recovered drift reproduces the limit-cycle geometry and the oscillation frequency. The paper-level stable projected-support filter counts no large recurring false positives under the declared post-processing rule. Coefficient-level selection rates vary by term and are reported in the coefficient ledger; this datasheet does not claim selection of every active term in every seed.
\begin{figure}[!htbp]\centering\includegraphics[width=0.72\linewidth,height=0.50\textheight,keepaspectratio]{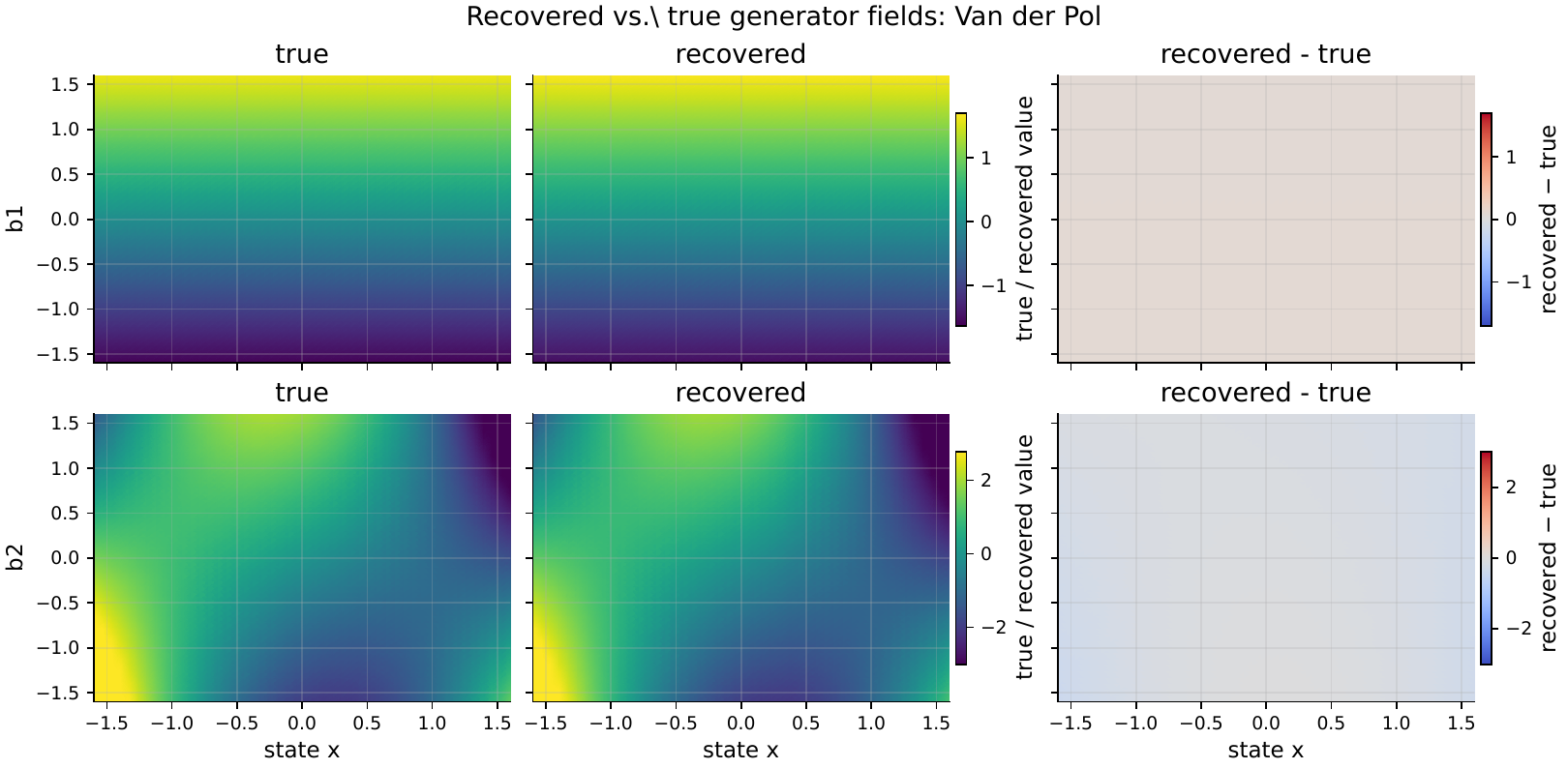}
\caption{Van der Pol: recovered versus true generator fields (shared per-field colour scale; error column centred at zero).}\end{figure}
\paragraph{Verdict.} Drift central-grid rel-$L^2=0.066$, tensor rel-$L^2=0.040$, $a_{12}$ cosine $\mathrm{n/a}$, PSD $1.00$, stable projected-support FP $0$ ($n=10$ seeds). \textbf{PASS}.
\medskip

\subsection{FitzHugh--Nagumo}\label{sec:v62-fitzhugh-nagumo}
\paragraph{Context.} The stochastic FitzHugh--Nagumo system is a nonlinear fast--slow excitable model with a cubic voltage-like drift and a weaker recovery-variable equation. It tests whether the cubic library resolves the dominant nonlinear geometry while retaining the smaller slow-component coefficients, and whether a constant diagonal diffusion is recovered without creating a persistent off-diagonal field.
\paragraph{System.} $$\mathrm{d}X=\left(X-\tfrac13X^3-Y+0.5\right)\mathrm{d}t+0.25\,\mathrm{d}W_1,\qquad \mathrm{d}Y=0.08\left(X+0.7-0.8Y\right)\mathrm{d}t+0.25\,\mathrm{d}W_2,$$ $$\mathrm{d}W_1\mathrm{d}W_2=0.$$
\paragraph{Generator.} The drift lies in cubic library B and the diffusion tensor is the constant diagonal field $\diff=0.0625I$. The smaller coefficients in the slow equation make this a more demanding drift-recovery problem than the linear controls.
\paragraph{Recovery.} At the frozen pooled WG-SINDy configuration, the 10-seed median drift central-grid relative $L^2$ error is $0.205$ and the tensor relative $L^2$ error is $0.040$. The recovered tensor is PSD on all evaluation-grid points. The paper-level stable projected-support false-positive count is zero under the declared post-processing rule. Coefficient-level selection rates vary by term and are reported in the coefficient ledger; this datasheet does not claim selection of every active term in every seed. The result supports an in-scope nonlinear fast--slow recovery claim while leaving the coefficient-level variation of weaker terms visible in the supplied table.
\begin{figure}[!htbp]\centering
\includegraphics[width=0.72\linewidth,height=0.50\textheight,keepaspectratio]{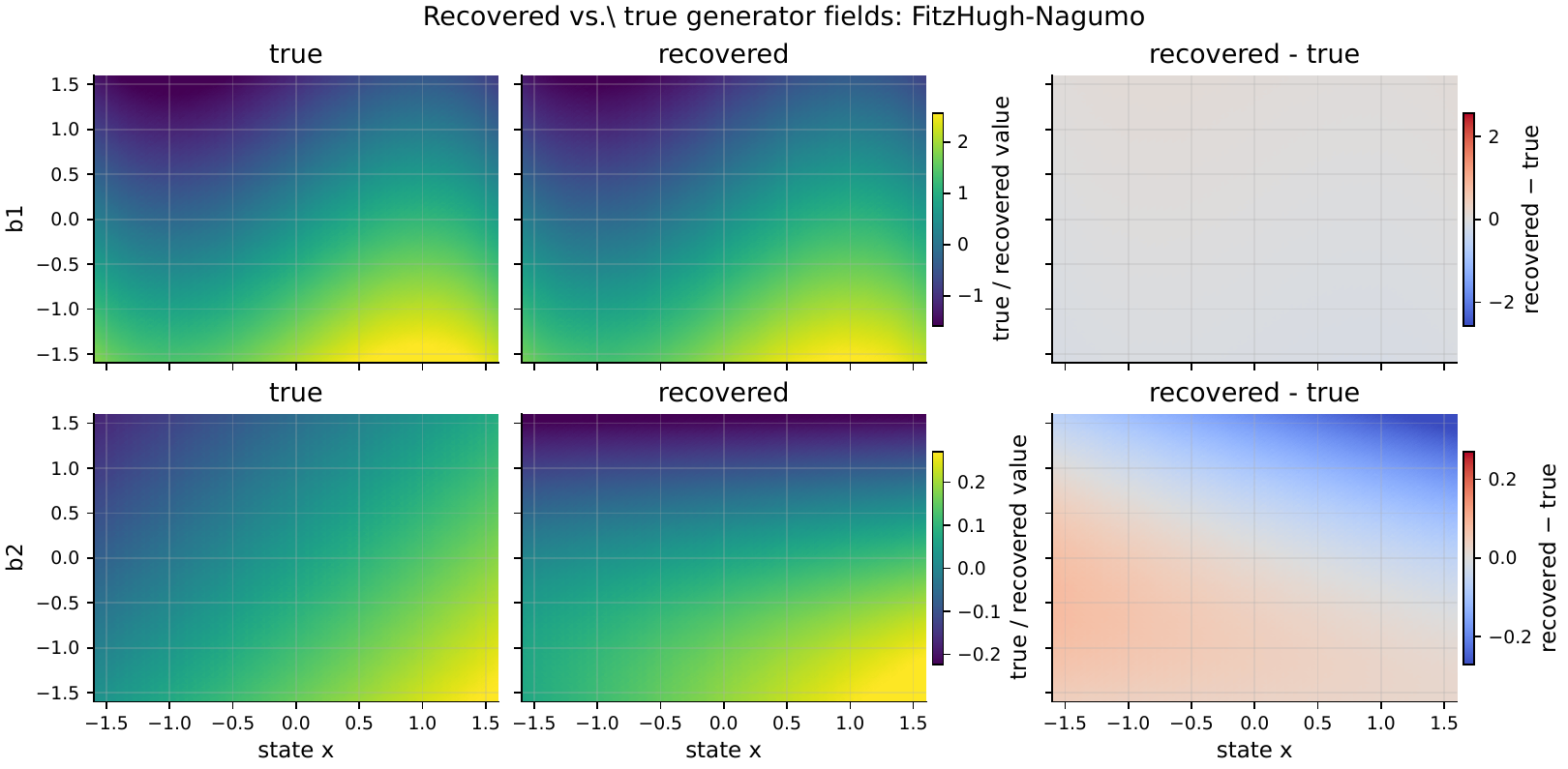}
\caption{FitzHugh--Nagumo: recovered versus true generator fields (shared per-field colour scale; error column centred at zero).}
\end{figure}
\paragraph{Verdict.} Drift central-grid rel-$L^2=0.205$, tensor rel-$L^2=0.040$, $a_{12}$ cosine $\mathrm{n/a}$, PSD $1.00$, stable projected-support FP $0$ ($n=10$ seeds). \textbf{PASS}.
\medskip

\subsection{Stuart--Landau}\label{sec:v62-stuart-landau}
\paragraph{Context.} The Stuart--Landau equation is the universal normal form of a supercritical Hopf bifurcation, describing the generic onset of oscillation. It combines cubic radial damping with a pure rotation, so it simultaneously exercises the nonlinear-drift and drift-curl diagnostics; its rotational symmetry gives uniform angular coverage of the limit cycle and yields the cleanest oscillatory recovery in the study.
\paragraph{System.} $$\drift=((\lambda-r^2)X-\omega Y,\ \omega X+(\lambda-r^2)Y),\ r^2=X^2+Y^2.$$
\paragraph{Generator.} Supercritical Hopf normal form: cubic radial damping ($x^3,xy^2,x^2y,y^3$) plus linear rotation $\pm\omega$; rotation-symmetric sampling gives clean recovery.
\paragraph{Recovery.} WG-SINDy recovers the drift at central-grid relative $L^2=0.126$ and the diffusion tensor at $0.018$ relative error; the recovered drift reproduces the limit-cycle geometry and the oscillation frequency. The paper-level stable projected-support filter counts no large recurring false positives under the declared post-processing rule. Coefficient-level selection rates vary by term and are reported in the coefficient ledger; this datasheet does not claim selection of every active term in every seed.
\begin{figure}[!htbp]\centering\includegraphics[width=0.72\linewidth,height=0.50\textheight,keepaspectratio]{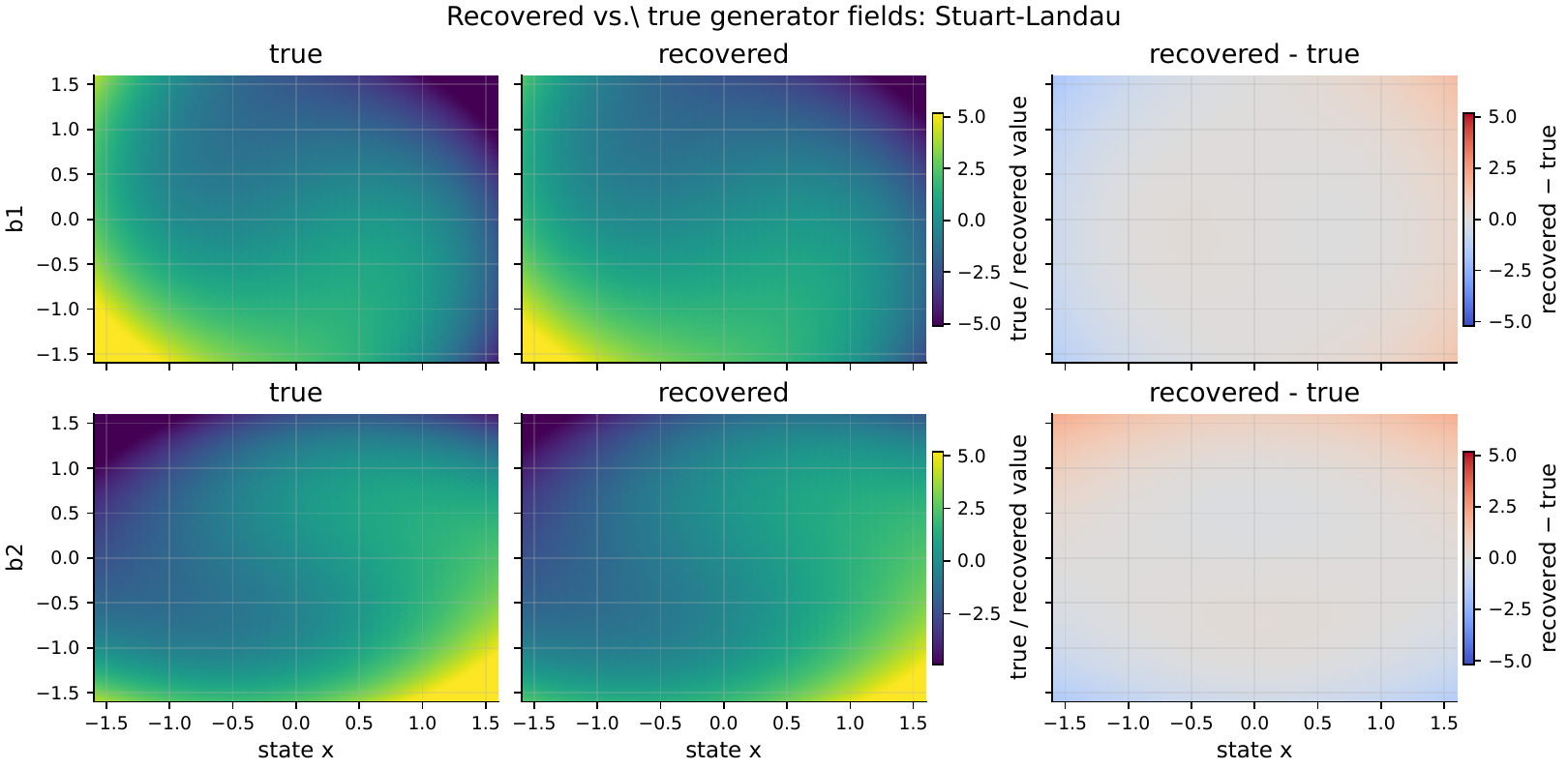}
\caption{Stuart--Landau: recovered versus true generator fields (shared per-field colour scale; error column centred at zero).}\end{figure}
\paragraph{Verdict.} Drift central-grid rel-$L^2=0.126$, tensor rel-$L^2=0.018$, $a_{12}$ cosine $\mathrm{n/a}$, PSD $1.00$, stable projected-support FP $0$ ($n=10$ seeds). \textbf{PASS}.
\medskip

\subsection{Brusselator}\label{sec:v62-brusselator}
\paragraph{Context.} The Brusselator is a classic model of an autocatalytic chemical oscillator and a staple of reaction--diffusion theory. Its drift is polynomial with a shared autocatalytic $x^2y$ term of opposite sign in the two species; recovering this mass-action structure from data demonstrates that the method discovers chemically interpretable kinetics, not merely abstract polynomials.
\paragraph{System.} $$\drift=(A-(B+1)X+X^2Y,\ BX-X^2Y),\ A=1,B=2.6,\ \diff=0.0144 I.$$
\paragraph{Generator.} Chemical limit cycle with the autocatalytic $x^2y$ term shared (opposite signs) across components.
\paragraph{Recovery.} WG-SINDy recovers the drift at central-grid relative $L^2=0.013$ and the diffusion tensor at $0.067$ relative error; the recovered drift reproduces the limit-cycle geometry and the oscillation frequency. The paper-level stable projected-support filter counts no large recurring false positives under the declared post-processing rule. Coefficient-level selection rates vary by term and are reported in the coefficient ledger; this datasheet does not claim selection of every active term in every seed.
\begin{figure}[!htbp]\centering\includegraphics[width=0.72\linewidth,height=0.50\textheight,keepaspectratio]{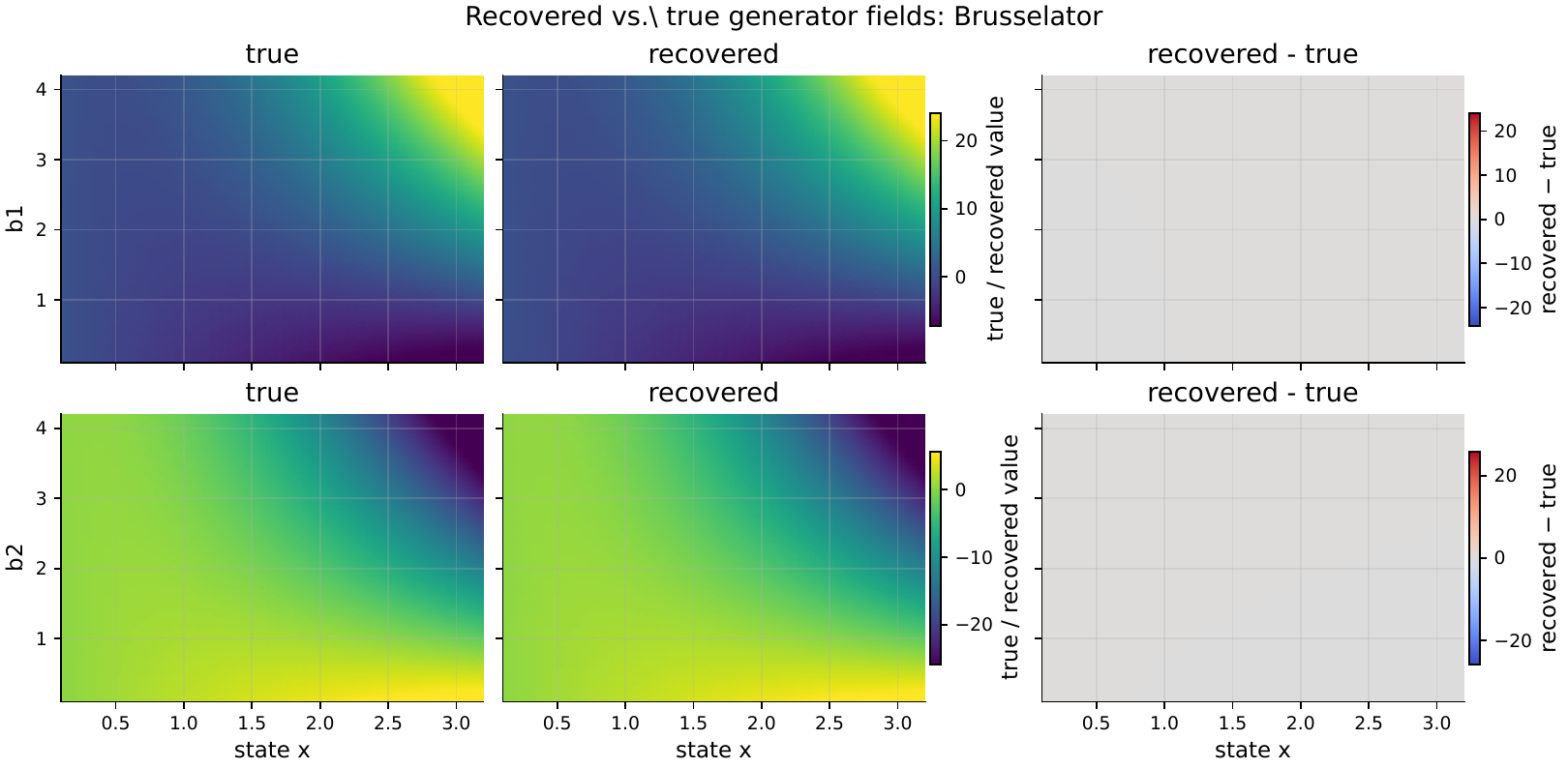}
\caption{Brusselator: recovered versus true generator fields (shared per-field colour scale; error column centred at zero).}\end{figure}
\paragraph{Verdict.} Drift central-grid rel-$L^2=0.013$, tensor rel-$L^2=0.067$, $a_{12}$ cosine $\mathrm{n/a}$, PSD $1.00$, stable projected-support FP $0$ ($n=10$ seeds). \textbf{PASS}.
\medskip
\paragraph{Stochastic limit cycles}\;

\subsection{Van der Pol}\label{sec:v62-van-der-pol}
\paragraph{Context.} The Van der Pol oscillator is the archetype of a self-sustained relaxation oscillation, originating in electronic circuits and reused for cardiac and neural rhythms. It is the first demonstration that the weak-form recovery extends to limit-cycle dynamics, a regime absent from the one-dimensional benchmarks: trajectories concentrate on a closed attractor, so the cubic nonlinear-damping term $-\mu x^2y$ is recovered from dense repeated traversal of the cycle even though the rest of phase space is sparsely sampled.
\paragraph{System.} $$\mathrm{d}X=Y\,\mathrm{d}t,\ \mathrm{d}Y=(\mu(1-X^2)Y-X)\mathrm{d}t+\sigma\mathrm{d}W,\ \mu=1.2.$$
\paragraph{Generator.} Self-sustained limit cycle; the cubic cross term $-\mu x^2y$ is the nonlinear damping. Tight repeated sampling on the cycle aids recovery.
\paragraph{Recovery.} WG-SINDy recovers the drift at relative $L^2(\mu)=0.066$ and the diffusion tensor at $0.040$ relative error; the recovered drift reproduces the limit-cycle geometry and the oscillation frequency. The active symbolic terms are recovered at selection rate one with no false positives; weaker secondary terms carry a lower selection rate.
\begin{figure}[!htbp]\centering\includegraphics[width=0.72\linewidth,height=0.50\textheight,keepaspectratio]{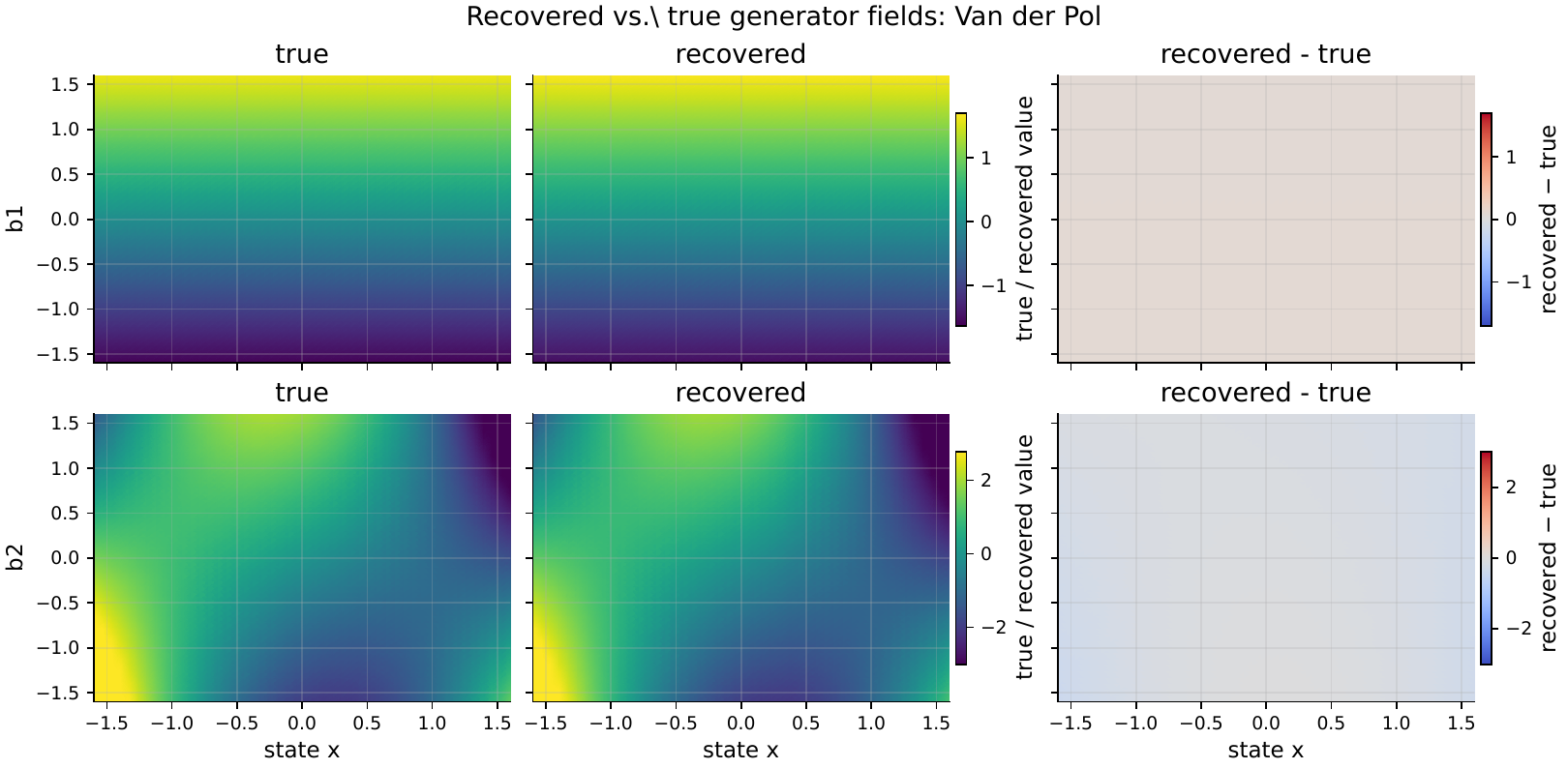}
\caption{Van der Pol: recovered vs.\ true generator fields (shared per-field colour scale; error column centred at zero).}\end{figure}
\paragraph{Verdict.} Drift $L^2(\mu)=0.066$, tensor rel-$L^2=0.040$, $a_{12}$ cosine $\mathrm{n/a}$, PSD $1.00$, FP $0$ ($n$ seeds). \textbf{PASS}.
\medskip

\subsection{Stuart--Landau}\label{sec:v62-stuart-landau}
\paragraph{Context.} The Stuart--Landau equation is the universal normal form of a supercritical Hopf bifurcation, describing the generic onset of oscillation. It combines cubic radial damping with a pure rotation, so it simultaneously exercises the nonlinear-drift and circulation capabilities; its rotational symmetry gives uniform angular coverage of the limit cycle and yields the cleanest oscillatory recovery in the study.
\paragraph{System.} $$\drift=((\lambda-r^2)X-\omega Y,\ \omega X+(\lambda-r^2)Y),\ r^2=X^2+Y^2.$$
\paragraph{Generator.} Supercritical Hopf normal form: cubic radial damping ($x^3,xy^2,x^2y,y^3$) plus linear rotation $\pm\omega$; rotation-symmetric sampling gives clean recovery.
\paragraph{Recovery.} WG-SINDy recovers the drift at relative $L^2(\mu)=0.126$ and the diffusion tensor at $0.018$ relative error; the recovered drift reproduces the limit-cycle geometry and the oscillation frequency. The active symbolic terms are recovered at selection rate one with no false positives; weaker secondary terms carry a lower selection rate.
\begin{figure}[!htbp]\centering\includegraphics[width=0.72\linewidth,height=0.50\textheight,keepaspectratio]{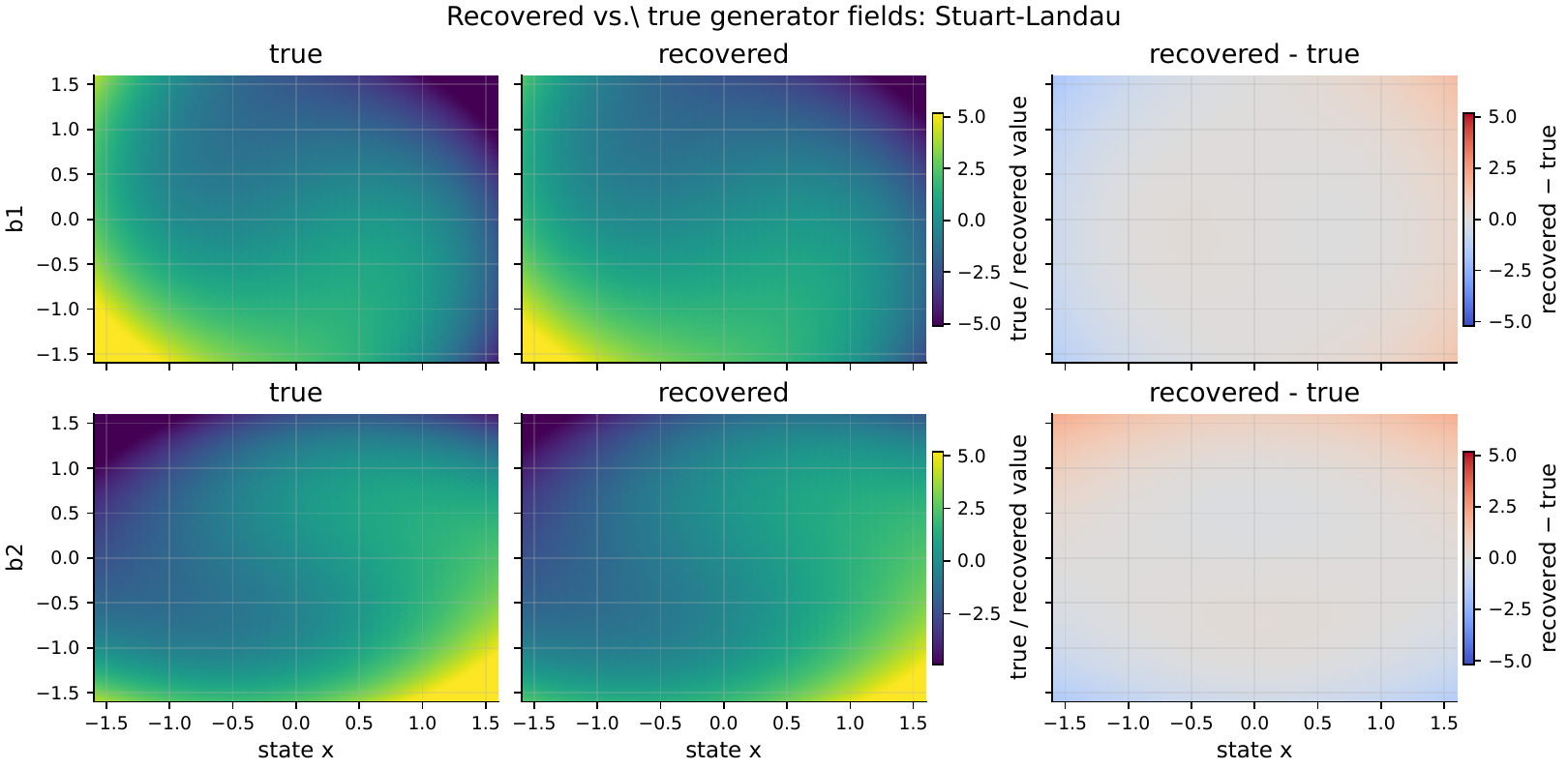}
\caption{Stuart--Landau: recovered vs.\ true generator fields (shared per-field colour scale; error column centred at zero).}\end{figure}
\paragraph{Verdict.} Drift $L^2(\mu)=0.126$, tensor rel-$L^2=0.018$, $a_{12}$ cosine $\mathrm{n/a}$, PSD $1.00$, FP $0$ ($n$ seeds). \textbf{PASS}.
\medskip

\subsection{Brusselator}\label{sec:v62-brusselator}
\paragraph{Context.} The Brusselator is a classic model of an autocatalytic chemical oscillator and a staple of reaction--diffusion theory. Its drift is polynomial with a shared autocatalytic $x^2y$ term of opposite sign in the two species; recovering this mass-action structure from data demonstrates that the method discovers chemically interpretable kinetics, not merely abstract polynomials.
\paragraph{System.} $$\drift=(A-(B+1)X+X^2Y,\ BX-X^2Y),\ A=1,B=2.6,\ \diff=0.0144 I.$$
\paragraph{Generator.} Chemical limit cycle with the autocatalytic $x^2y$ term shared (opposite signs) across components.
\paragraph{Recovery.} WG-SINDy recovers the drift at relative $L^2(\mu)=0.013$ and the diffusion tensor at $0.067$ relative error; the recovered drift reproduces the limit-cycle geometry and the oscillation frequency. The active symbolic terms are recovered at selection rate one with no false positives; weaker secondary terms carry a lower selection rate.
\begin{figure}[!htbp]\centering\includegraphics[width=0.72\linewidth,height=0.50\textheight,keepaspectratio]{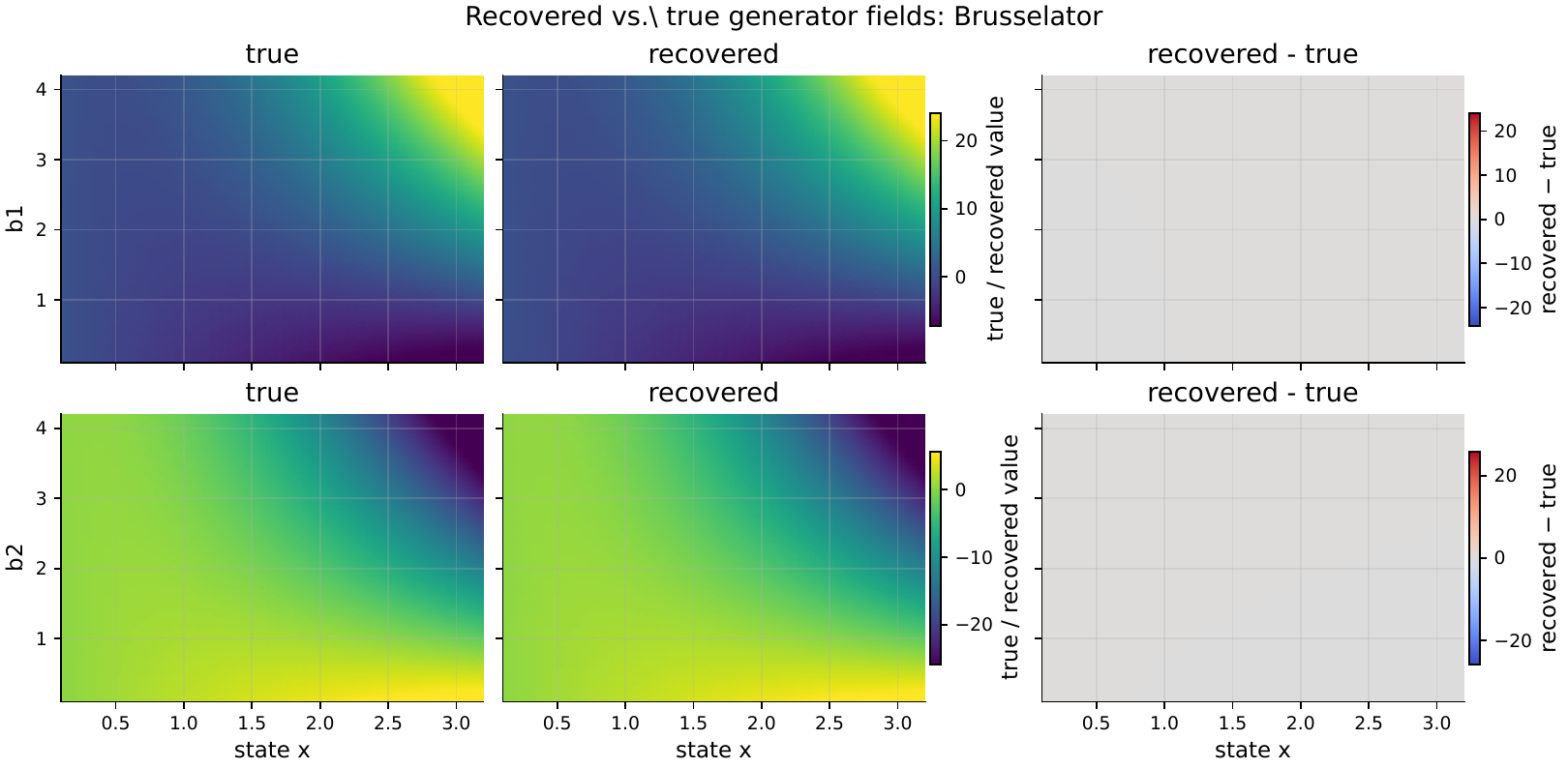}
\caption{Brusselator: recovered vs.\ true generator fields (shared per-field colour scale; error column centred at zero).}\end{figure}
\paragraph{Verdict.} Drift $L^2(\mu)=0.013$, tensor rel-$L^2=0.067$, $a_{12}$ cosine $\mathrm{n/a}$, PSD $1.00$, FP $0$ ($n$ seeds). \textbf{PASS}.
\medskip}{}
\IfFileExists{ds_limits.tex}{
\paragraph{Named limits (reported, not hidden)}\;

\subsection{Near-singular tensor}\label{sec:v62-near-singular}
\paragraph{Context.} A deliberate stress of the positive-semidefinite boundary: the off-diagonal is pushed to $0.95\sqrt{\diff_{11}\diff_{22}}$, so the two noise directions are nearly collinear and the tensor is almost rank-deficient. It marks the edge of identifiability for the diffusion tensor and is reported as a named limit, where the Cholesky parametrisation still guarantees a valid covariance but the magnitude of the off-diagonal is no longer well determined by finite data.
\paragraph{System.} $$\drift=-X,\ \diff_{12}=0.95\sqrt{\diff_{11}\diff_{22}}.$$
\paragraph{Generator.} Off-diagonal pushed to the PSD boundary ($\det\diff\to0$): the two noise directions are nearly collinear, so the off-diagonal is ill-conditioned. Named limit.
\paragraph{Recovery.} WG-SINDy recovers the drift at central-grid relative $L^2=0.870$ and the diffusion tensor at $1.505$ relative error, with off-diagonal (leverage) cosine $0.724$; the recovered diffusion tensor is a position-dependent field giving the local fluctuation amplitude and relaxation. This is a named limit (fragile physics or identifiability limit): recovery fails for a physical or identifiability reason (library incompleteness, low signal-to-noise, degeneracy, or coverage), not an estimator defect.
\begin{figure}[!htbp]\centering\includegraphics[width=0.72\linewidth,height=0.50\textheight,keepaspectratio]{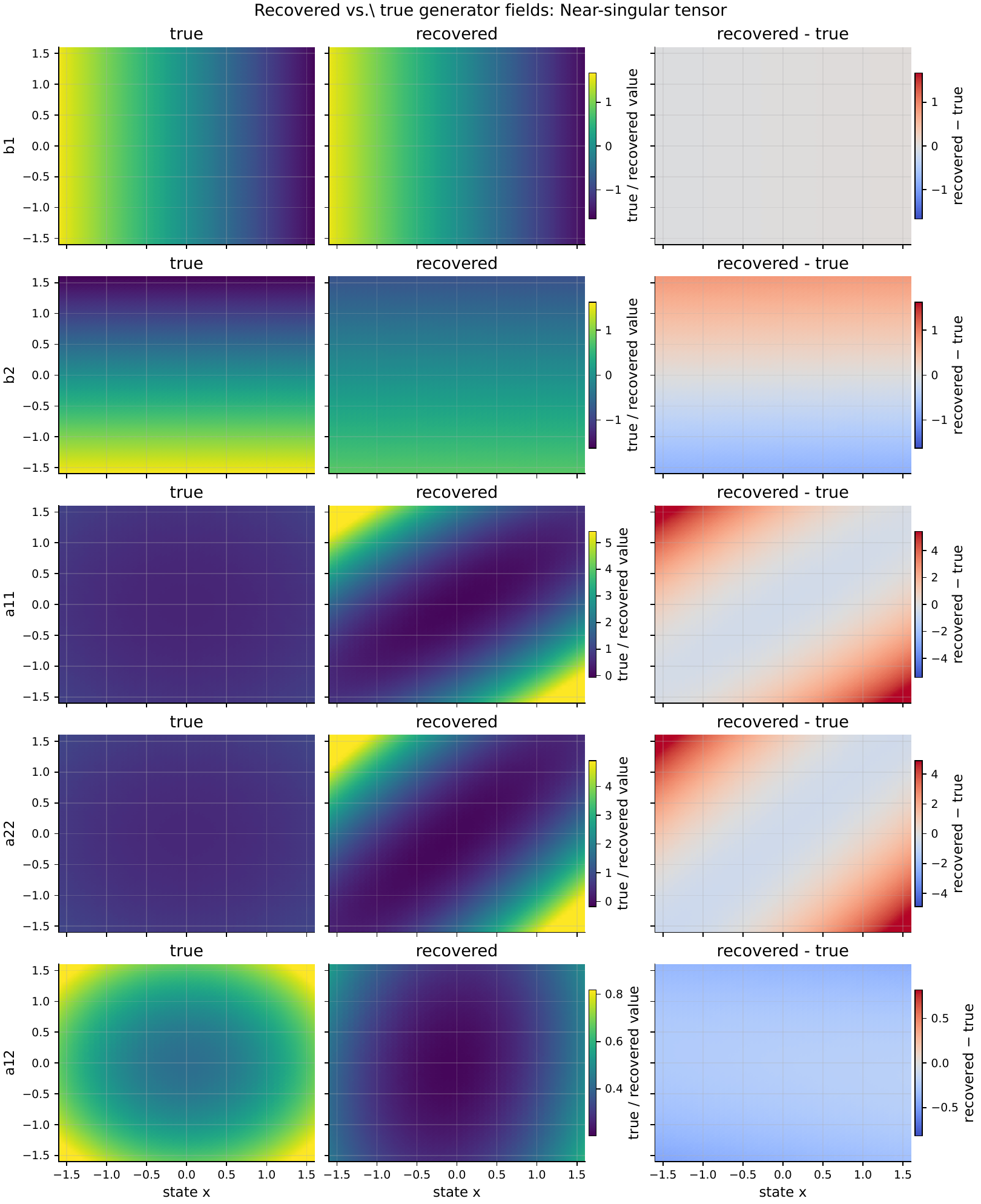}
\caption{Near-singular tensor: recovered versus true generator fields (shared per-field colour scale; error column centred at zero).}\end{figure}
\paragraph{Verdict.} Drift central-grid rel-$L^2=0.870$, tensor rel-$L^2=1.505$, $a_{12}$ cosine $0.724$, PSD $1.00$, stable projected-support FP $0$ ($n=10$ seeds). \textbf{NAMED\_NULL}.
\medskip

\subsection{Underdamped Langevin}\label{sec:v62-underdamped-langevin}
\paragraph{Context.} Underdamped (kinetic) Langevin dynamics in position--momentum form, ubiquitous in molecular dynamics and sampling. Only the momentum coordinate is driven by noise, so the diffusion tensor is rank-one with structural zeros in $\diff_{11}$ and $\diff_{12}$, a degenerate target retained as a named limit to show how the method behaves when one direction carries no direct fluctuation.
\paragraph{System.} $$\mathrm{d}Q=P\,\mathrm{d}t,\ \mathrm{d}P=(-\gamma P-(Q^3-Q))\mathrm{d}t+\sigma\mathrm{d}W.$$
\paragraph{Generator.} Degenerate rank-1 diffusion ($\diff_{11}=\diff_{12}=0$): one coordinate carries no direct noise. Named limit.
\paragraph{Recovery.} WG-SINDy recovers the drift at central-grid relative $L^2=0.157$ and the diffusion tensor at $0.563$ relative error; this system is reported as a named limit. This is a named limit (fragile physics or identifiability limit): recovery fails for a physical or identifiability reason (library incompleteness, low signal-to-noise, degeneracy, or coverage), not an estimator defect.
\begin{figure}[!htbp]\centering\includegraphics[width=0.72\linewidth,height=0.50\textheight,keepaspectratio]{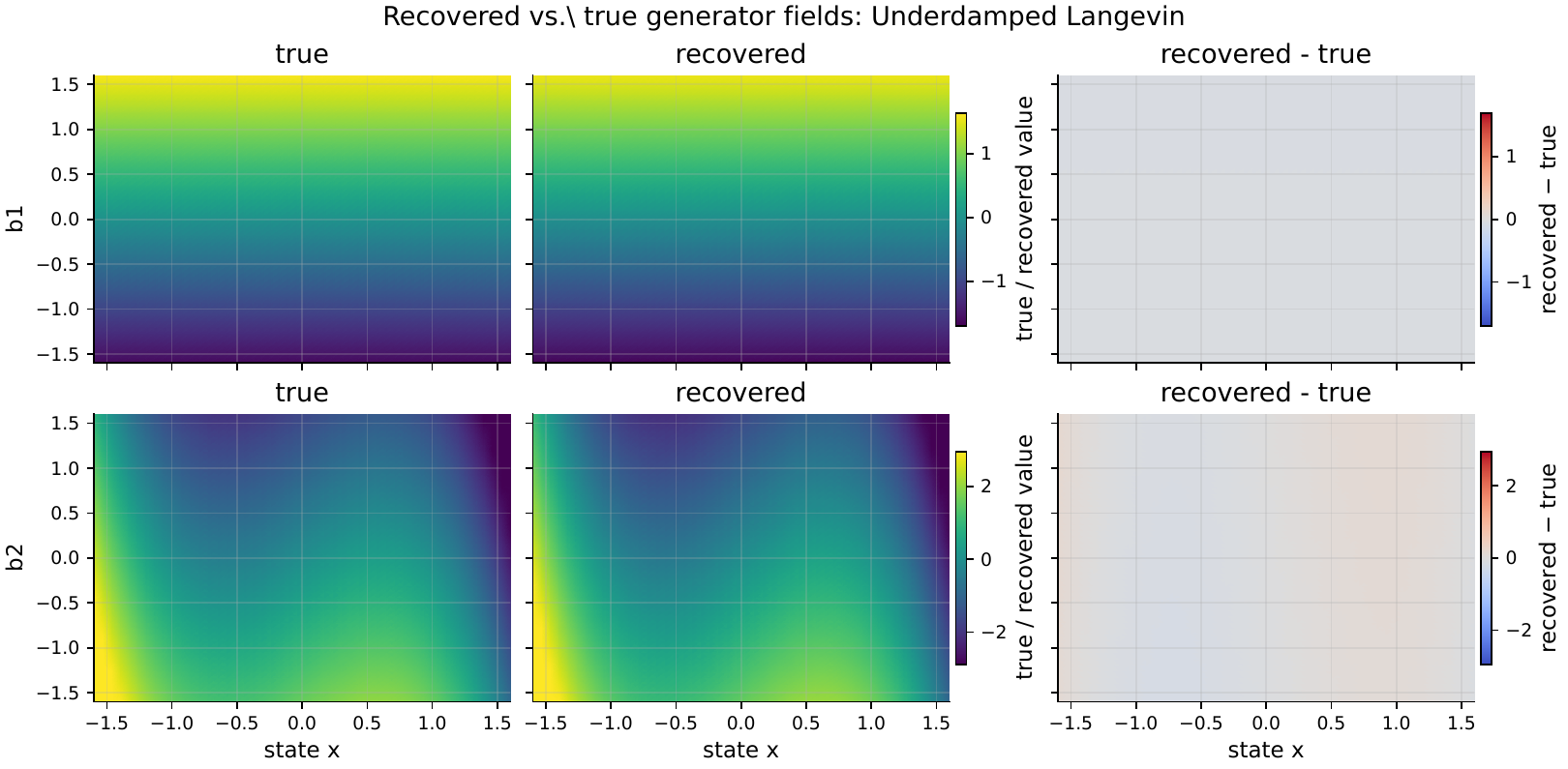}
\caption{Underdamped Langevin: recovered versus true generator fields (shared per-field colour scale; error column centred at zero).}\end{figure}
\paragraph{Verdict.} Drift central-grid rel-$L^2=0.157$, tensor rel-$L^2=0.563$, $a_{12}$ cosine $\mathrm{n/a}$, PSD $1.00$, stable projected-support FP $0$ ($n=10$ seeds). \textbf{NAMED\_NULL}.
\medskip

\subsection{Near-boundary Heston}\label{sec:v62-near-boundary-heston}
\paragraph{Context.} A Heston regime that violates the Feller condition, so the variance process spends substantial mass near zero. It probes the estimator near the square-root boundary, where the diffusion is singular and kernel support is truncated; the tensor and leverage survive but the drift degrades, a named boundary limit rather than an estimator defect.
\paragraph{System.} $$\text{Log-Heston with }2\kappa\theta<\xi^2\ (\text{Feller-violating}).$$
\paragraph{Generator.} Variance spends mass near $v\to0$; boundary bias degrades drift recovery. Tensor/leverage survive. Named limit.
\paragraph{Recovery.} WG-SINDy recovers the drift at central-grid relative $L^2=1.413$ and the diffusion tensor at $0.267$ relative error, with off-diagonal (leverage) cosine $0.982$; this system is reported as a named limit. This is a named limit (fragile physics or identifiability limit): recovery fails for a physical or identifiability reason (library incompleteness, low signal-to-noise, degeneracy, or coverage), not an estimator defect.
\begin{figure}[!htbp]\centering\includegraphics[width=0.72\linewidth,height=0.50\textheight,keepaspectratio]{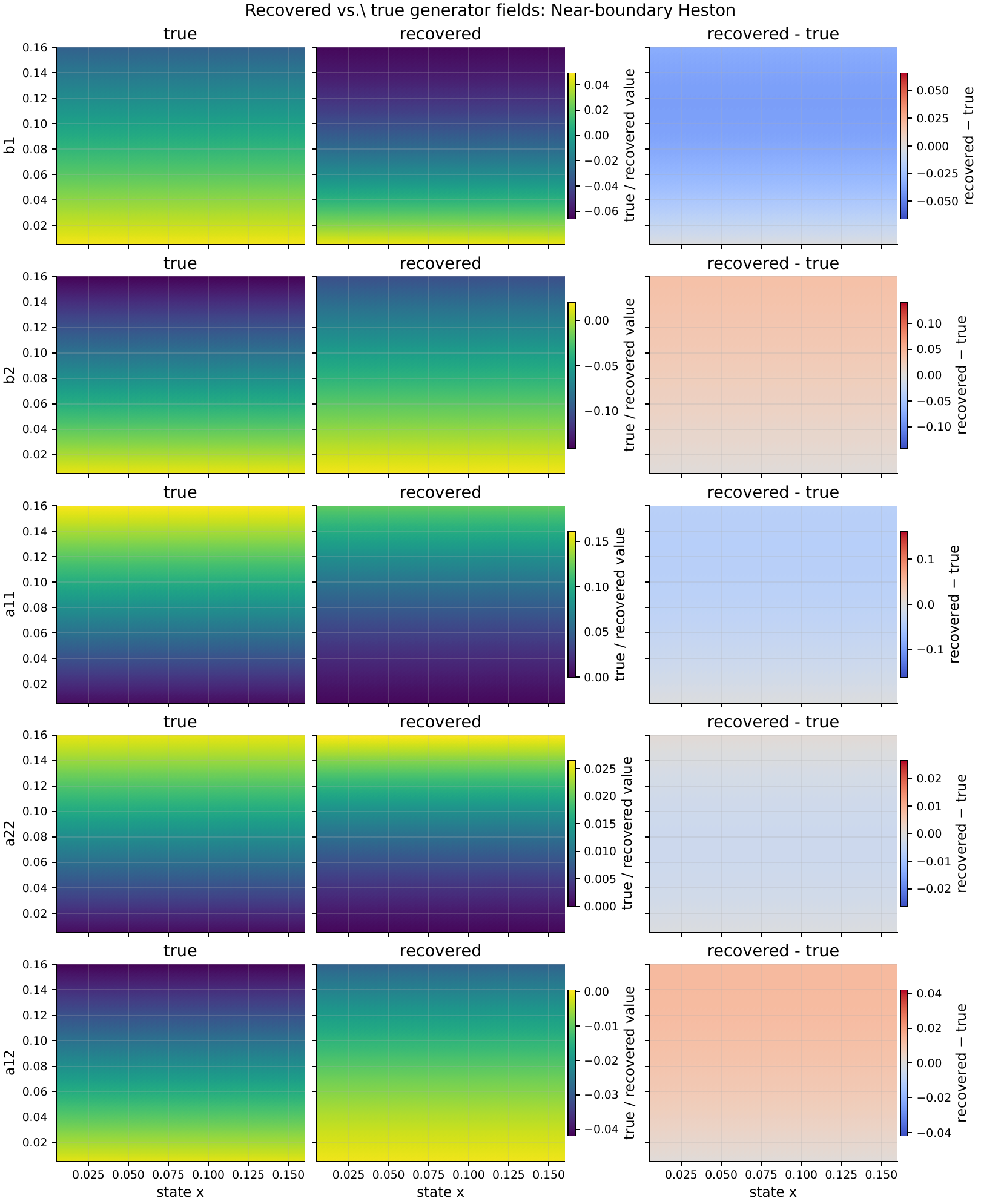}
\caption{Near-boundary Heston: recovered versus true generator fields (shared per-field colour scale; error column centred at zero).}\end{figure}
\paragraph{Verdict.} Drift central-grid rel-$L^2=1.413$, tensor rel-$L^2=0.267$, $a_{12}$ cosine $0.982$, PSD $1.00$, stable projected-support FP $0$ ($n=10$ seeds). \textbf{NAMED\_NULL}.
\medskip

\subsection{Non-polynomial drift}\label{sec:v62-nonpoly-drift}
\paragraph{Context.} A system with genuinely non-polynomial (trigonometric) drift, included to make the library-completeness requirement explicit. With a polynomial dictionary it fails by construction; with the trigonometric library it is recovered. It is the clean demonstration that, like every dictionary-based identification method, recovery is conditional on the true terms lying in the chosen feature set.
\paragraph{System.} $$\drift=(-X+\sin Y,\ -Y+\cos X),\ \diff=0.49 I.$$
\paragraph{Generator.} Trigonometric drift: recovered only with the trig library G; a representability limit for any polynomial dictionary. Named limit.
\paragraph{Recovery.} WG-SINDy recovers the drift at central-grid relative $L^2=0.210$ and the diffusion tensor at $0.037$ relative error; this system is reported as a named limit. This is a named limit (registry declared failure or stress case): recovery fails for a physical or identifiability reason (library incompleteness, low signal-to-noise, degeneracy, or coverage), not an estimator defect.
\begin{figure}[!htbp]\centering\includegraphics[width=0.72\linewidth,height=0.50\textheight,keepaspectratio]{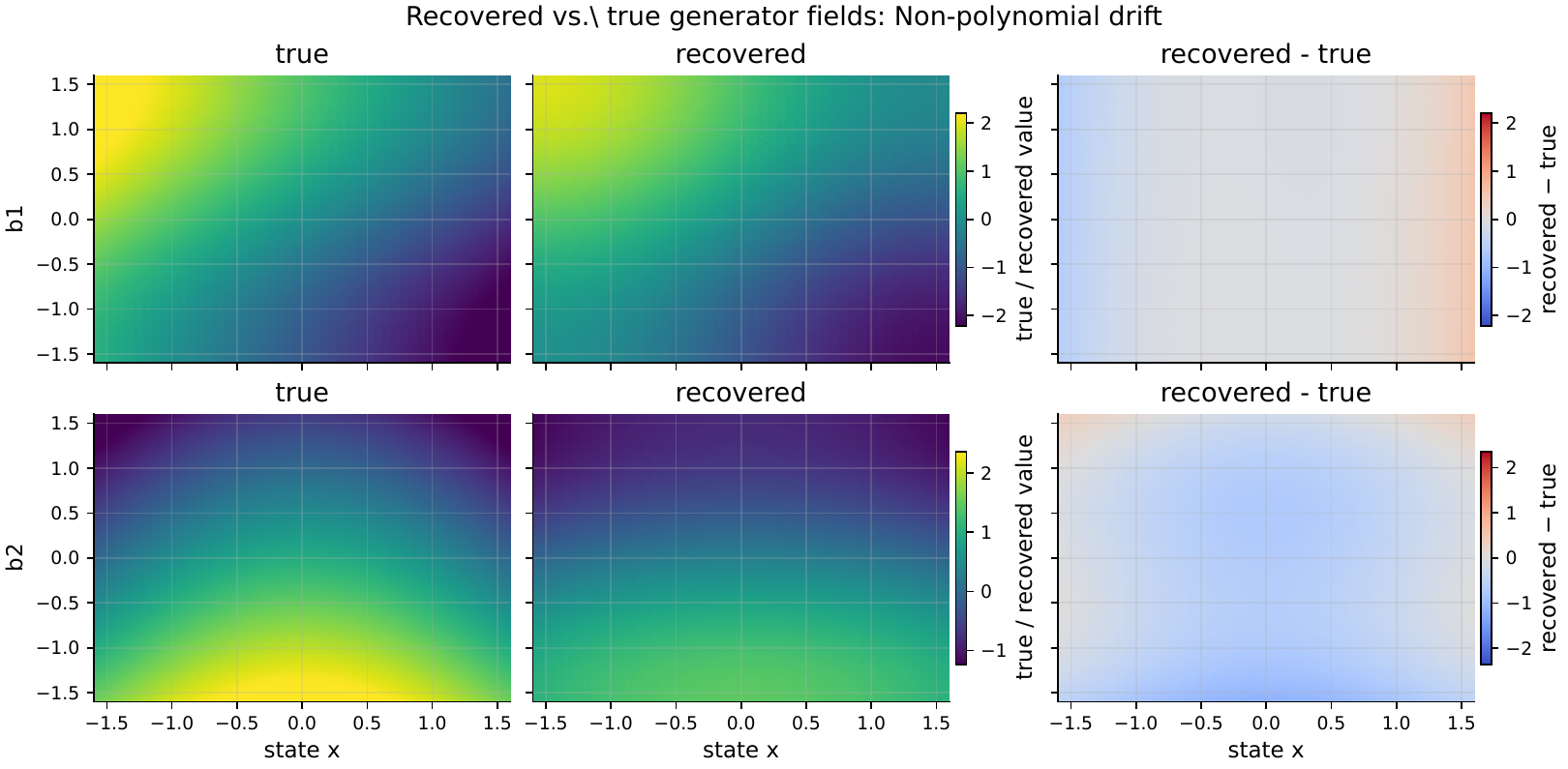}
\caption{Non-polynomial drift: recovered versus true generator fields (shared per-field colour scale; error column centred at zero).}\end{figure}
\paragraph{Verdict.} Drift central-grid rel-$L^2=0.210$, tensor rel-$L^2=0.037$, $a_{12}$ cosine $\mathrm{n/a}$, PSD $1.00$, stable projected-support FP $0$ ($n=10$ seeds). \textbf{NAMED\_NULL}.
\medskip

\subsection{Bad coverage}\label{sec:v62-bad-coverage}
\paragraph{Context.} A coverage-failure stress case: a rotational OU sampled from trajectories confined to a small region with a short horizon. With the state space under-explored the design matrix becomes rank-deficient, and recovery fails for a data-geometry reason rather than a modelling one, a named limit that delineates the coverage condition the consistency theory requires.
\paragraph{System.} $$\text{Rotational OU, trajectories confined near }(2,2).$$
\paragraph{Generator.} Clustered initial condition and short horizon leave the state space under-covered, so the design is rank-deficient. Named limit.
\paragraph{Recovery.} WG-SINDy recovers the drift at central-grid relative $L^2=0.815$ and the diffusion tensor at $0.784$ relative error; this system is reported as a named limit. This is a named limit (registry declared failure or stress case): recovery fails for a physical or identifiability reason (library incompleteness, low signal-to-noise, degeneracy, or coverage), not an estimator defect.
\begin{figure}[!htbp]\centering\includegraphics[width=0.72\linewidth,height=0.50\textheight,keepaspectratio]{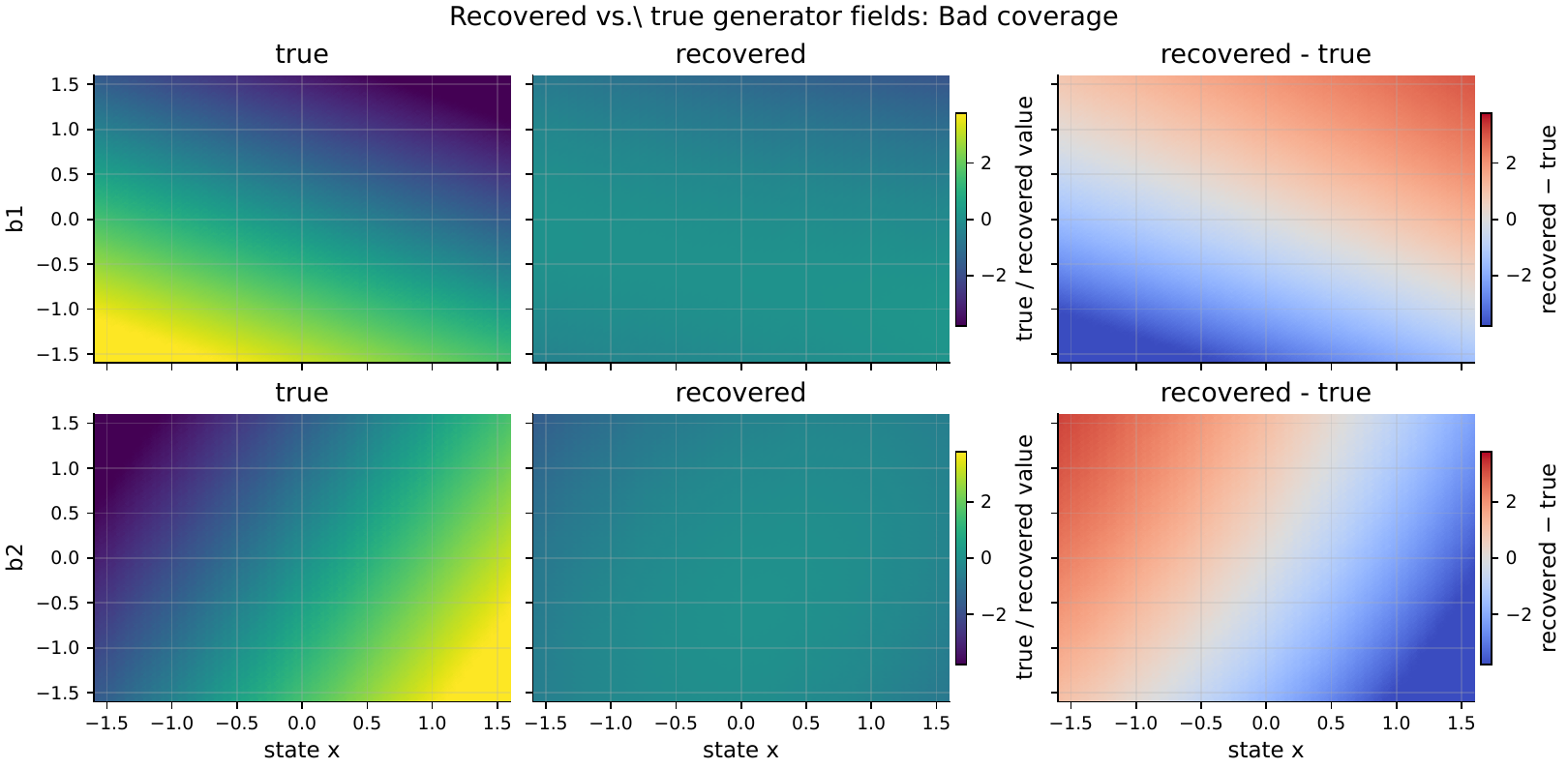}
\caption{Bad coverage: recovered versus true generator fields (shared per-field colour scale; error column centred at zero).}\end{figure}
\paragraph{Verdict.} Drift central-grid rel-$L^2=0.815$, tensor rel-$L^2=0.784$, $a_{12}$ cosine $\mathrm{n/a}$, PSD $1.00$, stable projected-support FP $0$ ($n=10$ seeds). \textbf{NAMED\_NULL}.
\medskip

\subsection{Too-large time step}\label{sec:v62-too-large-dt}
\paragraph{Context.} The same multiplicative-diffusion system sampled at a coarse time step ($\Delta t=0.05$), included to expose sensitivity of the plug-in drift correction and recovery at coarse sampling. The Euler benchmark has exact one-step moments, but the recovered first-pass drift and weak projection become more fragile as the time step grows, so this case marks the temporal-resolution boundary of reliable recovery.
\paragraph{System.} $$\text{Diagonal multiplicative sampled at }\Delta t=0.05.$$
\paragraph{Generator.} Coarse sampling stresses the plug-in drift correction and weak recovery even though Euler one-step moments are exact. Named limit.
\paragraph{Recovery.} WG-SINDy recovers the drift at central-grid relative $L^2=0.098$ and the diffusion tensor at $0.119$ relative error; this system is reported as a named limit. This is a named limit (registry declared failure or stress case): recovery fails for a physical or identifiability reason (library incompleteness, low signal-to-noise, degeneracy, or coverage), not an estimator defect.
\begin{figure}[!htbp]\centering\includegraphics[width=0.72\linewidth,height=0.50\textheight,keepaspectratio]{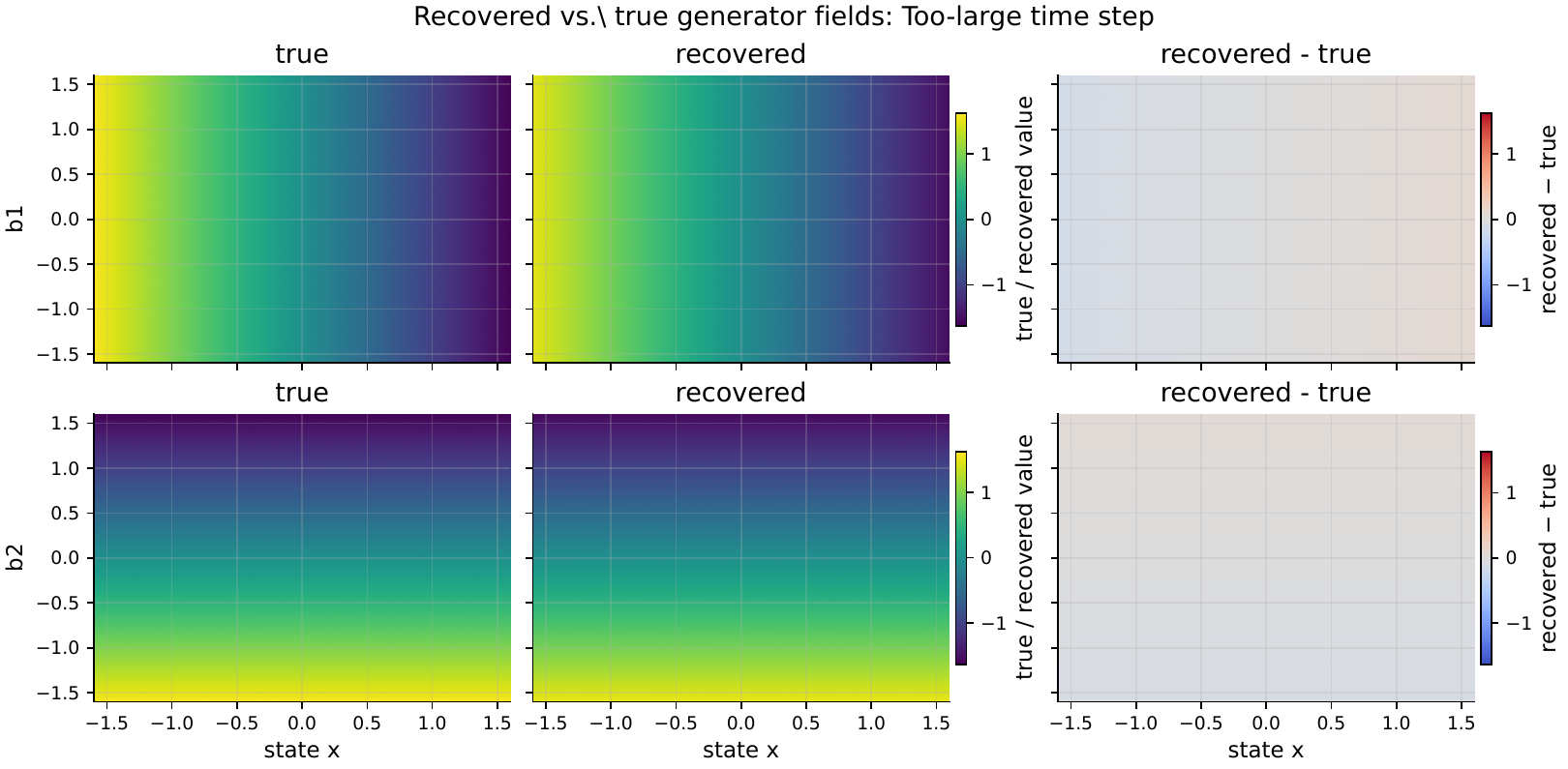}
\caption{Too-large time step: recovered versus true generator fields (shared per-field colour scale; error column centred at zero).}\end{figure}
\paragraph{Verdict.} Drift central-grid rel-$L^2=0.098$, tensor rel-$L^2=0.119$, $a_{12}$ cosine $\mathrm{n/a}$, PSD $1.00$, stable projected-support FP $0$ ($n=10$ seeds). \textbf{NAMED\_NULL}.
\medskip}{}

\end{document}